\documentclass[a4paper,11pt]{article}
\pdfoutput=1
\usepackage{jheppub}
\usepackage[utf8]{inputenc}
\usepackage{lmodern}
\usepackage{amsmath,amssymb,amsfonts}
\usepackage{mathrsfs}
\usepackage{booktabs,tabularx,array}
\usepackage{microtype}
\usepackage{bm}
\usepackage{placeins}
\usepackage{xcolor}
\numberwithin{equation}{section}

\makeatletter
\newcommand*\rel@kern[1]{\kern#1\dimexpr\macc@kerna}
\newcommand*\widebar[1]{%
  \begingroup
  \def\mathaccent##1##2{%
    \rel@kern{0.8}%
    \overline{\rel@kern{-0.8}\macc@nucleus\rel@kern{0.2}}%
    \rel@kern{-0.2}%
  }%
  \macc@depth\@ne
  \let\math@bgroup\@empty \let\math@egroup\macc@set@skewchar
  \mathsurround\z@ \frozen@everymath{\mathgroup\macc@group\relax}%
  \macc@set@skewchar\relax
  \let\mathaccentV\macc@nested@a
  \macc@nested@a\relax111{#1}%
  \endgroup
}
\makeatother

\preprint{
\begin{minipage}{5cm}
\small
\flushright
KYUSHU-HET-384
\end{minipage}}

\title{Geometric Bounds on Axion-Photon Couplings in Non-Supersymmetric Heterotic Compactifications}

\author{Yuichi Koga$^{1}$ and}
\author{Hajime Otsuka$^{2,3}$}
\affiliation{
$^1$Institute for Advanced Study, Kyushu University, 744 Motooka, Nishi-ku, Fukuoka 819-0395, Japan
}
\affiliation{
$^2$Department of Physics, Kyushu University, 744 Motooka, Nishi-ku, Fukuoka 819-0395, Japan
}
\affiliation{
$^3$Quantum and Spacetime Research Institute (QuaSR), Kyushu University, 744 Motooka, Nishi-ku, Fukuoka 819-0395, Japan}

\emailAdd{koga.yuichi@phys.kyushu-u.ac.jp}
\emailAdd{otsuka.hajime@phys.kyushu-u.ac.jp}

\abstract{
We study the photon couplings of QCD axions and axion-like particles in non-supersymmetric heterotic string theory on Calabi-Yau threefolds with line bundles. After the St\"uckelberg mechanism and anomaly matching are taken into account, the QCD axion direction is determined by the QCD axion-coupling coefficients together with the axion kinetic matrix. We focus on the electromagnetic-to-color axion coupling ratio, $E/N$. 
When the physical electromagnetic and QCD axion-coupling vectors are proportional with the ratio fixed by the gauge levels, the QCD axion has $E/N=8/3$ for $k_3=k_2=1$ and $k_Y=5/3$, independently of the axion kinetic matrix. 
It is found that direct flux breaking of the visible $SO(16)_{\rm v}$
can produce independent QCD and electromagnetic axion-coupling
vectors for the same level assignment. 
The Hermitian Yang-Mills equations and the heterotic Bianchi identity then restrict the departure from $8/3$. 
We derive analytic bounds for backgrounds with and without
five-brane charge under assumptions about the axion kinetic
terms and low-energy matching. 
An $SU(5)$-like model and a direct-flux Standard-Model-like model
illustrate proportional and independent Green-Schwarz coupling
vectors, respectively. 
For axions with no QCD coupling, we derive a geometric upper bound on their photon couplings and evaluate it in an explicit compactification.
}

\makeatletter
\gdef\@fpheader{}
\makeatother

\begin{document}

\maketitle

\section{Introduction}
\label{sec:introduction}

Axions are generic in string compactifications because higher-dimensional antisymmetric tensor fields lead to periodic scalars in four dimensions~\cite{Witten:1984dg,Choi:1985bz,Svrcek:2006yi}.
Their couplings to QCD and electromagnetism are determined by the
ten-dimensional gauge embedding, the internal gauge background, and the
St\"uckelberg mixings that remove some axionic directions. Measurements or
bounds on an axion mass and its photon coupling can therefore probe features
of the compactification that are not visible in the four-dimensional gauge
spectrum by itself.

Most quantitative studies of axion couplings have focused on supersymmetric
compactifications. Low-energy searches for supersymmetry test whether
supersymmetry survives to low energies, but they do not by themselves
determine whether the underlying string compactification was supersymmetric
or intrinsically non-supersymmetric. This motivates a complementary
question: can axion observables help distinguish these two classes of string
vacua? Recent studies have shown that the gauge embedding can restrict the axion-photon coupling in perturbative heterotic
compactifications \cite{Agrawal:2024ejr,Reig:2025dqb}. We therefore study how
the axion-photon coupling is constrained in a tachyon-free non-supersymmetric
heterotic string theory.

In the four-dimensional low-energy theory, we denote the QCD and
electromagnetic axion-coupling vectors by $n$ and $e$, respectively.
These coefficients include the effects of the St\"uckelberg
mechanism and low-energy anomaly matching. 
When QCD provides the dominant axion potential, the electromagnetic-to-color coupling ratio of the QCD axion is
\begin{align}
 \left(\frac EN\right)_{\rm eff}
 =\frac{e^TG^{-1}n}{n^TG^{-1}n}.
 \label{eq:intro-effective-ratio}
\end{align}
Here $G$ is the positive-definite axion kinetic matrix.\footnote{For axion kinetic and St\"uckelberg mixing, see, e.g. 
Refs.~\cite{Agrawal:2017cmd,Choi:1909.11685,Fraser:1910.11349}.} 
If the two coupling vectors satisfy $e=r n$, Eq.~\eqref{eq:intro-effective-ratio} is independent of $G$ and
gives $(E/N)_{\rm eff}=r$.

For the class of supersymmetric heterotic compactifications considered in
Ref.~\cite{Agrawal:2024ejr}, the Standard Model gauge factors are embedded in
one ten-dimensional $E_8$ factor and couple to the same axion combination.
The QCD and electromagnetic coefficients then satisfy
$e=[(k_2+k_Y)/k_3]n$, so Eq.~\eqref{eq:intro-effective-ratio} becomes
\begin{align}
 \left(\frac EN\right)_{\rm eff}
 =\frac{k_2+k_Y}{k_3},
 \label{eq:intro-susy-reference-ratio}
\end{align}
where $k_3$, $k_2$, and $k_Y$ denote the embedding levels of
$SU(3)_C$, $SU(2)_L$, and $U(1)_Y$, respectively, which fix the
relative normalization of their generators. For $k_3=k_2=1$ and $k_Y=5/3$, this gives $8/3$. Thus the value $8/3$ in
this class follows from the common axion dependence of the QCD and
electromagnetic topological terms, rather than from supersymmetry by itself.
Supersymmetric $E_8\times E_8$ compactifications with non-standard
hypercharge embeddings can evade this relation when electromagnetism
receives contributions from both gauge factors \cite{Reig:2025dqb}. 
The associated bounds concern the photon coupling relative to the
mass, rather than a universal upper limit $E/N\le8/3$. 
The mass spectra and gauge couplings of non-QCD axions in
perturbative $E_8\times E_8$ compactifications are studied in
Ref.~\cite{Leedom:2025mlr}. For the QCD axion, Ref.~\cite{Benabou:2026jtv} relates its mass to the unified gauge coupling and evaluates
it in ensembles of toric Calabi-Yau hypersurfaces and
complete-intersection Calabi-Yau manifolds.

We study the same question in the tachyon-free $SO(16)_{\rm v}\times SO(16)_{\rm h}$
heterotic string \cite{Dixon:1986iz,Alvarez-Gaume:1986ghj}, using smooth
Calabi-Yau line-bundle backgrounds in the framework of
Ref.~\cite{Blaszczyk:2015zta}. 
Our aim is to determine which relations between the QCD and electromagnetic
axion couplings follow from the gauge embedding and how direct
flux breaking can change them. We then use the compactification
conditions to constrain the departure of $E/N$ from $8/3$.
We distinguish these constraints from the dependence on the
axion kinetic terms and low-energy matching. 
Direct flux breaking to the Standard Model has also been studied in
supersymmetric heterotic Calabi-Yau compactifications and in $SO(32)$
heterotic line-bundle models
\cite{Anderson:1411.0034,Otsuka:1809.00838}.
Related three-generation $SO(32)$ constructions with multiple Abelian
magnetic fluxes or line bundles have been studied in
Refs.~\cite{Abe:2015mua,Otsuka:2018oyf}. 
Axion observables can then probe aspects of the string construction that are complementary to observables associated with low-energy supersymmetry breaking.

Our first result is that the level assignment $k_3=k_2=1$ and $k_Y=5/3$
inside one visible $SO(16)_{\rm v}$ factor does not by itself imply
$e=(8/3)n$. With direct flux breaking to the Standard Model, quartic
Green-Schwarz contractions can make the QCD and electromagnetic terms
depend on different combinations of physical axions. The ratio in
Eq.~\eqref{eq:intro-effective-ratio} can then depend on the axion kinetic
matrix. The departure from $8/3$ is nevertheless not arbitrary: the same
internal fluxes that determine the Green-Schwarz couplings must satisfy the
Hermitian Yang-Mills (HYM) equations and the heterotic Bianchi identity. These conditions lead to general bounds on $E/N$ without specifying a particular Calabi-Yau geometry.

The same flux constraints also bound the photon couplings of
axion-like particles (ALPs) with no QCD coupling. At fixed moduli, we derive
an upper bound on the axion-photon
couplings over all such directions, without specifying the potentials
that generate their masses. This quantity does not depend on the
choice of orthonormal basis within the QCD-neutral axion space.
In the direct-flux model, we calculate it for the two axions other
than the QCD axion and compare it with the geometric bound. 
Note that their individual masses and the distribution of the photon coupling
between mass eigenstates depend on the additional axion potentials.

The paper is organized as follows. Section~\ref{sec:physical-framework}
defines the physical axion degrees of freedom and derives
Eq.~\eqref{eq:intro-effective-ratio}, including the condition under which
$E/N$ is independent of the kinetic matrix. Section~\ref{sec:setup}
introduces the Calabi-Yau line-bundle data, consistency conditions, and
Green-Schwarz reduction in the non-supersymmetric $SO(16)_{\rm v}\times SO(16)_{\rm h}$ heterotic string theory. Section~\ref{sec:embedding-criterion} derives the flux condition
that controls the relation between the color and electromagnetic
Green-Schwarz coefficients. 
It also determines when this difference defines a physical axion
coupling and derives general bounds from the HYM equations and
the Bianchi identity for backgrounds with and without five-brane sources. 
Section~\ref{sec:mass-constraints} derives general conditions for
mass-coupling bounds without fixing a Calabi-Yau background. It also
compares these conditions with supersymmetric heterotic constructions.
Section~\ref{sec:models} evaluates the geometric quantities on complete intersection Calabi-Yau (CICY) threefolds~\cite{Candelas:1987kf,Anderson:2017aux} and
presents an $SU(5)$-like model and a direct-flux Standard-Model-like
model. Section~\ref{sec:conclusions} is devoted to the conclusions. 
Appendix~\ref{app:ns5-reduction} derives the contributions of Neveu-Schwarz five-branes (NS5-branes) to the axion couplings and charges from the local anomaly-inflow terms.

\section{The physical QCD axion and the relation \texorpdfstring{$e=r n$}{e = r n}}
\label{sec:physical-framework}

In section~\ref{sec:physical-axion}, we identify the QCD axion when several axions are present and
some axion directions are removed through the St\"uckelberg mechanism. We first derive
the kinetic metric on the remaining axion space and show how the relevant
inverse-metric contractions can still be evaluated in the original axion
coordinates. We then determine the QCD mass eigenstate and its photon
coupling. Section~\ref{sec:proportionality-criterion} gives the condition
under which $E/N$ is independent of the axion kinetic matrix. The
compactification-dependent Green-Schwarz coefficients are introduced in the
following sections; their relation to the low-energy coefficients $n$ and
$e$ requires the corresponding anomaly and threshold matching.

\subsection{From gauged shifts to the QCD axion direction}
\label{sec:physical-axion}

Let us consider four-dimensional St\"uckelberg couplings for the dimensionless
axions $\theta^A$ and the Abelian gauge fields $A_\mu^I$~\cite{Choi:1909.11685,Fraser:1910.11349}. Their two-derivative
kinetic term is
\begin{align}
 \mathcal L_{\rm kin}=-\frac12 G_{AB}
 \left(\partial_\mu\theta^A-Q^A{}_I A_\mu^I\right)
 \left(\partial^\mu\theta^B-Q^B{}_J A^{J\mu}\right).
 \label{eq:axion-kinetic-general}
\end{align}
Here $G_{AB}$ is the positive-definite axion kinetic matrix and $Q^A{}_I$
are the St\"uckelberg charges. The kinetic term is invariant under
\begin{align}
 \delta\theta^A=Q^A{}_I\lambda^I,
 \qquad
 \delta A_\mu^I=\partial_\mu\lambda^I,
 \label{eq:axion-gauge-shift}
\end{align}
where $\lambda^I$ are the Abelian gauge-transformation parameters.

We write the low-energy topological interactions relevant to the QCD axion as
\begin{align}
 \mathcal L_{\rm top}
 =
 \frac{\alpha_s}{8\pi} n_A\theta^A
 G^a_{\mu\nu}\widetilde G^{a\mu\nu}
 +
 \frac{\alpha_{\rm em}}{8\pi} e_A\theta^A
 F_{\mu\nu}\widetilde F^{\mu\nu}.
 \label{eq:low-energy-topological-couplings}
\end{align}
The tensors $G^a_{\mu\nu}$ and $F_{\mu\nu}$ are the canonically normalized QCD and electromagnetic field strengths.
We use
$\widetilde F^{\mu\nu}
=\frac12\epsilon^{\mu\nu\rho\sigma}F_{\rho\sigma}$,
and similarly for the QCD field strength. 
The constants $\alpha_s$ and $\alpha_{\rm em}$ are the QCD and
electromagnetic fine-structure constants.
Here $n_A$ and $e_A$ are the low-energy QCD and electromagnetic
axion-coupling coefficients, respectively.\footnote{If additional massless
Abelian gauge fields are present, their kinetic mixing must also be included
when identifying the observed photon~\cite{Holdom:1985ag}. We use $e_A$ for the coefficient after
this gauge-field matching; when no additional massless visible $U(1)$ remains,
it is the coefficient associated with $Q_{\rm em}=T_3+Y$.} Contributions
from integrating out heavy states, charged matter phases, and localized
sectors are included in these coefficients~\cite{Anastasopoulos:hep-th0605225,Choi:1909.11685}. The light-QCD meson contribution
is not included in $e_A$ and is represented below by $\mathcal C_{\rm QCD}$.

Under the St\"uckelberg transformation in
Eq.~\eqref{eq:axion-gauge-shift}, the two axion combinations in
Eq.~\eqref{eq:low-energy-topological-couplings} shift by $n^TQ\boldsymbol\lambda$ and
$e^TQ\boldsymbol\lambda$. Since $n$ and $e$ denote the low-energy couplings,
they must be invariant under these gauged shifts:
\begin{align}
 n^TQ=0,
 \qquad
 e^TQ=0.
 \label{eq:physical-gauge-invariance}
\end{align}
Eq.~\eqref{eq:physical-gauge-invariance} need not hold for a Green-Schwarz
coefficient obtained directly from dimensional reduction. Such a term can
have a gauge variation that participates in anomaly cancellation~\cite{Anastasopoulos:hep-th0605225,Choi:1909.11685}. The
condition applies to the QCD and electromagnetic axion couplings after the
corresponding low-energy matching has been performed.

We next integrate out the massive Abelian gauge fields to obtain the kinetic
term on the physical axion space. The full gauge-field equation of motion
contains derivatives from the gauge kinetic term. At momenta well below the
St\"uckelberg masses, these terms are suppressed by powers of momentum over
the heavy-vector mass and give higher-derivative corrections. The leading
equation is therefore
\begin{align}
 Q^TG(\partial_\mu\theta-QA_\mu)=0.
 \label{eq:heavy-vector-eom}
\end{align}
For the derivation through
Eq.~\eqref{eq:physical-quotient-metric}, we choose a gauge-field
basis that separates the massive and massless combinations and restrict
$Q$, $A_\mu$, and $\boldsymbol\lambda$ to the massive sector.
Outside this derivation, $Q$ denotes the full Abelian charge matrix.
Since
$G$ is positive definite, $Q^TGQ$ is then positive definite on this gauge
subspace, and Eq.~\eqref{eq:heavy-vector-eom} gives
\begin{align}
 A_\mu=(Q^TGQ)^{-1}Q^TG\,\partial_\mu\theta.
 \label{eq:heavy-vector-solution}
\end{align}

Two axion configurations related by $\theta\rightarrow\theta+Q\boldsymbol\lambda$
are gauge equivalent. Locally, we introduce coordinates $\theta_{\rm phys}$ for the
inequivalent axion directions and choose a matrix $R$ such that
\begin{align}
 \theta=Q\boldsymbol\lambda+R\theta_{\rm phys}.
 \label{eq:axion-local-decomposition}
\end{align}
Here a change of $\boldsymbol\lambda$ moves along a gauge orbit, whereas $\theta_{\rm phys}$ labels the
remaining physical axion degrees of freedom. The matrices $Q$ and $R$
together span the local axion field space. 
We choose the gauge $\boldsymbol\lambda=0$. Substituting $\theta=R\theta_{\rm phys}$ into
Eq.~\eqref{eq:heavy-vector-solution} and then substituting the resulting
$A_\mu$ into Eq.~\eqref{eq:axion-kinetic-general} gives
\begin{align}
 G_{\rm phys}
 =R^T\left[G-GQ(Q^TGQ)^{-1}Q^TG\right]R.
 \label{eq:physical-quotient-metric}
\end{align}
The QCD and electromagnetic coefficient vectors in the $\theta_{\rm phys}$
coordinates are given by
$n_{\rm phys}=R^Tn$ and $e_{\rm phys}=R^Te$. 
Using the gauge-invariance conditions $n^TQ=e^TQ=0$, we obtain
\begin{align}
 n_{\rm phys}^TG_{\rm phys}^{-1}n_{\rm phys}
 &=n^TG^{-1}n,
\nonumber\\
e_{\rm phys}^TG_{\rm phys}^{-1}n_{\rm phys}
 &=e^TG^{-1}n.
 \label{eq:qcd-inverse-metric-identity}
\end{align}
Thus the inverse-metric contractions needed for the QCD axion can be
calculated directly from $G$, $n$, and $e$, without explicitly constructing
$R$ or $G_{\rm phys}$. The result holds for a general positive kinetic matrix,
including bulk five-brane kinetic mixing. 
The effects of additional axion potentials on the mass eigenstates and their photon couplings are discussed in section~\ref{sec:mass-constraints}.

To identify the state whose mass is generated by QCD, we first include only
the QCD contribution to the axion potential. We denote its vacuum energy by
$\mathcal E_{\rm QCD}(\vartheta)$ and define the susceptibility at the
CP-conserving minimum by~\cite{GrillidiCortona:2015jxo}
\begin{align}
 \chi_{\rm QCD}=\left.
 \frac{\partial^2\mathcal E_{\rm QCD}}{\partial\vartheta^2}
 \right|_{\vartheta=0}.
 \label{eq:qcd-susceptibility-definition}
\end{align}
The QCD vacuum angle entering this potential is
$\vartheta=\bar\theta+n_{\rm phys}^T\theta_{\rm phys}$, where $\bar\theta$ is the axion-independent strong CP angle,
including the phase of the quark-mass determinant. 
Expanding around the minimum gives
the rank-one Hessian $\chi_{\rm QCD}n_{\rm phys}n_{\rm phys}^T$ for
$n_{\rm phys}\neq0$. We denote the canonically normalized field along the direction that acquires the QCD mass by $a_{\rm QCD}$. The corresponding axion configuration is
\begin{align}
 \theta_{\rm phys}=\frac{G_{\rm phys}^{-1}n_{\rm phys}}
 {\sqrt{n_{\rm phys}^TG_{\rm phys}^{-1}n_{\rm phys}}}\,
 a_{\rm QCD}.
 \label{eq:qcd-massive-direction}
\end{align}
All variations $\delta \theta_{\rm phys}$ satisfying
$n_{\rm phys}^T\delta \theta_{\rm phys}=0$ remain flat at this order. 
The QCD axion therefore need not coincide with the universal
axion or with one particular K\"ahler axion; its direction is determined by
the QCD coupling coefficients together with the kinetic matrix~\cite{Agrawal:2017cmd}.

With the convention that the coefficient of $G\widetilde G$ is
$a_{\rm QCD}/f_{\rm QCD}$, Eqs.~\eqref{eq:qcd-inverse-metric-identity} and
\eqref{eq:qcd-massive-direction} give
\begin{align}
 f_{\rm QCD}^{-2}=n^TG^{-1}n.
 \label{eq:qcd-decay-constant}
\end{align}
The mass generated by QCD is
\begin{align}
 m_a^2=\frac{\chi_{\rm QCD}}{f_{\rm QCD}^2}.
 \label{eq:qcd-mass}
\end{align}
Using Eq.~\eqref{eq:qcd-inverse-metric-identity}, the ratio of electromagnetic
and color couplings of the same QCD mass eigenstate is
\begin{align}
 \left(\frac EN\right)_{\rm eff}
 =\frac{e^TG^{-1}n}{n^TG^{-1}n}.
 \label{eq:qcd-observables-invariant}
\end{align}
We write this ratio as $(E/N)_{\rm eff}$ because the QCD mass
eigenstate depends on the axion kinetic matrix. It need not coincide with a ratio
of integer anomaly coefficients associated with a primitive periodic axion
direction; for a general kinetic matrix, the QCD eigenstate need not lie
along an integral lattice direction~\cite{Fraser:1910.11349}. Standard low-energy QCD matching then
gives
\begin{align}
 g_{a\gamma\gamma}=\frac{\alpha_{\rm em}}{2\pi f_{\rm QCD}}
 \left[\left(\frac EN\right)_{\rm eff}-\mathcal C_{\rm QCD}\right],
 \qquad
 \mathcal C_{\rm QCD}=1.92(4),
 \label{eq:qcd-photon-coupling}
\end{align}
in the convention $\mathcal L\supset
g_{a\gamma\gamma}a_{\rm QCD}F\widetilde F/4$~\cite{GrillidiCortona:2015jxo}.
The QCD susceptibility and $\mathcal C_{\rm QCD}$ are infrared inputs and
are not determined by the compactification calculation.

\subsection{Proportional axion couplings and kinetic-metric independence of
\texorpdfstring{$E/N$}{E/N}}
\label{sec:proportionality-criterion}

Eq.~\eqref{eq:qcd-observables-invariant} shows that the QCD axion coupling
ratio generally depends on both the QCD and electromagnetic coupling
coefficients and the axion kinetic matrix. We first ask when this metric
dependence is absent. If the physical electromagnetic and QCD couplings are
proportional,
\begin{align}
 e_{\rm phys}=r\,n_{\rm phys},
 \label{eq:physical-proportionality}
\end{align}
then
\begin{align}
 \left(\frac EN\right)_{\rm eff}=r
\end{align}
for any positive-definite axion kinetic matrix.

The converse holds if the ratio is required to be independent of every
positive-definite metric on the physical axion space. To see this, we choose
physical axion coordinates such that
$n_{\rm phys}=(n_1,0,\ldots)^T$. If any component $(e_{\rm phys})_i$ with
$i>1$ is nonzero, varying the off-diagonal inverse-metric component
$(G_{\rm phys}^{-1})_{i1}$ changes the numerator of
Eq.~\eqref{eq:qcd-observables-invariant} and hence changes the ratio. Thus
metric independence for arbitrary positive-definite metrics requires
$e_{\rm phys}$ to be proportional to $n_{\rm phys}$. The proportionality
condition is therefore both sufficient and necessary when the full space of
positive-definite kinetic metrics is considered.

For the level assignment used in this paper,
\begin{align}
 k_3=k_2=1,
 \qquad
 k_Y=\frac53,
\end{align}
the corresponding level ratio is~\cite{Agrawal:2022lsp,Agrawal:2024ejr}
\begin{align}
 r_{\rm level}\equiv\frac{k_2+k_Y}{k_3}=\frac83.
 \label{eq:level-ratio}
\end{align}
We define the departure from the relation $e=r_{\rm level}n$ by
\begin{align}
 p=e-r_{\rm level}n.
 \label{eq:physical-difference-vector}
\end{align}
On the physical axion space, the corresponding vector is
$p_{\rm phys}=R^Tp=e_{\rm phys}-r_{\rm level}n_{\rm phys}$.
Since $n^TQ=e^TQ=0$, the difference $p$ also satisfies $p^TQ=0$.
Eq.~\eqref{eq:qcd-observables-invariant} then becomes
\begin{align}
 \left(\frac EN\right)_{\rm eff}
 =r_{\rm level}
 +\frac{p^TG^{-1}n}{n^TG^{-1}n}.
 \label{eq:kinetic-departure}
\end{align}
If $p=0$, the ratio is $8/3$ independently of the kinetic matrix. More
generally, if $p=\rho n$, the two couplings remain proportional and the
ratio is the constant $8/3+\rho$. Dependence on the kinetic matrix is
possible only when $p$ and $n$ are linearly independent, equivalently when
the physical QCD and electromagnetic axion couplings are linearly
independent.

Linear independence does not by itself imply that $E/N$ varies within a
given compactification. The physical kinetic matrix is determined by the
compactification moduli and generally spans only a restricted set of
positive-definite matrices. Let $z$ parameterize a deformation along the
allowed locus in moduli space. The coupling ratio then becomes
\begin{align}
 \mathcal R(z)\equiv
 \left(\frac EN\right)_{\rm eff}
 =
 \frac{
 e_{\rm phys}^T G_{\rm phys}^{-1}(z)n_{\rm phys}
 }{
 n_{\rm phys}^T G_{\rm phys}^{-1}(z)n_{\rm phys}
 }.
\end{align}
For a deformation along which the low-energy topological coefficients are fixed, differentiating this expression with respect to $z$ gives
\begin{align}
 \frac{d\mathcal R}{dz}
 =
 \frac{
 (e_{\rm phys}-\mathcal R n_{\rm phys})^T
 \left(\frac{dG_{\rm phys}^{-1}}{dz}\right)n_{\rm phys}
 }{
 n_{\rm phys}^TG_{\rm phys}^{-1}n_{\rm phys}
 }.
 \label{eq:kinetic-tangent-test}
\end{align}
Thus linear independence of the QCD and electromagnetic axion couplings makes kinetic dependence possible, while an actual variation of $E/N$ along the allowed moduli locus requires the numerator in Eq.~\eqref{eq:kinetic-tangent-test} to be nonzero.

Section~\ref{sec:embedding-criterion} derives the flux-dependent
difference between the QCD and electromagnetic Green-Schwarz
coefficients and determines when it survives in the physical
couplings. The HYM equations and the Bianchi identity then bound
the departure of $E/N$ from $8/3$. In Model~B in section~\ref{sec:model-b}, the allowed
K\"ahler moduli give an explicit dependence of $E/N$ on the kinetic matrix, under the low-energy matching and QCD-dominance
assumptions of section~\ref{sec:model-b}.

\section{\texorpdfstring{$SO(16)_{\rm v}\times SO(16)_{\rm h}$}{SO(16)\_v x SO(16)\_h} heterotic string on CY threefolds with line bundles}
\label{sec:setup}

In this section, we connect the four-dimensional framework of
section~\ref{sec:physical-framework} to the ten-dimensional tachyon-free heterotic string
with gauge algebra
$\mathfrak{so}(16)_{\rm v}\oplus\mathfrak{so}(16)_{\rm h}$
\cite{Dixon:1986iz,Alvarez-Gaume:1986ghj}. We first review the terms in the ten-dimensional action that determine the
Bianchi identity and the Green-Schwarz axion couplings. We then introduce
the Calabi-Yau geometry and the internal gauge flux needed for the
dimensional reduction. Section~\ref{sec:consistency} describes the
consistency conditions and the resulting four-dimensional gauge and
matter content. Section~\ref{subsec:bulk-gs} derives the axion couplings
and axion charges. Section~\ref{subsec:gs-ns5} describes the additional terms
associated with magnetic five-brane sources. Loop-induced moduli
potentials and their stationary points are outside the present
calculation.\footnote{Determining the full moduli dependence of the one-loop
potential is challenging for the smooth Calabi--Yau line-bundle
backgrounds considered here. Explicit analyses of one-loop
potentials and moduli stability have been carried out in toroidal
and Scherk--Schwarz compactifications of non-supersymmetric
heterotic strings; see, e.g.,
Refs.~\cite{Koga:2022qch,Funakoshi:2025lxs,Honda:2026bvd}.}

\subsection{Ten-dimensional action and geometric data}
\label{subsec:10d-action}

The leading bosonic action of the
$SO(16)_{\rm v}\times SO(16)_{\rm h}$ heterotic string theory
has the same form as in the supersymmetric heterotic theories,
apart from the gauge group~\cite{Blaszczyk:2015zta}. 
In the string frame, the terms relevant to this compactification are
\begin{align}
 S_{10}
 =\frac{1}{2\kappa_{10}^{2}}
 \int d^{10}x\,\sqrt{-g_{10}}\,e^{-2\Phi}
 \bigg[
 R(\omega_+)
 +4\,\partial_M\Phi\,\partial^M\Phi
 -\frac12|H_3|^2
 -\frac{\alpha'}{4}
 \left(
 \operatorname{tr}_{16}|F_{\rm v}|^2
 +\operatorname{tr}_{16}|F_{\rm h}|^2
 \right)
 \bigg].
 \label{eq:10d-bosonic-action}
\end{align}
The gauge-field strengths $F_{\rm v}$ and $F_{\rm h}$ belong to the
ten-dimensional $SO(16)_{\rm v}$ and $SO(16)_{\rm h}$ gauge groups, respectively.\footnote{The subscripts ${\rm v}$ and ${\rm h}$
denote the visible and hidden factors, respectively.} 
The trace $\operatorname{tr}_{16}$ is taken in the vector
representation of the corresponding $SO(16)$ factor.
Here $g_{10}=\det g^{(10)}_{MN}$,
$\kappa_{10}^2=(2\pi)^7(\alpha')^4/2$,
$\Phi$ is the ten-dimensional dilaton, and
$R(\omega_+)$ is evaluated with the torsionful connection
$\omega_+=\omega+H_3/2$.
For a $p$-form $\Omega_p$, we use
$|\Omega_p|^2
=\Omega_{M_1\ldots M_p}\Omega^{M_1\ldots M_p}/p!$.
The non-supersymmetric $SO(16)_{\rm v}\times SO(16)_{\rm h}$ theory also has a
nonvanishing one-loop vacuum energy, which gives a dilaton potential and
backreacts on the ten-dimensional background~\cite{Dixon:1986iz,Alvarez-Gaume:1986ghj}. We work at leading order
in the string-loop expansion and do not include this loop-induced
potential in the compactification analysis.

The Kalb-Ramond field strength is
\begin{align}
 H_3
 &=dB_2-\frac{\alpha'}{4}X_3.
 \label{eq:10d-kalb-ramond}
\end{align}
The corresponding Chern-Simons form satisfies
\begin{align}
 dX_3
 &=
 \operatorname{tr}_{16}F_{\rm v}^2
 +\operatorname{tr}_{16}F_{\rm h}^2
 -\operatorname{tr}R(\omega_+)^2.
 \label{eq:10d-bianchi-form}
\end{align}
The cohomological Bianchi identity used below follows from
Eqs.~\eqref{eq:10d-kalb-ramond} and \eqref{eq:10d-bianchi-form}, with
additional magnetic sources included when NS5-branes are present. 
The ten-dimensional action also contains the Green-Schwarz term~\cite{Green:1984sg,Blaszczyk:2015zta}
\begin{align}
 S_{\rm GS}
 &=\frac{1}{24(2\pi)^5\alpha'}
 \int B_2\wedge\mathscr X_8.
 \label{eq:10d-gs-term}
\end{align}
For the non-supersymmetric $SO(16)_{\rm v}\times SO(16)_{\rm h}$ theory,
$\mathscr X_8$ is determined by the chiral ten-dimensional fermion
spectrum. 
This eight-form determines the four-dimensional Green-Schwarz
axion couplings after dimensional reduction.
Its explicit form is given in section~\ref{subsec:bulk-gs}.
In the following, we specify the internal geometry and gauge background needed for the dimensional reduction.

Let us take $X$ to be a Calabi-Yau threefold and $\omega_i$ to be a basis
of $H^{1,1}(X)$. The Poincar\'e-dual divisors are denoted by $D_i$.
Expanding the K\"ahler form in this basis gives
\begin{align}
 J
 &=t^i\omega_i.
 \label{eq:kahler-expansion}
\end{align}
We use $\ell_s=2\pi\sqrt{\alpha'}$ and measure areas in units of
$\ell_s^2$. The K\"ahler parameters $t^i$ are therefore dimensionless. We define the dimensionless Calabi-Yau volume $\mathcal V$ by
$\operatorname{Vol}(X)=\ell_s^6\mathcal V$.

The topological information needed below is encoded in the triple
intersection numbers and the second Chern class:
\begin{align}
 \kappa_{ijk}
 &=\int_X\omega_i\wedge\omega_j\wedge\omega_k,
 \qquad
 c_{2i}
 =\int_X\omega_i\wedge c_2(TX).
 \label{eq:intersection-data}
\end{align}
The intersection numbers determine the total Calabi-Yau volume,
\begin{align}
 \mathcal V
 &=\frac16\kappa_{ijk}t^it^jt^k,
 \label{eq:volume}
\end{align}
and the volumes of the divisors $D_i$ are
\begin{align}
 \tau_i
 &=\frac12\kappa_{ijk}t^jt^k.
 \label{eq:divisor-volumes}
\end{align}
Here, repeated geometric indices are summed. These quantities enter the
HYM conditions below and, through the K\"ahler moduli, the axion kinetic
matrix.

We introduce an Abelian gauge background in the Cartan subalgebra of
the $SO(16)_{\rm v}$ and $SO(16)_{\rm h}$ factors:
\begin{align}
 \frac{\overline F}{2\pi}
 &=\sum_I f_I H_I,
 \qquad\mathrm{with}\qquad
 f_I=V_i^I\omega_i,
 \qquad
 \operatorname{tr}_{16}(H_IH_J)=2\delta_{IJ}.
 \label{eq:flux-background}
\end{align}
Here $H_I$ are Cartan generators and $f_I$ are the corresponding internal
two-form fluxes. The coefficients $V_i^I$ specify how the gauge flux is
embedded in the cohomology of $X$. We distinguish the visible indices
$I=1,\ldots,8$ from the hidden indices $I=9,\ldots,16$. 
A nonzero Cartan flux breaks the ten-dimensional gauge algebra to the
subalgebra that commutes with the background. 
In particular, a root $\alpha$ gives an unbroken generator only if
\begin{align}
 \sum_I \alpha_I f_I=0.
\end{align}

The same background determines the line bundle seen by each charged
string state. For a state with weight $w$, the associated line bundle
$L_w$ has first Chern class
\begin{align}
 c_1(L_w)\equiv q_w
 &=\sum_I w_I f_I.
 \label{eq:physical-line-class}
\end{align}
The topological class $q_w$ determines the index, and hence the net
chirality, of the corresponding four-dimensional zero modes. 
The same flux that breaks the gauge symmetry also determines the chiral matter content and the anomalies that must be cancelled in the four-dimensional theory.

Flux quantization must be imposed on the charges of the physical string
states. Thus $q_w$ must be an integral cohomology class for every weight
$w$ in the string charge lattice. A half-integer entry of $V_i^I$ does
not by itself imply a fractional physical line bundle. What matters is whether the flux is integrally quantized on the full electric charge lattice, including the spinor sectors~\cite{Blaszczyk:2015zta}. It can be directly checked in a model of section~\ref{sec:model-b}. 
The same fluxes enter quadratically in the Bianchi identity and in the
Green-Schwarz coefficients. The consistency conditions introduced next
therefore constrain the same data that determine the low-energy gauge
symmetry, chiral spectrum, and axion couplings.

\subsection{Consistency conditions and four-dimensional spectrum}
\label{sec:consistency}

The flux background must satisfy the heterotic Bianchi identity.
To write its gauge contribution, we define the four-forms:
\begin{align}
 S_{\rm v}
 &=\sum_{I=1}^{8}f_I^2,
 \qquad
 S_{\rm h}
 =\sum_{I=9}^{16}f_I^2,
 \label{eq:flux-squares}
\end{align}
where products of internal forms include the wedge product. Reducing the
ten-dimensional identity in Eqs.~\eqref{eq:10d-kalb-ramond} and
\eqref{eq:10d-bianchi-form} with magnetic five-brane sources gives
\begin{align}
 [W]
 &=c_2(TX)+\frac12(S_{\rm v}+S_{\rm h})
 =\sum_{\ell}[C_{\ell}].
 \label{eq:bianchi}
\end{align}
The gauge flux is then constrained by the topology of the
Calabi-Yau manifold. If the tangent-bundle and gauge-bundle
contributions do not cancel, the remaining magnetic charge is
represented by the five-brane class $[W]$.
Note that each $C_{\ell}$ is a holomorphic curve wrapped by an NS5-brane with unit magnetic charge, and $W_{\ell,i}=\int_{C_{\ell}}\omega_i$ gives its degree with respect
to $D_i$. The components of the total magnetic class are
$W_i=\sum_{\ell}W_{\ell,i}$. Other trace conventions differ by an overall factor
of two in the components of the magnetic class. We adopt the
anomaly-inflow normalization of
Refs.~\cite{Basile:2023knk,Basile:2025mnj}.

The Bianchi identity gives a topological constraint on the background. The HYM equations impose an additional condition on the gauge field. For the line-bundle backgrounds considered here, the fluxes are taken to be of type $(1,1)$, so the holomorphy condition $(f_I)^{(0,2)}=0$ is already satisfied. The remaining HYM equation requires
$g^{m\bar n}(f_I)_{m\bar n}=0$, where
$g^{m\bar n}$ is the inverse internal metric in complex coordinates.
This condition is called primitivity and can also be written as
$J^2\wedge f_I=0$. For a holomorphic line bundle, a gauge field satisfying
this local equation exists if and only if the flux obeys the
zero-slope condition~\cite{Donaldson:1985zz,Uhlenbeck:1986ntc}. On a Calabi-Yau threefold, this is
\begin{align}
 \int_XJ^2\wedge f_I
 &=2V_i^I\tau_i
 =0.
 \label{eq:hym}
\end{align}
We use this HYM solution below for each Abelian flux component. We require a solution in the interior of the K\"ahler cone. 
When $[W]\neq0$, we require $[W]$ to lie in the Mori cone of $X$ so that it can be represented by effective holomorphic curves wrapped by NS5-branes. 
The explicit Standard-Model-like direct-flux model studied below satisfies the Bianchi identity with $W=0$. No charged
bundle modulus expectation value is used.

The same flux background determines the net chiral matter content. For
the associated line bundle $L_w$ introduced above, the index is given by
\begin{align}
 \chi(X,L_w)
 &=\frac16\int_X q_w^3
 +\frac1{12}\int_Xc_2(TX)\wedge q_w.
 \label{eq:line-index}
\end{align}
This index gives the net chirality of the four-dimensional zero modes. Its overall sign also depends on the ten-dimensional chirality of the representation from which the state originates. We express the Standard Model spectrum in terms of left-handed four-dimensional fermions. The ten-dimensional Majorana-Weyl condition identifies weights related by conjugation, and this identification must be taken into account when converting sums over weights into four-dimensional multiplicities. 
Eq.~\eqref{eq:line-index} determines only the net chiral spectrum; the multiplicities of index-zero vector-like pairs require the corresponding cohomology groups.

We next determine the massless gauge symmetry. For a root $\alpha$, the
corresponding non-Abelian generator survives when
\begin{align}
 \sum_I\alpha_I V_i^I
 &=0
\end{align}
for every $i$. If the compactification contains a freely acting quotient
with Wilson vector $w_{\rm W}$, the surviving root must also satisfy
\begin{align}
 \alpha\cdot w_{\rm W}
 &\in\mathbb Z.
\end{align}
These conditions determine the non-Abelian gauge algebra that commutes
with the internal background. A Wilson line imposes an additional
projection when it is present. 
For Abelian gauge fields, commuting with the internal background is not
sufficient. A Cartan gauge field can acquire a St\"uckelberg mass
through its coupling to the axions. This is particularly important for
hypercharge.

Let us write the hypercharge generator as
$Y=\sum_I y_IH_I$ and denote its Cartan coefficients by $y=(y_I)^T$. 
With the hypercharge convention $Q_{\rm em}=T_L^3+Y$ fixed,
the Cartan normalization in Eq.~\eqref{eq:flux-background}
determines its embedding level through the quadratic traces:
\begin{align}
 k_Y
 &=\operatorname{tr}_{16}(Y_{\rm v}^2)
  +\operatorname{tr}_{16}(Y_{\rm h}^2)
 =2y^Ty
 =2\sum_I y_I^2.
 \label{eq:hypercharge-level}
\end{align}
Here $Y_{\rm v}$ and $Y_{\rm h}$ are the projections of $Y$ onto
the visible and hidden factors.
The non-Abelian embedding levels are positive integers and coincide
with the affine current-algebra levels because both $SO(16)$ factors
have level one.
The Abelian level depends on the chosen charge normalization and
need not be an integer.
Recall that the matrix $Q$ denotes the axion charges under the Abelian gauge
transformations, as in $\delta\theta=Q\lambda$ in
section~\ref{sec:physical-axion}. 
Since a hypercharge transformation
corresponds to $\lambda=y\lambda_Y$, 
we find
\begin{align}
 \delta\theta
 &=(Qy)\lambda_Y.
\end{align}
If $Qy\neq0$, the hypercharge gauge field participates in a
St\"uckelberg coupling and becomes massive. Massless hypercharge
therefore requires
\begin{align}
 Qy
 &=0.
 \label{eq:massless-hypercharge}
\end{align}
The rows of $Q$ include the universal
axion and, when present, the five-brane axions. 
Eq.~\eqref{eq:massless-hypercharge} must therefore hold for all of
these axionic directions. Both explicit examples considered below have
$k_Y=5/3$ and level-one color and weak gauge factors.

These conditions determine the admissible flux backgrounds and their
four-dimensional gauge and net chiral matter content. We do not
calculate Yukawa textures or the masses of index-zero vector-like pairs.
We now use the same backgrounds to derive the axion couplings and axion
charges.

\subsection{Green-Schwarz couplings and axion charges}
\label{subsec:bulk-gs}

We review the Green-Schwarz polynomial and axion charges in the
non-supersymmetric $SO(16)_{\rm v}\times SO(16)_{\rm h}$ theory.
We then derive the flux dependence of the axion couplings to
each gauge factor.

The ten-dimensional Green-Schwarz polynomial is
\begin{align}
 \mathscr X_8={}&-\operatorname{tr}_{16}F_{\rm v}^4
                 -\operatorname{tr}_{16}F_{\rm h}^4
 +\frac14\bigl(\operatorname{tr}_{16}F_{\rm v}^2\bigr)^2
 +\frac14\bigl(\operatorname{tr}_{16}F_{\rm h}^2\bigr)^2
 \nonumber\\[-2pt]
 &-\frac14\operatorname{tr}_{16}F_{\rm v}^2\,
                     \operatorname{tr}_{16}F_{\rm h}^2.
 \label{eq:x8-polynomial}
\end{align}
The common normalization will be fixed by a unit universal-axion
coupling to a level-one non-Abelian factor. This form of the polynomial
and the corresponding reduction follow from the ten-dimensional chiral
spectrum \cite{Blaszczyk:2015zta,Basile:2025mnj}.
The eight-form $\mathscr X_8$ contains no curvature terms.
This differs from the supersymmetric $SO(32)$ and $E_8\times E_8$
heterotic theories, whose Green-Schwarz eight-forms contain mixed
gauge-gravity and purely gravitational terms~\cite{Blumenhagen:2005ga,Blumenhagen:2005pm}.
The curvature contribution to the Bianchi identity in
Eq.~\eqref{eq:10d-bianchi-form} is nevertheless present.

The internal components of $B_2$ in Eq.~\eqref{eq:10d-gs-term} give the K\"ahler axions $b^i$, while dualizing its four-dimensional two-form gives the universal axion $a_0$~\cite{Svrcek:2006yi}. We use the axion coordinates
$\Theta=(a_0,b^i)^T$ and normalize $a_0$ so that its coefficient is $C_0=1$ for a level-one non-Abelian gauge factor. All axion couplings and axion charges below are written in these coordinates.

We first determine the coupling of these axions to a four-dimensional gauge field. 
Let $T=\sum_I c_I^{\rm gen}H_I$ be a Cartan generator, with $c^{\rm gen}=(c_I^{\rm gen})^T$.
We denote its four-dimensional field-strength two-form before canonical
normalization by $\mathcal F$. For this fluctuation, the ten-dimensional
field strength is $F_a=\overline F_a+T_a\mathcal F$, where $T_a$ is the
part of $T$ in the gauge factor $a={\rm v},{\rm h}$. Factors of $2\pi$
are written explicitly; $\mathcal F$ has not been divided by $2\pi$. 
Specifically, the terms containing two powers of $\mathcal F$ have the form:
\begin{align}
 \left.\mathscr X_8\right|_{\mathcal F^2}
 &=6(2\pi)^2\,\mathcal C(T)\wedge\mathcal F\wedge\mathcal F,
\end{align}
where $\mathcal C(T)$ is an internal four-form. After integration over the Calabi-Yau manifold, this four-form determines the coupling of the K\"ahler axions to the four-dimensional gauge field associated with $T$. Explicitly,
\begin{align}
 \mathcal C(T)={}&k_{\rm v}(T)B_{\rm v}+k_{\rm h}(T)B_{\rm h}
                 -2\bigl[U_{\rm v}(T)+U_{\rm h}(T)\bigr]
 \nonumber\\
 &+\frac23\bigl[\ell_{\rm v}(T)^2+\ell_{\rm h}(T)^2
                         -\ell_{\rm v}(T)\ell_{\rm h}(T)\bigr].
 \label{eq:general-cartan-coupling}
\end{align}
Here the quantities appearing in the reduction are defined by
\begin{align}
 k_a(T)
 &=2\sum_{I\in a}(c_I^{\rm gen})^2,
 \qquad
 \ell_a(T)=\sum_{I\in a}c_I^{\rm gen} f_I,
 \qquad
 U_a(T)=\sum_{I\in a}(c_I^{\rm gen})^2 f_I^2,
 \label{eq:cartan-definitions}
\end{align}
where $a={\rm v},{\rm h}$ denotes the visible or hidden sector. We also define
\begin{align}
 B_{\rm v}
 &=\frac16S_{\rm v}-\frac1{12}S_{\rm h},
 \qquad
 B_{\rm h}
 =\frac16S_{\rm h}-\frac1{12}S_{\rm v}.
 \label{eq:common-four-forms}
\end{align}

To see how these quantities arise, let us consider either the $SO(16)_{\rm v}$ or the $SO(16)_{\rm h}$ factor. Expanding the quadratic trace gives
\begin{align}
 \operatorname{tr}_{16}(\overline F_a+T_a\mathcal F)^2
 &=2(2\pi)^2S_a+4(2\pi)\ell_a(T)\mathcal F+k_a(T)\mathcal F^2.
\end{align}
The quartic trace can be expanded in the same way. Extracting the terms containing two powers of the four-dimensional field strength from the full Green-Schwarz polynomial gives Eq.~\eqref{eq:general-cartan-coupling}. 
No use has been made of the condition $W=0$ in this reduction, so Eq.~\eqref{eq:general-cartan-coupling} also applies when five-brane charge is present.

The relation between $\mathcal C(T)$ and the K\"ahler-axion couplings follows from the decomposition of the Kalb-Ramond field,
\begin{align}
 B_2
 &=B_2^{(4)}+2\pi\alpha' b^i\omega_i,
\end{align}
where $B_2^{(4)}$ is the four-dimensional two-form and $b^i$ are dimensionless K\"ahler axions. Using the $\mathcal F^2$ term above and the internal component
$2\pi\alpha'b^i\omega_i$ of $B_2$, integration over $X$ gives
\begin{align}
\frac1{4(2\pi)^2} b^i
 \left(
 \int_X\omega_i\wedge\mathcal C(T)
 \right)
 \mathcal F\wedge\mathcal F.
\end{align}
After canonical normalization of the four-dimensional gauge field,
this prefactor gives the $\alpha_a/(8\pi)$ normalization used for the four-dimensional topological interactions. 
The K\"ahler-axion coefficient is then given by
\begin{align}
 C_i(T)
 &=\int_X\omega_i\wedge\mathcal C(T).
\end{align}

The universal axion has a different origin. It is obtained by dualizing the four-dimensional two-form $B_2^{(4)}$. The four-dimensional component of the Kalb-Ramond field strength satisfies the Bianchi identity inherited from
\begin{align}
 dH_3
 &=-\frac{\alpha'}{4}
 \left(
 \operatorname{tr}_{16}F_{\rm v}^2
 +\operatorname{tr}_{16}F_{\rm h}^2
 -\operatorname{tr}R^2
 \right).
\end{align}
For a four-dimensional gauge field associated with the Cartan generator $T$, the four-dimensional gauge contribution is
\begin{align}
 \left.dH_3^{(4)}\right|_{\mathcal F^2}
 &=
 -\frac{\alpha'}{4}
 \left[
 \operatorname{tr}_{16}(T_{\rm v}^2)
 +\operatorname{tr}_{16}(T_{\rm h}^2)
 \right]
 \mathcal F\wedge\mathcal F.
\end{align}
Using $\operatorname{tr}_{16}(H_IH_J)=2\delta_{IJ}$, we obtain
\begin{align}
 \operatorname{tr}_{16}(T_a^2)
 &=2\sum_{I\in a}(c_I^{\rm gen})^2
 =k_a(T),
 \qquad
 a={\rm v},{\rm h}.
\end{align}
Hence
\begin{align}
 \left.dH_3^{(4)}\right|_{\mathcal F^2}
 &=
 -\frac{\alpha'}{4}
 \left[
 k_{\rm v}(T)+k_{\rm h}(T)
 \right]
 \mathcal F\wedge\mathcal F.
\end{align}

In the four-dimensional dual description, the universal axion $a_0$ imposes this Bianchi identity. The term proportional to $\mathcal F\wedge\mathcal F$ becomes an $a_0\mathcal F\wedge\mathcal F$ coupling with a coefficient proportional to
$k_{\rm v}(T)+k_{\rm h}(T)$. We normalize $a_0$ so that a level-one non-Abelian gauge factor has unit coefficient. With this convention, the universal axion coefficient is thus
\begin{align}
 C_0(T)
 &=k_{\rm v}(T)+k_{\rm h}(T).
 \label{eq:cartan-coupling-row}
\end{align}

We now specialize the general result to a level-one $SU(m)$ subgroup contained in the visible $SO(16)_{\rm v}$ factor. 
Suppose that the $m$ Cartan directions associated with this
$SU(m)$ subgroup carry the same internal flux $A$. For an $SU(m)$ Cartan generator $T$, tracelessness gives
\begin{align}
 \sum_{I\in SU(m)} c_I^{\rm gen}
 &=0.
\end{align}
Since these flux entries are equal, this implies
\begin{align}
 \ell_{\rm v}(T)
 &=\sum_{I\in SU(m)}c_I^{\rm gen} A
 =A\sum_{I\in SU(m)}c_I^{\rm gen}
 =0.
\end{align}
We normalize the $SU(m)$ Cartan generator by
$\operatorname{tr}_{\boldsymbol m}T^2=1/2$.
For this regular embedding, the vector representation of $SO(16)$
decomposes as
$\boldsymbol m\oplus\overline{\boldsymbol m}\oplus
(16-2m)\boldsymbol 1$ under $SU(m)$.
Consequently,
$\operatorname{tr}_{16}T^2=2\operatorname{tr}_{\boldsymbol m}T^2=1$,
which is the index-one normalization.
In the Cartan basis of Eq.~\eqref{eq:flux-background}, this gives
\begin{align}
 k_{\rm v}(T)
 &=2\sum_{I\in SU(m)}(c_I^{\rm gen})^2
 =1,
\end{align}
while $k_{\rm h}(T)=0$ because the generator lies entirely in the visible factor. It follows that
\begin{align}
 U_{\rm v}(T)
 &=\sum_{I\in SU(m)}(c_I^{\rm gen})^2 A^2
 =\frac{A^2}{2},
 \qquad
 U_{\rm h}(T)=0.
\end{align}
Substituting these relations into Eq.~\eqref{eq:general-cartan-coupling}, we obtain
\begin{align}
 \mathcal C(T)
 &=B_{\rm v}
 -2\left(\frac{A^2}{2}\right)
 =B_{\rm v}-A^2.
\end{align}
The universal-axion coefficient follows in the same way from
\begin{align}
 C_0(T)
 &=k_{\rm v}(T)+k_{\rm h}(T)
 =1.
\end{align}
Using Eq.~\eqref{eq:cartan-coupling-row}, the axion coefficients for the level-one $SU(m)$ factor are given by
\begin{align}
 C_0^{SU(m)}
 &=1,
 \qquad
 C_i^{SU(m)}
 =\int_X\omega_i\wedge(B_{\rm v}-A^2).
 \label{eq:nonabelian-row}
\end{align}
Note that the term $-A^2$ originates from the quartic contraction
$\operatorname{tr}(\overline F^2T^2)$. It remains even though the common flux commutes with the $SU(m)$ generators. 
Different non-Abelian gauge factors can therefore have different axion couplings when the corresponding Cartan directions carry different fluxes.

We next determine the axion charges under the Abelian gauge transformations. These charges are needed to identify the axion directions that participate in the St\"uckelberg mechanism. In the notation of section~\ref{sec:physical-axion},
\begin{align}
 \delta\Theta^A
 &=Q^A{}_I\lambda^I.
 \label{eq:gs-axion-transformation}
\end{align}
For the K\"ahler axions, the charges follow directly from the reduction of the modified Kalb-Ramond field strength. Expanding
$B_2=B_2^{(4)}+2\pi\alpha'b^i\omega_i$ 
and using the internal flux
$\frac{\overline F}{2\pi}=V_i^I\omega_i H_I$, the components of $H_3$ along $\omega_i$ contain the four-dimensional gauge fields with coefficients $V_i^I$. Then, gauge invariance of $H_3$ requires the K\"ahler axions to transform as
\begin{align}
 \delta b^i
 &=V_i^I\lambda^I
\end{align}
in the convention of Eq.~\eqref{eq:gs-axion-transformation}. Hence
\begin{align}
 Q_i{}^I
 &=V_i^I.
 \label{eq:internal-axion-charges}
\end{align}

The universal-axion charge arises differently. The universal axion $a_0$ is obtained by dualizing the four-dimensional two-form $B_2^{(4)}$. Its gauging is determined by the part of the ten-dimensional Green-Schwarz polynomial that contains one external Abelian field strength and three internal field strengths. 
For an Abelian generator $H_I$ belonging to the gauge factor
$a={\rm v},{\rm h}$, we write
\begin{align}
 F_a&=\overline F_a+H_I \mathcal F^I+\cdots,
\end{align}
where $\mathcal F^I$ is the four-dimensional field-strength two-form
along $H_I$, before canonical normalization. For a fluctuation along
$T=\sum_I c_I^{\rm gen}H_I$, the two notations are related by
$\mathcal F^I=c_I^{\rm gen}\mathcal F$.

Let us extract the terms linear in $\mathcal F^I$ from $\mathscr X_8$. Using
$\overline F_a=2\pi\sum_{J\in a}f_JH_J$ and
$\operatorname{tr}_{16}(H_IH_J)=2\delta_{IJ}$, the quadratic trace gives
\begin{align}
 \left.\operatorname{tr}_{16}F_a^2\right|_{\mathcal F^I}
 &=4(2\pi)f_I\wedge \mathcal F^I,
\end{align}
while the purely internal part is
\begin{align}
 \left.\operatorname{tr}_{16}F_a^2\right|_{\rm int}
 &=2(2\pi)^2S_a.
\end{align}
The quartic trace gives
\begin{align}
 \left.\operatorname{tr}_{16}F_a^4\right|_{\mathcal F^I}
 &=8(2\pi)^3f_I^3\wedge \mathcal F^I.
\end{align}
Substituting these expressions into Eq.~\eqref{eq:x8-polynomial},
the terms linear in $\mathcal F^I$ become
\begin{align}
 \left.\mathscr X_8\right|_{\mathcal F^I}
 &=(2\pi)^3 \mathcal F^I\wedge
 \left(4f_I S_a-2f_I S_{\bar a}-8f_I^3\right),
\end{align}
where $\bar a$ labels the other gauge factor $SO(16)_{\bar a}$. Using
\begin{align}
 B_a&=\frac16S_a-\frac1{12}S_{\bar a},
\end{align}
this can be written as
\begin{align}
 \left.\mathscr X_8\right|_{\mathcal F^I}
 &=24(2\pi)^3\mathcal F^I\wedge\left(f_I B_a-\frac13f_I^3\right).
\end{align}

After integration over the Calabi-Yau manifold, this term determines the gauging of the four-dimensional two-form $B_2^{(4)}$. Dualizing $B_2^{(4)}$ to $a_0$ therefore gives
\begin{align}
 Q_0{}^I
 &=\int_X
 \left(
 f_I B_a-\frac13f_I^3
 \right),
 \qquad I\in a.
 \label{eq:universal-gauging}
\end{align}
Thus $Q_i{}^I$ is linear in the internal flux and follows from the reduction of $H_3$, whereas $Q_0{}^I$ is cubic in the flux and follows from the component of the Green-Schwarz polynomial linear in the four-dimensional Abelian field strength. These expressions agree with the non-supersymmetric $SO(16)_{\rm v}\times SO(16)_{\rm h}$ reduction of Ref.~\cite{Blaszczyk:2015zta}.

Note that the coefficients $C_A(T)$ and the charges $Q^A{}_I$ have different roles. The former determine the axion couplings to gauge-field topological densities. The latter determine which axion directions are gauged. Both are required before the Green-Schwarz coefficients can be related to the physical axion couplings discussed in section~\ref{sec:physical-framework}.

We now apply these results to the Standard Model gauge factors. The color and weak coefficients follow from Eq.~\eqref{eq:nonabelian-row}. 
For $a=3,2,Y,\gamma$, we write $C_a$ as a column vector
in axion space, with components $C_{Aa}\equiv(C_a)_A$,
where $A$ labels the axion coordinates.
The symmetric Cartan coefficients $C^{IJ}=C^{JI}$ are
also axion-space vectors, defined by
$C_A(T)=c_I^{\rm gen}(C^{IJ})_A c_J^{\rm gen}$. 
Setting $T=Y$ therefore gives
\begin{align}
 C_Y
 &=y_I C^{IJ}y_J.
\end{align}
The factors $y_I$ already account for the embedding of hypercharge in the Cartan subalgebra; canonical normalization of the hypercharge gauge field does not introduce an additional contraction with $y_I$. 
For the electroweak convention $Q_{\rm em}=T_L^3+Y$, the electromagnetic coefficient is
\begin{align}
 C_\gamma
 &=C_2+C_Y.
 \label{eq:electromagnetic-row}
\end{align}
The corresponding topological interactions are
\begin{align}
 \mathcal L_{\rm top}
 ={}&
 \frac{\alpha_s}{8\pi}(C_3^T\Theta)
 G^a_{\mu\nu}\widetilde G^{a\mu\nu}
 +
 \frac{\alpha_{\rm em}}{8\pi}(C_\gamma^T\Theta)
 F_{\mu\nu}\widetilde F^{\mu\nu}.
 \label{eq:topological-lagrangian}
\end{align}
If additional massless Abelian gauge fields remain, their kinetic mixing must also be included when identifying the observed photon.

The quantities relevant for comparing the QCD and electromagnetic axion couplings are
\begin{align}
 \Delta_{23}
 &=C_2-C_3,
\end{align}
and
\begin{align}
 P
 &=C_\gamma-r_{\rm level}C_3,
 \qquad\mathrm{with}\qquad
 r_{\rm level}=\frac{k_2+k_Y}{k_3}.
 \label{eq:coupling-differences}
\end{align}
The first quantity measures the difference between the weak and color Green-Schwarz coefficients. The second measures the departure from the relation
\begin{align}
 C_\gamma
 &=\frac{k_2+k_Y}{k_3}C_3,
\end{align}
which would follow if the electromagnetic and QCD coefficients were proportional to their gauge levels. For $k_3=k_2=1$ and $k_Y=5/3$, this ratio is $8/3$.

At this stage, $C_3$, $C_\gamma$, $\Delta_{23}$, and $P$ are Green-Schwarz coefficients obtained from dimensional reduction. They should not yet be identified with the physical low-energy axion couplings. The latter must be invariant under the gauged axion shifts and must include the required anomaly matching. We impose these conditions in the following section.

\subsection{Magnetic five-brane sources}
\label{subsec:gs-ns5}

When the Bianchi identity requires $W\neq0$, wrapped NS5-branes introduce additional axionic degrees of freedom and modify both the Green-Schwarz coefficients and the axion charges. We include these contributions to determine whether the NS5-brane sector can change the relation between the QCD and electromagnetic axion couplings.

We use the local worldvolume couplings proposed in Refs.~\cite{Basile:2023knk,Basile:2025mnj}. Upon reduction on the wrapped curve, each worldvolume tensor field gives one
four-dimensional axion $\sigma_{\ell}$. 
For color in the visible factor, the corresponding contributions to the Green-Schwarz coefficients are
\begin{align}
 \delta C_{i3}
 &=\frac{W_i}{24},
 \qquad
 C_{\sigma_{\ell}3}
 =\frac12.
 \label{eq:ns5-color-shifts}
\end{align}
Here $W_i=\sum_{\ell} W_{\ell,i}$ denotes the total magnetic charge introduced in section~\ref{sec:consistency}. The first term modifies the coupling of the K\"ahler axions, while the second gives the direct coupling of the tensor axion $\sigma_{\ell}$ to the color gauge fields.

The NS5-brane sector also modifies the axion charges under Abelian gauge transformations. The corresponding contributions are
\begin{align}
 \delta Q_0{}^I
 &=\frac1{24}\sum_{\ell}\int_{C_{\ell}}f_I,
 \qquad
 Q_{\sigma_{\ell}}{}^I
 =\frac12\int_{C_{\ell}}
 \begin{cases}
  f_I,&I\in{\rm v},\\
  -f_I,&I\in{\rm h}.
 \end{cases}
 \label{eq:ns5-charge-shifts}
\end{align}
The first term shifts the charge of the universal axion, while the second gives the charge of the tensor axion associated with the curve $C_{\ell}$. The relative normalization of Eqs.~\eqref{eq:ns5-color-shifts} and \eqref{eq:ns5-charge-shifts} follows from the anomaly-inflow analysis of Refs.~\cite{Basile:2023knk,Basile:2025mnj}. The scalar and its dual two-form from the chiral tensor describe the same four-dimensional degree of freedom. Appendix~\ref{app:ns5-reduction} derives these coefficients by
reducing the local anomaly-inflow terms over $C_{\ell}$.

We now examine whether these localized contributions can change the relative QCD and electromagnetic axion couplings. Suppose that color, weak isospin, and hypercharge are all embedded in the visible $SO(16)_{\rm v}$ factor. 
The local NS5-brane couplings are then proportional to the corresponding gauge levels. For $k_3=k_2=1$ and $k_Y=5/3$, their electromagnetic and color contributions therefore satisfy
\begin{align}
 \delta C_{i\gamma}
 &=\frac83\,\delta C_{i3},
\end{align}
and the tensor axion satisfies the same relation,
\begin{align}
 C_{\sigma_{\ell}\gamma}
 &=\frac83\,C_{\sigma_{\ell}3}.
\end{align}
It follows that
\begin{align}
 \delta P_i
 &=0,
 \qquad
 P_{\sigma_{\ell}}
 =0,
 \label{eq:ns5-proportionality}
\end{align}
where $P=C_\gamma-(8/3)C_3$ for the gauge levels considered here.

Thus the local NS5-brane contribution does not generate a departure from
\begin{align}
 C_\gamma
 &=\frac83 C_3
\end{align}
when the Standard Model gauge factors are embedded in the visible $SO(16)_{\rm v}$. The NS5-branes can be required to satisfy the Bianchi identity and can enlarge the axion sector, but their local anomaly-inflow couplings do not by themselves generate a nonzero $P$. Any nonzero value of the corresponding physical difference
\begin{align}
 p
 &=e-\frac83 n
\end{align}
must therefore arise from the bulk Green-Schwarz couplings or from the subsequent low-energy matching.

The local anomaly polynomial determines the anomaly-related
couplings but not the primitive periodicity of the tensor axion.
We next apply the bulk and localized couplings to the Standard
Model embedding.

\section{Standard Model embeddings and general bounds from consistency conditions}
\label{sec:embedding-criterion}

We next connect the Standard Model gauge embedding to the possible
departure of $E/N$ from $8/3$ and to the compactification conditions that
restrict this departure. We consider a visible
$SU(3)_C\times SU(2)_L$ embedding with
$k_3=k_2=1$ and hypercharge level $k_Y=5/3$.
Section~\ref{subsec:embedding-generators} determines the Cartan-flux
background compatible with the chosen color and weak generators and
derives the conditions for massless hypercharge.
Section~\ref{subsec:embedding-proportionality} derives the
Green-Schwarz difference between the electromagnetic and QCD axion
couplings.
Section~\ref{subsec:embedding-physical-rank} then determines when this
difference survives as an independent coupling on the physical axion
space after the axion charges and anomaly matching are taken into
account.
Finally, Section~\ref{subsec:general-bounds} combines these results with
the HYM equations and the Bianchi identity to derive bounds on the
departure of $E/N$ from $8/3$. The chiral spectrum is imposed later in
the explicit models.

\subsection{Gauge generators and massless hypercharge}
\label{subsec:embedding-generators}

Let us collect the eight visible Cartan fluxes introduced in
section~\ref{subsec:10d-action} into
$f_{\rm v}\equiv(f_1,\ldots,f_8)$. In an orthonormal Cartan basis, the
roots of the visible $\mathfrak{so}(16)_{\rm v}$ are
$\pm \mathbf e_i\pm \mathbf e_j$, with $i<j$. For a root $\alpha$, the corresponding
generator $E_\alpha$ commutes with the background flux precisely when
$\alpha\cdot f_{\rm v}=0$. We choose the color roots
$\mathbf e_1-\mathbf e_2$ and $\mathbf e_2-\mathbf e_3$ and the weak root $\mathbf e_4-\mathbf e_5$. 
For these generators to remain unbroken, the flux must satisfy
\begin{align}
 (\mathbf e_1-\mathbf e_2)\cdot f_{\rm v}
 =(\mathbf e_2-\mathbf e_3)\cdot f_{\rm v}
 =(\mathbf e_4-\mathbf e_5)\cdot f_{\rm v}
 =0
 \quad\Longleftrightarrow\quad
 f_{\rm v}=(A,A,A,B,B,C,D,F).
 \label{eq:embedding-block-form}
\end{align}
Here $A,B,C,D,F$ are internal two-form classes, with $F=f_8$. Thus
Eq.~\eqref{eq:embedding-block-form} is the most general Cartan-flux form
that preserves these specified color and weak generators. It does not
classify all possible Standard Model embeddings in $SO(16)_{\rm v}$.

Additional non-Abelian generators can remain unbroken for special
relations among $A,B,C,D,F$. To keep track of possible gauge enhancement,
let us define the set of roots that commute with the background flux,
\begin{align}
 \mathcal R_{\rm v}(f)=
 \left\{
 \alpha\in\{\pm \mathbf e_i\pm \mathbf e_j\mid i<j\}:
 \alpha\cdot f_{\rm v}=0
 \right\}.
 \label{eq:embedding-root-set}
\end{align}
In the absence of additional Wilson lines, the unbroken non-Abelian gauge algebra is exactly
$\mathfrak{su}(3)\oplus\mathfrak{su}(2)$ when
\begin{align}
 \mathcal R_{\rm v}(f)=
 \{\mathbf e_i-\mathbf e_j\mid i\ne j,\ i,j\in\{1,2,3\}\}
 \ \cup
 \{\pm(\mathbf e_4-\mathbf e_5)\}.
 \label{eq:embedding-no-enhancement}
\end{align}
If the compactification also contains a flat Wilson line, a root in
Eq.~\eqref{eq:embedding-root-set} survives only when it is invariant
under that holonomy. For a Cartan Wilson line $w_{\rm W}$ this requires
$\alpha\cdot w_{\rm W}\in\mathbb Z$. These root conditions determine the
unbroken non-Abelian generators. They do not determine which Abelian
gauge fields acquire masses through St\"uckelberg couplings, nor do they
determine the four-dimensional matter spectrum.

For the embedding considered here, the hypercharge generator is
\begin{align}
 y=\left(
 -\frac13,-\frac13,-\frac13,0,0,
 \frac12,\frac12,0;\,0^8
 \right)^T,
 \qquad
 Y=\sum_I y_IH_I.
 \label{eq:embedding-hypercharge}
\end{align}
Using Eq.~\eqref{eq:hypercharge-level}, this generator has
$k_Y=5/3$.
With the generator normalization described in
section~\ref{subsec:bulk-gs}, the regular color and weak embeddings
specified above have $k_3=k_2=1$.

Masslessness of hypercharge requires all axions to be neutral under the
corresponding Abelian gauge transformation. 
For the internal axions, $Q_i{}^I=V_i^I$, so their charges
under a hypercharge transformation are $Q_i{}^Iy_I$. 
It implies that they are neutral under hypercharge when
\begin{align}
 \sum_Iy_If_I
 =-A+\frac12(C+D)=0.
 \label{eq:embedding-internal-y}
\end{align}
This condition can be parameterized as
\begin{align}
 C=A+L,
 \qquad
 D=A-L,
 \qquad
 f_{\rm v}=(A,A,A,B,B,A+L,A-L,F).
 \label{eq:embedding-massless-blocks}
\end{align}

Eq.~\eqref{eq:embedding-massless-blocks} only implements the
hypercharge condition on the internal axions. Flux quantization must
still be imposed using the full sixteen-component gauge-lattice
embedding, including the hidden $SO(16)_{\rm h}$ factor. As explained in
section~\ref{subsec:10d-action}, individual Cartan entries need
not themselves be integral line-bundle classes when the full string
charge lattice contains spinor weights. The combinations seen by physical
string states must nevertheless be integral. In particular,
$L=C-A$ is the line class seen by the visible root $\mathbf e_6-\mathbf e_1$.

The universal axion gives an additional condition. Using
Eq.~\eqref{eq:universal-gauging} together with
$\sum_Iy_If_I=0$, its charge under hypercharge is
\begin{align}
 Q_0y
 =-\frac13\int_X
 \left[
 -A^3+\frac12(A+L)^3+\frac12(A-L)^3
 \right]
 =-\int_XA L^2.
 \label{eq:embedding-universal-y}
\end{align}
The first equality follows because the term in
$Q_0{}^I$ proportional to $B_{\rm v}$ cancels after contraction with
$y_I$. Therefore, within the bulk axion sector, hypercharge remains
massless when Eq.~\eqref{eq:embedding-internal-y} holds together with
\begin{align}
 \int_XA L^2=0.
 \label{eq:embedding-full-y}
\end{align}

The localized axion charges in
Eq.~\eqref{eq:ns5-charge-shifts} have vanishing hypercharge projection
when $\sum_Iy_If_I=0$. 
Hence the same hypercharge condition applies when the local NS5-brane couplings of section~\ref{subsec:gs-ns5} are included. 
There is also a useful consistency
check with the physical QCD axion coupling. Once
$n^TQ=0$ and $n_0=1$ hold, the internal and localized hypercharge
projections vanish, and
$n^TQy=0$ reduces to $Q_0y=0$.
These conditions establish the masslessness of the hypercharge gauge
field.

\subsection{Flux dependence of
\texorpdfstring{$C_\gamma-(8/3)C_3$}{Cgamma - (8/3) C3}}
\label{subsec:embedding-proportionality}

We next determine how the same flux background affects the relative QCD
and electromagnetic Green-Schwarz couplings.

The Green-Schwarz term relates the internal four-forms to the
four-dimensional axion couplings. Keeping the universal axion
$a_0$ and the K\"ahler axions $b^i$, we obtain
\begin{align}
 \mathcal L_{\rm GS}^{(4)}
 \supset{}&
 \frac{\alpha_s}{8\pi}
 \left(C_{03}a_0+C_{i3}b^i\right)
 G^a_{\mu\nu}\widetilde G^{a\mu\nu}
 \nonumber\\
 &+
 \frac{\alpha_2}{8\pi}
 \left(C_{02}a_0+C_{i2}b^i\right)
 W^A_{\mu\nu}\widetilde W^{A\mu\nu}
 \nonumber\\
 &+
 \frac{\alpha_Y}{8\pi}
 \left(C_{0Y}a_0+C_{iY}b^i\right)
 B_{\mu\nu}\widetilde B^{\mu\nu}.
 \label{eq:embedding-gs-lagrangian}
\end{align}
The tensors $W^A_{\mu\nu}$, with $A=1,2,3$, and $B_{\mu\nu}$
are the canonically normalized $SU(2)_L$ and $U(1)_Y$
field strengths, respectively.
The corresponding fine-structure constants are
$\alpha_2=g_2^2/(4\pi)$ and $\alpha_Y=g_Y^2/(4\pi)$,
where $g_2$ and $g_Y$ are the four-dimensional gauge couplings.
The K\"ahler-axion coefficients are obtained by pairing the
internal four-forms $\mathcal C_a$ with $\omega_i$:
\begin{align}
 C_{ia}
 =\int_X\omega_i\wedge\mathcal C_a,
 \qquad
 a=3,2,Y.
 \label{eq:embedding-four-form-reduction}
\end{align}
Hence the differences between the internal four-forms directly determine
the corresponding differences between the K\"ahler-axion coupling
coefficients.

For the flux in Eq.~\eqref{eq:embedding-massless-blocks},
Eq.~\eqref{eq:nonabelian-row} gives
\begin{align}
 \mathcal C_3=B_{\rm v}-A^2,
 \qquad
 \mathcal C_2=B_{\rm v}-B^2,
 \label{eq:embedding-nonabelian-forms}
\end{align}
where
$B_{\rm v}=S_{\rm v}/6-S_{\rm h}/12$ was defined in
Eq.~\eqref{eq:common-four-forms}.

The internal hypercharge condition
$\sum_Iy_If_I=0$ implies $\ell_{\rm v}(Y)=0$, while
$\ell_{\rm h}(Y)=0$ because the hypercharge generator has no hidden
component. Eq.~\eqref{eq:general-cartan-coupling} therefore gives
\begin{align}
 \mathcal C_Y
 =\frac53B_{\rm v}
 -\frac23A^2
 -\frac12\left[(A+L)^2+(A-L)^2\right]
 =\frac53(B_{\rm v}-A^2)-L^2.
 \label{eq:embedding-hypercharge-form}
\end{align}

It is useful to distinguish the weak-color nonuniversality from the
departure relevant to the QCD and electromagnetic couplings. The former
is
\begin{align}
 \mathcal C_2-\mathcal C_3
 =A^2-B^2.
 \label{eq:embedding-delta-form}
\end{align}
Hence the difference between the weak and color couplings of the
K\"ahler axions is
\begin{align}
 C_{i2}-C_{i3}
 =
 \int_X\omega_i\wedge(A^2-B^2),
 \label{eq:embedding-weak-color-lagrangian}
\end{align}
which measures the nonuniversality of the weak and color
Green-Schwarz couplings. For the QCD and electromagnetic couplings, the relevant combination is instead
$C_\gamma-(8/3)C_3$.

We define the corresponding internal four-form by
\begin{align}
 P^{(4)}
 \equiv
 \mathcal C_2+\mathcal C_Y-\frac83\mathcal C_3
 =A^2-B^2-L^2.
 \label{eq:embedding-p-form}
\end{align}
The superscript $(4)$ indicates that $P^{(4)}$ is an internal four-form,
as opposed to the vector of four-dimensional axion-coupling
coefficients obtained after integration over $X$. 
Using the electroweak relation
$C_{i\gamma}=C_{i2}+C_{iY}$, the departure of the K\"ahler-axion
coupling from the $8/3$ relation is
\begin{align}
 P_i
 \equiv
 C_{i\gamma}-\frac83C_{i3}
 =
 \int_X\omega_i\wedge P^{(4)}.
 \label{eq:embedding-p-row}
\end{align}
For the universal axion, the Green-Schwarz coefficients satisfy
\begin{align}
 C_{0\gamma}
 =\frac83C_{03},
 \label{eq:embedding-universal-relation}
\end{align}
and hence
\begin{align}
 P_0=0.
 \label{eq:embedding-universal-p}
\end{align}

Then, the QCD and electromagnetic terms involving the universal and
K\"ahler axions can be written as
\begin{align}
 \mathcal L_{\rm top}
 \supset{}&
 \frac{\alpha_s}{8\pi}
 \left(C_{03}a_0+C_{i3}b^i\right)
 G^a_{\mu\nu}\widetilde G^{a\mu\nu}
+
 \frac{\alpha_{\rm em}}{8\pi}
 \left[
 \frac83\left(C_{03}a_0+C_{i3}b^i\right)
 +P_i b^i
 \right]
 F_{\mu\nu}\widetilde F^{\mu\nu}.
 \label{eq:embedding-qcd-photon-lagrangian}
\end{align}
Thus $P_i$ directly measures the additional K\"ahler-axion coupling to
the electromagnetic topological density beyond the part proportional to
the QCD coupling.

These relations follow directly from the Green-Schwarz reduction. They
do not require $W=0$ and do not involve the axion kinetic matrix. 
The hidden flux and the internal two-form class $F$ cancel
from the differences in
Eqs.~\eqref{eq:embedding-delta-form} and
\eqref{eq:embedding-p-form}.
The tangent-bundle class enters the Bianchi identity but does
not appear explicitly in the bulk Green-Schwarz polynomial.

For the local NS5-brane couplings in
section~\ref{subsec:gs-ns5}, the couplings to color, weak isospin,
and hypercharge follow the same gauge-level ratios because all
three Standard Model gauge factors lie in the visible $SO(16)_{\rm v}$. 
Hence the NS5
contribution does not modify $P_i$, and the corresponding
tensor-axion component of $P$ vanishes.

A simple sufficient condition for
$C_\gamma=(8/3)C_3$ in the Green-Schwarz sector is
$B=A$ and $L=0$. In this limit
$\mathcal C_2-\mathcal C_3=0$ and $P^{(4)}=0$. The first seven visible
flux entries then coincide, and the non-Abelian gauge symmetry is
enhanced. It therefore does not satisfy the no-enhancement condition in
Eq.~\eqref{eq:embedding-no-enhancement}.

However, the $8/3$ relation does not require this enhanced-symmetry
limit. If the Standard Model generators descend from a common $SU(5)$
factor with $k_Y=5/3$, the Green-Schwarz coefficients satisfy
$C_2=C_3$ and $C_Y=(5/3)C_3$. Model~A in
section~\ref{sec:model-a} realizes this situation through a different
gauge embedding followed by a Wilson-line projection. Thus the
$SU(5)$-like mechanism should not be identified with
$B=A$, $L=0$ in
Eq.~\eqref{eq:embedding-massless-blocks}. More generally,
\begin{align}
 A^2-B^2-L^2=0
 \qquad\text{in }H^4(X,\mathbb R)
\end{align}
is sufficient for the Green-Schwarz relation
$C_\gamma=(8/3)C_3$ even when the individual flux classes are different.

\subsection{From Green-Schwarz coefficients to physical axion couplings}
\label{subsec:embedding-physical-rank}

The four-form $P^{(4)}$ determines a difference between coefficients in
the Green-Schwarz action. This does not yet imply a corresponding
difference between the physical QCD and electromagnetic axion couplings.
As discussed in section~\ref{sec:physical-framework}, the latter must be
defined after the axion charges and the required low-energy anomaly
matching have been taken into account.

Using $p$ defined in Eq.~\eqref{eq:physical-difference-vector},
the gauge-invariance conditions of
Eq.~\eqref{eq:physical-gauge-invariance} become
\begin{align}
 n^TQ=0,
 \qquad
 p^TQ=0.
 \label{eq:embedding-gauge-invariance}
\end{align}
These conditions allow $n$ and $p$ to define couplings on the
physical axion space. Since $e=(8/3)n+p$, the vectors $n_{\rm phys}$ and
$e_{\rm phys}$ span the same subspace as $n_{\rm phys}$ and $p_{\rm phys}$.

We now specialize to the sector considered here. After the required
low-energy contributions have been included, we assume that the
difference takes the form
\begin{align}
 p=(0,P_i,0,\ldots,0)^T,
 \qquad
 n_0=1.
 \label{eq:embedding-coupling-rows}
\end{align}
For the Green-Schwarz and localized contributions, the
departure from the $8/3$ relation is carried only by the K\"ahler
axions. The universal axion satisfies
$C_{0\gamma}=(8/3)C_{03}$ and therefore does not contribute to $p$,
while the NS5-brane setup adopted in
section~\ref{subsec:gs-ns5} gives $P_{\sigma_{\ell}}=0$ for the localized
axions. This motivates the form in
Eq.~\eqref{eq:embedding-coupling-rows}, with $p_0=0$ and vanishing
localized components. If additional finite contributions from the
low-energy theory modify this structure, they must be included in $p$ before determining whether the physical coupling vectors
are linearly independent.

For the form of $p$ in Eq.~\eqref{eq:embedding-coupling-rows}, the
second condition in Eq.~\eqref{eq:embedding-gauge-invariance} becomes
\begin{align}
 p^TQ^I
 =\int_XP^{(4)}\wedge f_I
 =0,
 \label{eq:embedding-p-invariance}
\end{align}
for every Cartan component $I$. 
Here we used $Q_i{}^I=V_i^I$, $P_0=0$, and the absence of a localized
component of $P$ in the NS5-brane setup adopted in
section~\ref{subsec:gs-ns5}.
Note that eq.~\eqref{eq:embedding-coupling-rows} also determines
whether $n$ and $p$ are linearly independent. 
If $p$ were proportional to $n$, so
that $p=\rho n$, comparison of the universal-axion components would
give $0=\rho n_0=\rho$. 
Hence a nonzero $p$ cannot be proportional to $n$. The physical QCD and electromagnetic coupling vectors are therefore linearly independent
whenever $p\neq0$.

We next relate this condition to the internal four-form $P^{(4)}$. 
The internal fluxes are real closed forms of type $(1,1)$.
Their products then give real classes of type $(2,2)$, including
$P^{(4)}$. We write
$H_{\mathbb R}^{p,p}(X)=H^{p,p}(X)\cap H^{2p}(X,\mathbb R)$ for the
space of real classes. The forms $\omega_i$ are a basis of
$H_{\mathbb R}^{1,1}(X)$. 
By Poincar\'e duality, the integral of a real $(1,1)$ class
against a real $(2,2)$ class is nondegenerate. We therefore choose
a dual basis $\{\widetilde\omega^{\,i}\}$ of $H_{\mathbb R}^{2,2}(X)$ satisfying
\begin{align}
 \int_X\omega_i\wedge\widetilde\omega^{\,j}
 =\delta_i{}^j.
 \label{eq:embedding-dual-basis}
\end{align}
For that reason, one can expand the cohomology class of $P^{(4)}$ as
\begin{align}
 P^{(4)}=P_j\,\widetilde\omega^{\,j}
 \qquad
 \mathrm{with}
 \qquad
 P_i
 =\int_X\omega_i\wedge P^{(4)}.
 \label{eq:embedding-p-pairing}
\end{align}
The coefficients $P_i$ vanish for all $i$ if and only if the
cohomology class of $P^{(4)}$ vanishes. For a Calabi-Yau threefold the
class considered here lies in $H_{\mathbb R}^{2,2}(X)\subset H^4(X,\mathbb R)$, so
\begin{align}
 P_i=0
 \quad\text{for all }i
 \quad\Longleftrightarrow\quad
 P^{(4)}=0
 \quad\text{in }H^4(X,\mathbb R).
 \label{eq:embedding-poincare-criterion}
\end{align}
Since $p=(0,P_i,0,\ldots,0)^T$, this equivalence and the
linear-independence argument above give
\begin{align}
 \operatorname{rank}
 \begin{pmatrix}
  n_{\rm phys}^T\\
  e_{\rm phys}^T
 \end{pmatrix}
 =2
 \quad\Longleftrightarrow\quad
 P^{(4)}\ne0
 \quad\text{in }H^4(X,\mathbb R),
 \label{eq:embedding-rank-criterion}
\end{align}
provided that the low-energy couplings satisfy
Eqs.~\eqref{eq:embedding-gauge-invariance} and
\eqref{eq:embedding-coupling-rows}.

We denote the K\"ahler-axion components of $n$ and $p$
by $n_K$ and $p_K$.
In Model~B in section~\ref{sec:model-b}, 
$P^{(4)}=A^2-B^2-L^2$ gives the internal coefficient vector
$p_K=(8,8,8,8,8,8)^T$. The corresponding difference vector and the QCD
coefficient vector both annihilate the axion-charge matrix. The QCD and
electromagnetic Green-Schwarz couplings are therefore linearly
independent before a K\"ahler metric is chosen. The allowed HYM locus
then determines which kinetic matrices are realized by the
compactification.

There is also a simple restriction on the number of K\"ahler axions
in this sector. For $W=0$ with only bulk axions, $n_0=1$ and $n^TQ=0$
imply $Q_0=-n_K^TQ_K$, where $Q_K$ is the internal axion-charge matrix.
The HYM equations give $\tau^TQ_K=0$, so
\begin{align}
 \operatorname{rank}Q=\operatorname{rank}Q_K\le h^{1,1}-1.
 \label{eq:embedding-bulk-rank-bound}
\end{align}
If all four extra visible Abelian generators are to become massive
through this mechanism, then $h^{1,1}\ge5$ is necessary. This condition
uses the bulk sector and the gauge-invariant couplings assumed here;
additional charged-matter phases or five-brane axions change the rank
counting. Geometries with fewer K\"ahler moduli can still be used to
evaluate the geometric norms below, but they do not realize this
particular massless-gauge-group requirement.

\subsection{General bounds from consistency conditions}
\label{subsec:general-bounds}

A nonzero $P^{(4)}$ makes dependence on the axion kinetic matrix
possible, but it does not determine the size of the resulting departure
from $8/3$. We now combine the flux relations above with the HYM
conditions and the Bianchi identity.

The bounds below use the flux expressions for the low-energy
coefficients, with $n_0=1$, $p_0=0$, and $n^TQ=p^TQ=0$. Only the
K\"ahler-axion components of $p$ can be nonzero. We assume that additional matter contributions leave these
coefficients unchanged on the physical axion space, as in the
numerical analysis of Model~B in section~\ref{sec:model-b}.
If these contributions change the coefficients or additional
matter phases remain light, the bounds must be reexamined.

Let us first specify the geometric norm used in the derivation. 
Up to the common normalization of the axion kinetic term,
\begin{align}
 \mathcal H_{ij}
 =\partial_i\partial_j[-\log\mathcal V]
 \label{eq:general-bound-kahler-metric}
\end{align}
is the tree-level metric for the K\"ahler axions~\cite{Svrcek:2006yi,Blaszczyk:2015zta}. For a
four-form $Z$, we denote its coefficient vector by
\begin{align}
 Z_i=\int_X\omega_i\wedge Z
\end{align}
when it appears inside a norm, and define
\begin{align}
 \lVert Z\rVert_{\mathcal H^{-1}}^2
 =Z_i(\mathcal H^{-1})^{ij}Z_j.
 \label{eq:general-bound-norm}
\end{align}
The notation $\mathcal H$ is used here to distinguish this K\"ahler
metric from $G_{\rm phys}^{-1}$ introduced in
section~\ref{sec:proportionality-criterion}.

The HYM equations imply that every flux class appearing in the background
is primitive with respect to $J$. For a primitive nonzero $(1,1)$ class
$\xi$,
\begin{align}
 \int_XJ^2\wedge\xi=0,
\end{align}
and the Hodge-Riemann bilinear relations imply~\cite{Dinh:math0501449}
\begin{align}
 q_J(\xi)
 \equiv
 -\int_XJ\wedge\xi^2
 >0.
 \label{eq:general-bound-q}
\end{align}
We use this positivity to define
\begin{align}
 \mu(J)
 =
 \sup_{\substack{\xi\ne0\\ \int_XJ^2\wedge\xi=0}}
 \frac{\lVert\xi^2\rVert_{\mathcal H^{-1}}}
      {q_J(\xi)}.
 \label{eq:general-bound-mu}
\end{align}
The ratio is homogeneous of degree zero in $\xi$.
For $J=\lambda J_0$ with fixed $J_0$, $\mu(J)$ is finite and
independent of an overall rescaling of $J$, i.e., $\lambda$.

We also define
\begin{align}
 c_J
 \equiv
 \int_XJ\wedge c_2(TX)=t^ic_{2i},
 \qquad
 s\equiv\alpha_{\rm tree}^{-1}
 =\frac{\mathcal V}{g_s^2}.
 \label{eq:general-bound-Cs}
\end{align}
Thus $c_J$ depends on the K\"ahler moduli through $J$, even though
$c_2(TX)$ and the components $c_{2i}$ are fixed. 
We assume $c_J>0$, as in the Calabi-Yau backgrounds considered below.
Here $g_s=e^{\langle\Phi\rangle}$ is the background string coupling and
the quantity $s$ is the tree-level inverse gauge coupling for a level-one non-Abelian factor in the normalization used here.

Let us write the hidden Cartan fluxes as
\begin{align}
 f_{\rm h}=(h_1,\ldots,h_8).
\end{align}
For the visible background
\begin{align}
 f_{\rm v}=(A,A, A, B,B, A+L,A-L,F),
\end{align}
the Bianchi identity becomes
\begin{align}
 W
 =c_2(TX)
 +\frac12
 \left[
 5A^2+2B^2+2L^2+F^2+\sum_a h_a^2
 \right].
 \label{eq:general-bound-bianchi}
\end{align}
All classes $A,B,L,F,h_a$ are primitive when the HYM equations are
satisfied. Pairing Eq.~\eqref{eq:general-bound-bianchi} with $J$ then
gives
\begin{align}
 \mathcal E_{\rm flux}
 &\equiv
 \frac52q_J(A)+q_J(B)+q_J(L)
 +\frac12q_J(F)
 +\frac12\sum_aq_J(h_a)
 \nonumber\\
 &=c_J-\int_XJ\wedge W.
 \label{eq:general-bound-flux-budget}
\end{align}
This relation will be used below to constrain the flux background. It is important that the
same classes which determine $P^{(4)}$ and the QCD Green-Schwarz
coefficient also appear in
Eq.~\eqref{eq:general-bound-flux-budget}. The Bianchi identity therefore
restricts the possible size of the axion-coupling coefficients rather
than imposing an unrelated condition.\footnote{Related bounds that combine the
heterotic tadpole condition, the HYM equations, the observed gauge coupling,
and the regime of validity of the four-dimensional description have been studied in Ref.~\cite{Ishiguro:2308.12421}.}

\subsubsection{Backgrounds without five-brane sources}
\label{subsec:general-bound-W0}

Let us first consider $W=0$. Eq.~\eqref{eq:general-bound-flux-budget}
then gives
\begin{align}
 \mathcal E_{\rm flux}=c_J.
\end{align}
The K\"ahler-axion components of
$p=e-(8/3)n$ are
\begin{align}
 (p_K)_i
 =
 \int_X\omega_i\wedge
 \left(A^2-B^2-L^2\right),
 \label{eq:general-bound-pk}
\end{align}
while the corresponding QCD coefficients are
\begin{align}
 (n_K)_i
 =
 \int_X\omega_i\wedge
 \left[
 -\frac16A^2+\frac13(B^2+L^2)
 +\frac16F^2
 -\frac1{12}\sum_a h_a^2
 \right].
 \label{eq:general-bound-nk-W0}
\end{align}
The expression for $n_K$ follows from
$\mathcal C_3=B_{\rm v}-A^2$ together with the visible and hidden flux
squares in Eq.~\eqref{eq:general-bound-bianchi}. These two coefficient
vectors satisfy
\begin{align}
 p_K+3n_K
 =
 \left[
 \frac12(A^2+F^2)
 -\frac14\sum_a h_a^2
 \right]_K,
 \label{eq:general-bound-correlation}
\end{align}
where the subscript $K$ denotes the vector obtained by pairing the
four-form in brackets with $\omega_i$.

For any primitive class $\xi$,
Eq.~\eqref{eq:general-bound-mu} implies
$\lVert\xi^2\rVert_{\mathcal H^{-1}}
\le\mu q_J(\xi)$.
The triangle inequality then gives
\begin{align}
 \lVert p_K\rVert_{\mathcal H^{-1}}
 \le
 \mu\left[
 q_J(A)+q_J(B)+q_J(L)
 \right],
 \label{eq:general-bound-p-norm}
\end{align}
and
\begin{align}
 \lVert p_K+3n_K\rVert_{\mathcal H^{-1}}
 \le
 \mu\left[
 \frac12q_J(A)+\frac12q_J(F)
 +\frac14\sum_aq_J(h_a)
 \right].
 \label{eq:general-bound-q-norm}
\end{align}
Adding the two inequalities and using
Eq.~\eqref{eq:general-bound-flux-budget} yields
\begin{align}
 \lVert p_K\rVert_{\mathcal H^{-1}}
 +\lVert p_K+3n_K\rVert_{\mathcal H^{-1}}
 \le\mu c_J.
 \label{eq:general-bound-correlated-norm}
\end{align}
The point of this relation is that the QCD coefficient and its departure
from the $8/3$ electromagnetic relation are not bounded independently;
the Bianchi identity correlates them.

At fixed moduli, we write the axion kinetic matrix as
\begin{align}
 G
 &=F_*^2\widehat G.
 \label{eq:general-bound-normalized-metric}
\end{align}
The common normalization $F_*$ is fixed by the ten-dimensional action
and the axion coordinates used in section~\ref{sec:setup}. In the leading
unwarped reduction, the kinetic term of the internal components of $B_2$
and the dualization of its four-dimensional two-form give
\begin{align}
 G_{\rm bulk}
 &=
 \frac{M_P^2}{8\pi^2}
 \operatorname{diag}(s^{-2},\mathcal H)
 =
 F_*^2\operatorname{diag}(s^{-2},\mathcal H),
 \label{eq:general-bound-leading-metric}
\end{align}
where $M_P$ is the reduced four-dimensional Planck mass. Thus
$F_*^2=M_P^2/(8\pi^2)$ in the convention specified in
section~\ref{subsec:10d-action}.

Both inverse-metric contractions in $E/N$ contain $F_*^{-2}$, so this
common factor cancels from the ratio. The relative normalization between the universal axion and the K\"ahler axions remains as shown above. We use the
dimensionless matrix $\widehat G$ when comparing the contractions with
the geometric norms, while $G$ retains its dimensionful definition in
section~\ref{sec:physical-axion}.
Moreover, the identities derived in section~\ref{sec:physical-axion} allow the relevant
inverse-metric contractions to be evaluated before explicitly
constructing the physical axion quotient. Since $n_0=1$ and $p_0=0$,
Eq.~\eqref{eq:qcd-observables-invariant} becomes
\begin{align}
 \Delta
 \equiv
 \left(\frac EN\right)_{\rm eff}-\frac83
 =
 \frac{
 p_K^T\mathcal H^{-1}n_K
 }{
 s^2+n_K^T\mathcal H^{-1}n_K
 }.
 \label{eq:general-bound-delta-W0}
\end{align}

The optimization leading to the bound can be reduced to three
dimensionless variables. Let
\begin{align}
 q_K\equiv p_K+3n_K,
\end{align}
and define
\begin{align}
 \rho_p=\frac{\lVert p_K\rVert_{\mathcal H^{-1}}}{s},
 \qquad
 \rho_q=\frac{\lVert q_K\rVert_{\mathcal H^{-1}}}{s},
 \qquad
 \cos\psi
 =
 \frac{
 p_K^T\mathcal H^{-1}q_K
 }{
 \lVert p_K\rVert_{\mathcal H^{-1}}
 \lVert q_K\rVert_{\mathcal H^{-1}}
 }.
\end{align}
Eq.~\eqref{eq:general-bound-correlated-norm} implies
\begin{align}
 \rho_p+\rho_q\le x,
 \qquad
 x\equiv\frac{\mu c_J}{s}.
 \label{eq:general-bound-x}
\end{align}
Since $n_K=(q_K-p_K)/3$,
Eq.~\eqref{eq:general-bound-delta-W0} can be written as
\begin{align}
 \Delta
 =
 \frac{
 3\rho_p(\rho_q\cos\psi-\rho_p)
 }{
 9+\rho_p^2+\rho_q^2-2\rho_p\rho_q\cos\psi
 }.
 \label{eq:general-bound-scalar-optimization}
\end{align}
Thus the problem reduces to extremizing
Eq.~\eqref{eq:general-bound-scalar-optimization} over
$\rho_p,\rho_q\ge0$, $-1\le\cos\psi\le1$, and $\rho_p+\rho_q\le x$.
This gives
\begin{align}
 \frac83-L(x)
 \le
 \left(\frac EN\right)_{\rm eff}
 \le
 \frac83+U(x),
 \label{eq:general-bound-W0-envelope}
\end{align}
with
\begin{align}
 U(x)
 =
 \frac{\sqrt{9+x^2}-3}{4},
 \label{eq:general-bound-U}
\end{align}
and
\begin{align}
 L(x)
 =
 \begin{cases}
 \displaystyle\frac{3x^2}{9+x^2},
 &x\le3\sqrt3,\\[6pt]
 \displaystyle\frac{3+\sqrt{9+x^2}}{4},
 &x\ge3\sqrt3.
 \end{cases}
 \label{eq:general-bound-L}
\end{align}

The two expressions for $L(x)$ arise because the values of $\rho_p$ and $\rho_q$
that maximize $-\Delta$ change at $x=3\sqrt3$. Below this value the
maximum occurs at $\rho_q=0$, whereas above it the maximum has $\rho_q>0$.
To show this, let $c=\cos\psi$. At fixed $\rho_p,\rho_q$, differentiation gives
\begin{align}
 \frac{\partial\Delta}{\partial c}
 &=\frac{3\rho_p\rho_q(9+\rho_q^2-\rho_p^2)}{(9+\rho_p^2+\rho_q^2-2\rho_p\rho_qc)^2}.
 \label{eq:general-bound-angular-derivative}
\end{align}
Thus it suffices to compare $c=\pm1$ and the boundary cases. For
$c=-1$, writing $\rho_\Sigma=\rho_p+\rho_q$ gives
$-\Delta=3\rho_p\rho_\Sigma/(9+\rho_\Sigma^2)\le3\rho_\Sigma^2/(9+\rho_\Sigma^2)$. Its maximum for $\rho_\Sigma\le x$ is
$3x^2/(9+x^2)$ at $\rho_p=x$, $\rho_q=0$.
For $c=1$ and a negative $\Delta$, let $d_{-}=\rho_p-\rho_q\ge0$. At fixed $d_{-}$, the
largest $-\Delta$ has $\rho_p+\rho_q=x$, so
\begin{align}
 -\Delta&=\frac{3d_{-}(x+d_{-})}{2(9+d_{-}^2)},
 \qquad 0\le d_{-}\le x.
 \label{eq:general-bound-lower-reduced}
\end{align}
The stationary point is
\begin{align}
 d_{-,*}&=\frac{9+3\sqrt{9+x^2}}{x}.
 \label{eq:general-bound-lower-stationary}
\end{align}
This stationary point satisfies $d_{-,*}\le x$ only when $x\ge3\sqrt3$.
For smaller $x$, the maximum is at $d_{-}=x$, which means $\rho_q=0$.
For larger $x$, the maximum is at $d_{-}=d_{-,*}$ and equals
$(3+\sqrt{9+x^2})/4$. These two cases give
Eq.~\eqref{eq:general-bound-L}. For the upper bound, $c=1$ and
$d_{+}=\rho_q-\rho_p\ge0$ give $\Delta=3d_{+}(x-d_{+})/[2(9+d_{+}^2)]$. Its stationary point
$d_{+}=3(\sqrt{9+x^2}-3)/x$ lies in $[0,x]$ and gives
Eq.~\eqref{eq:general-bound-U}. The case $x=0$ follows by continuity. 
The optimization is performed over a continuous set larger than the set
of quantized flux backgrounds. The interval in Eq.~\eqref{eq:general-bound-W0-envelope} therefore gives a necessary range; its endpoints need not be realized by an actual compactification.

For $x\ll1$,
\begin{align}
 -\frac{x^2}{3}
 +\frac{x^4}{27}
 +O(x^6)
 \le
 \Delta
 \le
 \frac{x^2}{24}
 -\frac{x^4}{864}
 +O(x^6).
 \label{eq:general-bound-W0-expansion}
\end{align}
Under an overall rescaling $J\rightarrow\lambda J$, with the ratios of the K\"ahler parameters fixed, $\mu$ is unchanged, while $c_J\propto\lambda$ and $s\propto\lambda^3/g_s^2$. 
Thus $x$ decreases both at large radius and at weak string coupling.
At fixed $s$, however, the volume and the string coupling cannot be
varied independently.
Note that these bounds use the leading bulk kinetic metric. Corrections that mix the universal and K\"ahler axions can affect the small departure from 8/3, as illustrated for the model in section~\ref{sec:model-b}. 

\begin{figure}[t]
 \centering \includegraphics[width=0.86\textwidth]{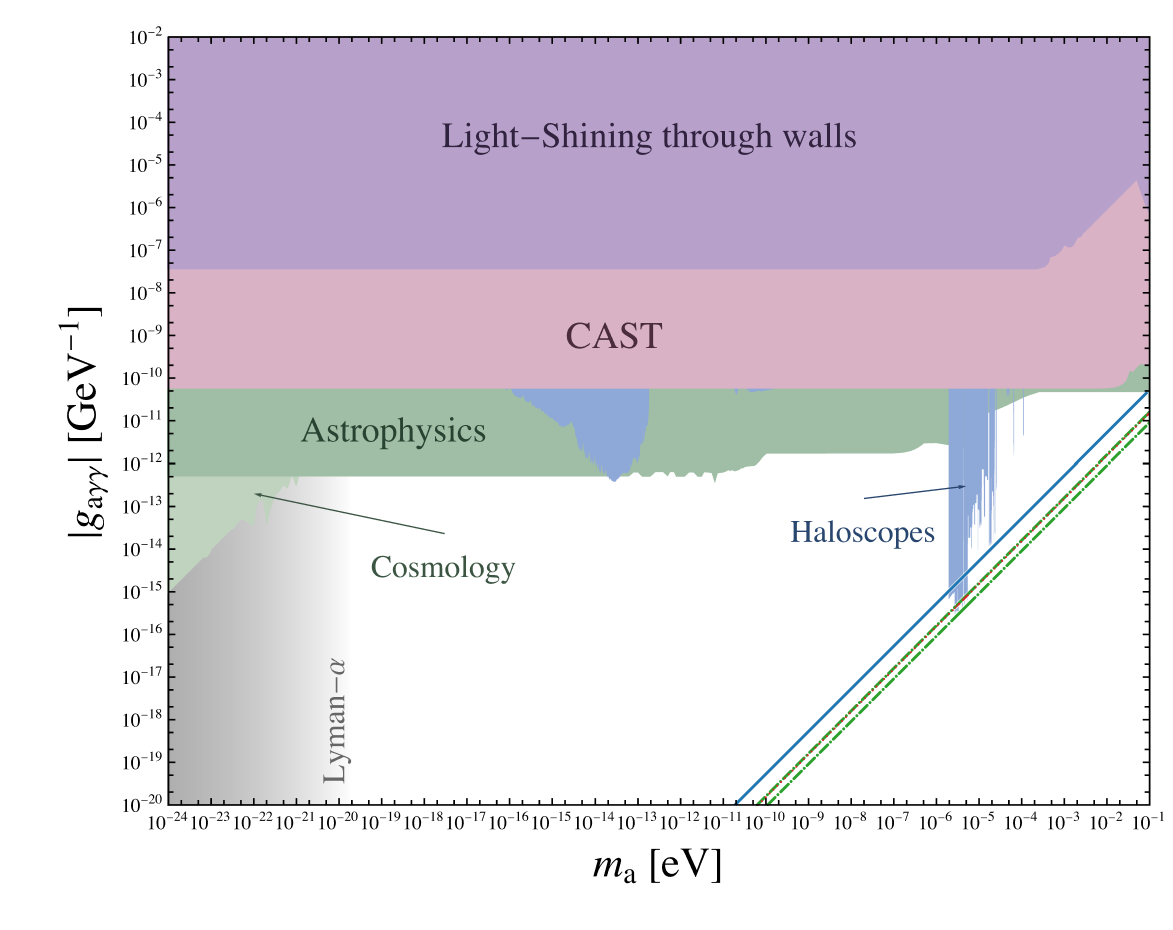}
 \caption{Bounds in the axion mass-photon-coupling plane obtained from
 Eqs.~\eqref{eq:general-bound-W0-envelope}-\eqref{eq:general-bound-L}
 for $W=0$, using the leading bulk kinetic metric. The experimental and observational limits are taken from Ref.~\cite{AxionLimits}. The curves apply to the axion whose mass and direction are determined only by the QCD potential. They are not general
 exclusion limits for ALPs.
 The solid blue and dash-dotted green upper boundaries
 correspond to $x_{\max}=10$ and $1$, respectively, where
 $x=\mu(J)c_J/s\le x_{\max}$. For $x_{\max}=1$, the allowed region lies
 between the two green lines. For $x_{\max}=10$, it extends from
 zero to the blue upper boundary because the coefficient interval
 contains $(E/N)_{\rm eff}=\mathcal C_{\rm QCD}$. Zero cannot be shown
 on the logarithmic axis. The dotted red line is the $x=0$ limit,
 where $L(0)=U(0)=0$ and the interval reduces to the $E/N=8/3$ QCD line. It nearly coincides with the upper green line. The conversion
 uses $|g_{a\gamma\gamma}|=\alpha_{\rm em}m_a
 |(E/N)_{\rm eff}-\mathcal C_{\rm QCD}|/(2\pi\sqrt{\chi_{\rm QCD}})$,
 with $\alpha_{\rm em}^{-1}=137.036$,
 $\chi_{\rm QCD}^{1/4}=75.5\,\mathrm{MeV}$, and
 $\mathcal C_{\rm QCD}=1.92$~\cite{GrillidiCortona:2015jxo}.
 QCD-input uncertainties are not included. The values of $x_{\max}$
 specify illustrative classes, not universal bounds on all Calabi-Yau
 compactifications.}
 \label{fig:general-cy-mass-photon}
\end{figure}

Combining Eqs.~\eqref{eq:qcd-mass} and
\eqref{eq:qcd-photon-coupling} maps the interval in
Eq.~\eqref{eq:general-bound-W0-envelope} into the
$(m_a,|g_{a\gamma\gamma}|)$ plane. Figure~\ref{fig:general-cy-mass-photon}
shows its boundaries for backgrounds satisfying $0\le x\le x_{\max}$
with $x_{\max}=1,10$, without choosing a particular Calabi-Yau geometry. The geometric inputs in these bounds will be evaluated for a specified
set of CICYs in section~\ref{subsec:cicy-geometric-inputs}.

\subsubsection{Backgrounds with five-brane sources}
\label{subsec:general-bound-Wnonzero}

Let us next take $W\neq0$ and assume that the magnetic class is
represented by effective five-brane curves, without anti five-branes.
We denote its K\"ahler-axion coefficient vector by
\begin{align}
 (W_K)_i\equiv\int_X\omega_i\wedge W.
\end{align}
It is useful to define
\begin{align}
 w_5
 \equiv
 \frac{\int_XJ\wedge W}{c_J},
 \qquad
 \rho_W
 \equiv
 \frac{\lVert W_K\rVert_{\mathcal H^{-1}}}{c_J}.
 \label{eq:general-bound-source-parameters}
\end{align}
Eq.~\eqref{eq:general-bound-flux-budget} then gives
\begin{align}
 \mathcal E_{\rm flux}=(1-w_5)c_J.
\end{align}
For a nonzero effective source,
$\int_XJ\wedge W>0$, while
$\mathcal E_{\rm flux}\ge0$; hence
$0\le (1-w_5)<1$.

The same argument used for $W=0$ gives the bulk correlation:
\begin{align}
 \lVert p_K\rVert_{\mathcal H^{-1}}
 +
 \lVert p_K+3n_{K,\rm bulk}\rVert_{\mathcal H^{-1}}
 \le
 \mu (1-w_5)c_J.
 \label{eq:general-bound-source-correlation}
\end{align}
For the local NS5 prescription used in this paper, the K\"ahler-axion
part of the QCD coefficient is shifted by
\begin{align}
 n_K
 =
 n_{K,\rm bulk}
 +\frac{W_K}{24}.
 \label{eq:general-bound-source-shift}
\end{align}
The source contribution to $p=e-(8/3)n$ vanishes, since both the shift of the K\"ahler-axion coefficients and the localized tensor-axion couplings obey the same $8/3$ relation.  
As shown in Eq.~\eqref{eq:ns5-proportionality}, the source sector gives $\delta P_i=0$ and $P_{\sigma_{\ell}}=0$.

To include the NS5-brane tensor axions in the bound, we assume that their kinetic matrix is positive definite and that there is no kinetic mixing with the bulk axions. In the normalization of
Eq.~\eqref{eq:general-bound-normalized-metric}, the full matrix is
\begin{align}
 \widehat G
 &=\operatorname{diag}(s^{-2},\mathcal H,\widehat G_5),
 \qquad \widehat G_5>0.
 \label{eq:general-bound-source-metric}
\end{align}
Let $n_5$ denote the components of $n$ along the tensor axions. Since
$p_0=0$ and $p_{\sigma_{\ell}}=0$, the numerator of the departure is
$p_K^T\mathcal H^{-1}n_K$ in these dimensionless units, whereas
\begin{align}
 n^T\widehat G^{-1}n
 &=s^2+n_K^T\mathcal H^{-1}n_K
   +n_5^T\widehat G_5^{-1}n_5.
 \label{eq:general-bound-source-denominator}
\end{align}
We define
\begin{align}
 \epsilon_{\rm eff}
 \equiv
 \frac{c_J}{\sqrt{n^T\widehat G^{-1}n}}
 =\frac{c_J}{F_*\sqrt{n^TG^{-1}n}}.
 \label{eq:general-bound-epsilon-effective}
\end{align}
This dimensionless quantity includes the contribution of the NS5-brane tensor axions to the QCD normalization. Its numerical value requires $\widehat G_5$. 
The bounds below therefore depend on the kinetic matrix $\widehat G_5$ of the NS5-brane tensor axions, which is not fixed by anomaly inflow.

The bounds in the presence of NS5-branes can be derived more explicitly as follows. Let us define
\begin{align}
 q_K\equiv p_K+3n_{K,\rm bulk},
 \qquad
 \mathcal B_{\rm flux}\equiv\mu (1-w_5)c_J,
 \qquad
 R_W\equiv\lVert W_K\rVert_{\mathcal H^{-1}}
 =\rho_Wc_J.
\end{align}
Eq.~\eqref{eq:general-bound-source-correlation} gives
$\lVert p_K\rVert+\lVert q_K\rVert\le \mathcal B_{\rm flux}$.
Using
$n_{K,\rm bulk}=(q_K-p_K)/3$, the numerator of the departure from
$8/3$ is
\begin{align}
 p_K^T\mathcal H^{-1}n_K
 =
 \frac{
 p_K^T\mathcal H^{-1}q_K
 -\lVert p_K\rVert_{\mathcal H^{-1}}^2
 }{3}
 +
 \frac{
 p_K^T\mathcal H^{-1}W_K
 }{24}.
 \label{eq:general-bound-source-numerator}
\end{align}
The Cauchy-Schwarz inequality therefore gives
\begin{align}
 p_K^T\mathcal H^{-1}n_K
 \ge
 -\frac{\mathcal B_{\rm flux}^2}{3}
 -\frac{\mathcal B_{\rm flux}R_W}{24}.
\end{align}
Dividing by the positive dimensionless denominator $n^T\widehat G^{-1}n$ gives
\begin{align}
 \left(\frac EN\right)_{\rm eff}
 \ge
 \frac83
 -A_-\epsilon_{\rm eff}^2,
 \label{eq:general-bound-source-lower}
\end{align}
where
\begin{align}
 A_-
 =
 \frac{\mu^2(1-w_5)^2}{3}
 +\frac{\mu (1-w_5)\rho_W}{24}.
 \label{eq:general-bound-Aminus}
\end{align}

For the upper bound, let
$\mathcal N_p=\lVert p_K\rVert_{\mathcal H^{-1}}$.
The maximum occurs when $p_K$, $q_K$, and $W_K$ are aligned so as to
maximize the numerator. Since
$\lVert q_K\rVert\le \mathcal B_{\rm flux}-\mathcal N_p$,
Eq.~\eqref{eq:general-bound-source-numerator} gives
\begin{align}
 p_K^T\mathcal H^{-1}n_K
 \le
 \frac{\mathcal N_p(\mathcal B_{\rm flux}-2\mathcal N_p)}{3}
 +\frac{\mathcal N_pR_W}{24},
 \qquad
 0\le \mathcal N_p\le \mathcal B_{\rm flux}.
 \label{eq:general-bound-source-upper-reduction}
\end{align}
Maximizing the right-hand side over $\mathcal N_p$ gives
\begin{align}
 A_+
 =
 \begin{cases}
 \displaystyle
 \frac{(\mu (1-w_5)+\rho_W/8)^2}{24},
 &\rho_W\le24\mu (1-w_5),\\[7pt]
 \displaystyle
 \frac{\mu (1-w_5)\rho_W}{24}
 -\frac{\mu^2(1-w_5)^2}{3},
 &\rho_W\ge24\mu (1-w_5).
 \end{cases}
 \label{eq:general-bound-Aplus}
\end{align}
The resulting two-sided bound is
\begin{align}
 \frac83-A_-\epsilon_{\rm eff}^2
 \le
 \left(\frac EN\right)_{\rm eff}
 \le
 \frac83+A_+\epsilon_{\rm eff}^2.
 \label{eq:general-bound-source-envelope}
\end{align}

The parameter $\rho_W$, which measures the size of the NS5-brane charge, is itself constrained by the Bianchi identity. Let us further define
\begin{align}
 (c_{2,K})_i\equiv c_{2i},
 \qquad
 \rho_{c_2}
 \equiv
 \frac{
 \lVert c_{2,K}\rVert_{\mathcal H^{-1}}
 }{c_J}.
\end{align}
Taking the norm of
Eq.~\eqref{eq:general-bound-bianchi}, using the triangle inequality and
the definition of $\mu$, gives
\begin{align}
 \rho_W
 \le
 \rho_{c_2}+\mu (1-w_5).
 \label{eq:general-bound-source-omega}
\end{align}
Thus the fraction of the Bianchi class carried by five-branes and the
norm of that class cannot be varied independently.

Eqs.~\eqref{eq:general-bound-source-lower}-\eqref{eq:general-bound-source-envelope}
separate three effects. First, $(1-w_5)<1$ reduces the allowed bulk-flux
contribution at fixed $c_J$. Second, the five-brane class shifts the QCD
axion coefficient through $W_K/24$ while leaving the electromagnetic-color
difference unchanged in the local NS5-brane prescription.
Third, assuming a positive kinetic matrix for the NS5-brane tensor axions
and no kinetic mixing between the bulk and NS5-brane axions, the
NS5-brane tensor axions increase $n^TG^{-1}n$ and therefore reduce the
possible departure from $8/3$.

The last statement is conditional. Bulk-five-brane kinetic mixing can
modify the inverse-metric contractions. In supersymmetric heterotic
M-theory reductions, five-brane moduli and bulk moduli enter the same
four-dimensional effective action and their kinetic terms are generically
not block diagonal \cite{Derendinger:hep-th0009054,Brandle:hep-th0109173}.
These results do not determine the kinetic matrix of the present
non-supersymmetric weakly coupled compactification, for which the microscopic
five-brane kinetic action is not known in sufficient detail to give an
unconditional numerical bound. The inequalities above
also do not replace flux quantization, the full differential Bianchi
identity, the chiral-index constraints, or the infrared matching needed
to identify the observed photon. They provide a necessary geometric test
of whether the compactification data can support a sizable departure
from the level value $8/3$ before a complete model is specified. 
Although the local NS5 terms leave $P$ unchanged, they modify the QCD
coefficients and the axion charges. Their kinetic terms can also change
$E/N$ when the bulk contribution gives $p\ne0$.

\section{General constraints from axion masses}
\label{sec:mass-constraints}

This section examines how axion masses constrain axion-photon couplings
without specifying a particular Calabi-Yau background.
Section~\ref{subsec:susy-qcd-alp-scope} derives a bound on the photon coupling relative to the axion mass when the QCD and electromagnetic axion couplings are proportional. 
Section~\ref{subsec:general-mass-constraints} gives a mass-coupling bound
for additional potentials and a geometric bound on the photon coupling
of an ALP with no QCD coupling. Section~\ref{sec:susy-comparison}
compares these results with supersymmetric heterotic compactifications.

\subsection{Photon-coupling bound from the QCD axion relation}
\label{subsec:susy-qcd-alp-scope}

The ratio in Eq.~\eqref{eq:qcd-observables-invariant} applies to the
axion direction whose mass is generated by the QCD potential when other
axion potentials are neglected. By contrast, the bounds on
$|g_{a\gamma\gamma}|/m_a$ in
Refs.~\cite{Agrawal:2022lsp,Agrawal:2024ejr,Reig:2025dqb} are formulated
for the physical mass eigenstates after the relevant axion potentials
have been included. The distribution of the QCD contribution among
mixed axion mass eigenstates has also been studied in
Ref.~\cite{Gavela:2023tzu}. We first derive a sufficient condition for bounding the photon coupling relative to the axion mass when
$e_{\rm phys}=r_{\rm level}n_{\rm phys}$.

At a CP-conserving vacuum, let $\theta_{\rm phys}$ denote the physical axion coordinates
introduced in section~\ref{sec:physical-axion}. 
We take $\theta_{\rm phys}=0$ at this vacuum and evaluate the kinetic matrix
there. The canonically normalized fields are
\begin{align}
 \phi=G_{\rm phys}^{1/2}\theta_{\rm phys}.
 \label{eq:comparison-canonical-axions}
\end{align}
We define
$\nu=G_{\rm phys}^{-1/2}n_{\rm phys}$, so that the QCD angle is
$n_{\rm phys}^T\theta_{\rm phys}=\nu^T\phi$. The QCD contribution to the mass matrix is
therefore $\chi_{\rm QCD}\nu\nu^T$. Including the remaining axion
potentials, we write
\begin{align}
 M^2
 &=
 \chi_{\rm QCD}\nu\nu^T+M_{\rm other}^2,
 \label{eq:proportional-positive-hessian}
\end{align}
where $M_{\rm other}^2$ is the Hessian of all non-QCD
contributions in the canonically normalized basis. 
We assume that $M_{\rm other}^2$ is positive semidefinite,
$M_{\rm other}^2\succeq0$, so these terms give a non-negative
contribution to the squared mass along every axion direction. 
The positivity condition is imposed on the combined non-QCD contribution,
rather than on each contribution separately. It is an additional
assumption beyond the stability of the full mass matrix $M^2$.

Let $\widehat u_j$ be a normalized eigenvector of $M^2$ with eigenvalue $m_j^2$.
Taking the expectation value of
Eq.~\eqref{eq:proportional-positive-hessian} along $\widehat u_j$ gives
\begin{align}
 m_j^2
 &=\chi_{\rm QCD}(\nu^T\widehat u_j)^2+\widehat u_j^TM_{\rm other}^2\widehat u_j
 \geq\chi_{\rm QCD}(\nu^T\widehat u_j)^2.
 \label{eq:proportional-mass-inequality}
\end{align}
This inequality bounds the QCD coupling of each mass
eigenstate in terms of its physical mass.

For masses and momenta below the QCD hadronic scales, ordinary
QCD matching gives the photon coefficient
$e_{\rm phys}-\mathcal C_{\rm QCD}n_{\rm phys}$ when no further
infrared contribution is present. If
$e_{\rm phys}=r_{\rm level}n_{\rm phys}$, the photon coupling therefore
depends on the same projection $\nu^T\widehat u_j$:
\begin{align}
 g_{j\gamma\gamma}
 &=
 \frac{\alpha_{\rm em}}{2\pi}
 \left(r_{\rm level}-\mathcal C_{\rm QCD}\right)\nu^T\widehat u_j.
 \label{eq:proportional-eigenstate-photon-coupling}
\end{align}
Combining this expression with
Eq.~\eqref{eq:proportional-mass-inequality} yields, for $m_j>0$,
\begin{align}
 \frac{|g_{j\gamma\gamma}|}{m_j}
 &\leq
 \frac{\alpha_{\rm em}}{2\pi\sqrt{\chi_{\rm QCD}}}
 \left|r_{\rm level}-\mathcal C_{\rm QCD}\right|.
 \label{eq:proportional-mass-coupling-bound}
\end{align}
For $r_{\rm level}=8/3$, the right-hand side gives the slope of the
QCD line in the $(m_a,|g_{a\gamma\gamma}|)$ plane. This result then gives a bound on the photon coupling relative to the
physical axion mass, rather than a universal upper bound on $E/N$. 
If $m_j=0$,
Eq.~\eqref{eq:proportional-mass-inequality} requires $\nu^T\widehat u_j=0$,
and Eq.~\eqref{eq:proportional-eigenstate-photon-coupling} then gives
$g_{j\gamma\gamma}=0$ within the same approximation.

The situation changes when $n_{\rm phys}$ and $e_{\rm phys}$ are
linearly independent. In that case there exists a physical axion
direction $\boldsymbol v$ such that
\begin{align}
 n_{\rm phys}^T\boldsymbol v=0,
 \qquad
 e_{\rm phys}^T\boldsymbol v\ne0.
 \label{eq:comparison-qcd-neutral-direction}
\end{align}
The first condition means that this direction does not appear in the
QCD vacuum angle and therefore remains flat if QCD is the only source
of an axion potential. The second condition shows that the same
direction nevertheless couples to photons. Once non-QCD potentials are
included, however, this direction can mix with the other axions and
need not coincide with a physical mass eigenstate. The existence of
independent QCD and electromagnetic couplings does not by
itself imply the existence of a light ALP above the QCD line.
Its position in the $(m_a,|g_{a\gamma\gamma}|)$ plane depends on the
additional axion potentials.
In F-theory GUTs, hypercharge flux can also produce ALPs
that couple to photons but not to QCD. Ref.~\cite{Nee:2026msa}
derives bounds on their photon coupling relative to their mass
by combining gauge-coupling unification with D-instanton
contributions to the axion potential.

Additional non-QCD potentials can lift these QCD-neutral axion
directions and modify the physical mass eigenstates. In the next
subsection, we parameterize such contributions and derive conditions
under which the resulting masses constrain the axion-photon couplings.

\subsection{Axion-like particles and additional potentials}
\label{subsec:general-mass-constraints}

In the previous subsection, the effects of all additional axion
potentials were included together in $M_{\rm other}^2$. To study how
these potentials constrain the photon coupling, we now specify the
axion combination that appears in each potential and the corresponding
curvature.

Let us work at the same vacuum and assume that QCD and the potentials
treated individually are minimized there. For $r=1,\ldots,N_{\rm pot}$, let
the quadratic term of the $r$th non-QCD potential be
$\chi_r((k_r^{\rm pot})^T\theta_{\rm phys})^2/2$. The dimensionless vector $k_r^{\rm pot}$ specifies the
axion combination appearing in this potential, while $\chi_r>0$
has mass dimension four and determines its quadratic coefficient.
In the canonically normalized basis of the previous subsection,
we write
\begin{align}
 M_{\rm other}^2
 &=\sum_{r=1}^{N_{\rm pot}}\chi_r\nu_r\nu_r^T+M_{\rm rem}^2,
 \qquad\mathrm{with}\qquad
 \nu_r=G_{\rm phys}^{-1/2}k_r^{\rm pot}.
 \label{eq:general-nonqcd-hessian-decomposition}
\end{align}
Here $M_{\rm rem}^2$ represents the part of $M_{\rm other}^2$ that is
not explicitly parameterized by the rank-one terms in the sum. It need
not be small. If there are no further contributions beyond those written
explicitly, then $M_{\rm rem}^2=0$. 
Including QCD as $k_0^{\rm pot}=n_{\rm phys}$ and
$\chi_0=\chi_{\rm QCD}$, so that $\nu_0=\nu$, the full mass matrix
becomes
\begin{align}
 M^2
 &=\sum_{r=0}^{N_{\rm pot}}\chi_r\nu_r\nu_r^T+M_{\rm rem}^2.
 \label{eq:general-mass-hessian}
\end{align}
To derive the mass-coupling bound, we assume $M_{\rm rem}^2\succeq0$. Together with $\chi_r>0$, this allows the
contributions parameterized by $k_r^{\rm pot}$ and $\chi_r$ to give a lower
bound on the mass squared of each axion eigenstate.

Let $c_\gamma$ denote the coefficient vector that appears in the
low-energy axion-photon interaction after the light-QCD contribution
and any further infrared matching contributions have been included.
With only the standard light-QCD contribution,
\begin{align}
 c_\gamma
 &=e_{\rm phys}-\mathcal C_{\rm QCD}n_{\rm phys}.
\end{align}
To relate the photon coupling to the masses generated by these
potentials, we first assume that the coefficient vector
$c_\gamma$ can be written as
\begin{align}
 c_\gamma
 &=\sum_{r=0}^{N_{\rm pot}}\beta_r k_r^{\rm pot}.
\end{align}
Then the photon coupling of a mass eigenstate depends on the same
projections $\nu_r^T\widehat u_j$ that enter its mass. 
For a normalized mass eigenvector $\widehat u_j$,
Eq.~\eqref{eq:general-mass-hessian} gives
\begin{align}
 m_j^2
 &=\sum_{r=0}^{N_{\rm pot}}\chi_r(\nu_r^T\widehat u_j)^2
   +\widehat u_j^TM_{\rm rem}^2\widehat u_j
\geq\sum_{r=0}^{N_{\rm pot}}\chi_r(\nu_r^T\widehat u_j)^2,
\end{align}
while its photon coupling is
\begin{align}
 g_{j\gamma\gamma}
 &=
 \frac{\alpha_{\rm em}}{2\pi}
 \sum_{r=0}^{N_{\rm pot}}\beta_r\nu_r^T\widehat u_j.
\end{align}
Applying the Cauchy-Schwarz inequality to
$\beta_r/\sqrt{\chi_r}$ and
$\sqrt{\chi_r}\nu_r^T\widehat u_j$ yields
\begin{align}
 \frac{|g_{j\gamma\gamma}|}{m_j}
 &\leq\frac{\alpha_{\rm em}}{2\pi}
       \sqrt{\sum_{r=0}^{N_{\rm pot}}\frac{\beta_r^2}{\chi_r}},
 \label{eq:general-mass-photon-bound}
\end{align}
with $m_j>0$. 
The axion kinetic matrix has cancelled from the right-hand side. The
bound instead depends on the strengths of the potentials through the
$\chi_r$ and on the relation between the photon coupling and the
corresponding axion combinations through the coefficients $\beta_r$.
If the vectors $k_r^{\rm pot}$ are linearly dependent, the coefficients $\beta_r$
in the decomposition of $c_\gamma$ need not be unique. In that case,
the smallest upper bound in
Eq.~\eqref{eq:general-mass-photon-bound} is obtained by minimizing
the right-hand side over all sets of $\beta_r$ that reproduce the
same $c_\gamma$.

For comparison with the QCD line, a sufficient condition is
\begin{align}
 \sum_{r=0}^{N_{\rm pot}}\frac{\beta_r^2}{\chi_r}
 &\leq
 \frac{(8/3-\mathcal C_{\rm QCD})^2}{\chi_{\rm QCD}}.
 \label{eq:general-qcd-line-condition}
\end{align}
Under this condition, Eq.~\eqref{eq:general-mass-photon-bound} gives a bound on the photon coupling relative to the axion mass that is at least as strong as the bound obtained for $r_{\rm level}=8/3$ with proportional QCD and electromagnetic couplings. 
Thus every massive state satisfying the assumptions above lies on or
below the QCD line. This condition is sufficient but not necessary.

Note that the bound above applies when the axion direction that couples to
photons is also constrained by the potentials parameterized by
$k_r^{\rm pot}$ and $\chi_r$. If there is a direction $\boldsymbol v$ satisfying
\begin{align}
 (k_r^{\rm pot})^T\boldsymbol v&=0
 \qquad\text{for all }r,
 \qquad
 c_\gamma^T\boldsymbol v\ne0,
\end{align}
these potentials do not give a mass to that direction even though it
couples to photons. Contributions contained in $M_{\rm rem}^2$ may
still lift it, but their size must then be included in deriving a
mass-coupling bound. 
A numerical bound therefore requires the potentials that give masses
to the photon-coupled axion directions to be identified and their
curvatures to be estimated. These quantities are not fixed by the
Green-Schwarz coefficients and depend on the compactification.

The geometric bound in
section~\ref{subsec:general-bounds} instead constrains the magnitude
of an ALP photon coupling without using the decomposition of
$M_{\rm other}^2$ or the values of $\chi_r$. 
Let $\widehat u_j$ be a normalized mass eigenvector with vanishing QCD coupling,
\begin{align}
 N_j
 &\equiv
 n_{\rm phys}^TG_{\rm phys}^{-1/2}\widehat u_j
 =0.
\end{align}
For the low-energy photon coefficient
$c_\gamma=e_{\rm phys}-\mathcal C_{\rm QCD}n_{\rm phys}$, with no
additional photon matching contribution, the term proportional to
$n_{\rm phys}$ vanishes upon projection onto this state. Its photon
coupling is then determined by
$p_{\rm phys}=e_{\rm phys}-(8/3)n_{\rm phys}$. 
For backgrounds without five-brane charge, using the flux relations and
the leading bulk metric of section~\ref{subsec:general-bounds}, we obtain
\begin{align}
 |g_{j\gamma\gamma}|
 &\leq\frac{\alpha_{\rm em}}{2\pi}
 \sqrt{p_{\rm phys}^TG_{\rm phys}^{-1}p_{\rm phys}}
 \leq\frac{\alpha_{\rm em}}{2\pi F_*}\,\mu(J)c_J,
 \label{eq:alp-geometric-photon-bound}
\end{align}
where the first inequality follows from Cauchy-Schwarz and
$\widehat u_j^T\widehat u_j=1$. The second uses $p_0=0$, the inverse-metric identity
of section~\ref{sec:physical-axion}, and
Eq.~\eqref{eq:general-bound-correlated-norm}. 

The same upper bound also constrains the combined photon coupling
of all QCD-neutral axion directions. Let $d$ be the number of physical
axions and assume $n_{\rm phys}\ne0$. In the canonical coordinates
$\phi=G_{\rm phys}^{1/2}\theta_{\rm phys}$, we choose an orthonormal basis
$\widehat u_j$, $j=1,\ldots,d-1$, satisfying $\nu^T\widehat u_j=0$, where
$\nu=G_{\rm phys}^{-1/2}n_{\rm phys}$. Each basis direction has
$g_{j\gamma\gamma}=(\alpha_{\rm em}/2\pi)
 p_{\rm phys}^TG_{\rm phys}^{-1/2}\widehat u_j$.
Since these vectors span all directions orthogonal to $\nu$,
$\sum_j \widehat u_j\widehat u_j^T=\mathbf1-\nu\nu^T/(\nu^T\nu)$.
Their squared couplings therefore satisfy
\begin{align}
 g_\perp^2
 &\equiv\sum_{j=1}^{d-1}|g_{j\gamma\gamma}|^2
 \nonumber\\
 &=\left(\frac{\alpha_{\rm em}}{2\pi}\right)^2
 \left[
 p_{\rm phys}^TG_{\rm phys}^{-1}p_{\rm phys}
 -\frac{\left(p_{\rm phys}^TG_{\rm phys}^{-1}n_{\rm phys}\right)^2}
 {n_{\rm phys}^TG_{\rm phys}^{-1}n_{\rm phys}}
 \right].
 \label{eq:alp-total-photon-coupling}
\end{align}
We take $g_\perp\ge0$. The second term subtracts the squared projection of
$G_{\rm phys}^{-1/2}p_{\rm phys}$ onto the canonical QCD direction. 
The sum of squared photon couplings is independent of the choice of orthonormal basis, although its value still depends
on the kinetic metric and the moduli. It vanishes when the physical
QCD and electromagnetic coefficient vectors are proportional. 

Sum rules relating axion masses and couplings have been
studied in Ref.~\cite{Gavela:2023tzu}. A more general treatment,
allowing independent QCD and electromagnetic anomaly vectors and
additional axion potentials, is given in
Ref.~\cite{FernandezNavarro:2026cyu}, in which the axion space is decomposed into the QCD direction and its orthogonal
directions. Eq.~\eqref{eq:alp-total-photon-coupling}
measures the photon coupling in the QCD-neutral space without
using the axion masses. We combine this projection identity with
the HYM equations and the Bianchi identity to obtain a geometric
upper bound on the combined photon coupling.

Since the subtracted term in Eq.~\eqref{eq:alp-total-photon-coupling} is nonnegative, 
the same $W=0$ assumptions
as in Eq.~\eqref{eq:alp-geometric-photon-bound} lead to 
\begin{align}
 \left(\sum_{j=1}^{d-1}|g_{j\gamma\gamma}|^2\right)^{1/2}
 &=g_\perp
 \le\frac{\alpha_{\rm em}}{2\pi F_*}\,\mu(J)c_J.
 \label{eq:alp-total-geometric-bound}
\end{align}
Thus at fixed geometry and moduli, the upper bound on the combined
coupling has no additional factor depending on the number of axion
directions.
Ref.~\cite{Halverson:2019cmy} studies how the effective
ALP-photon coupling varies with the number of axions in ensembles
of string compactifications. The bound above applies at fixed
geometry and moduli; its right-hand side can change when the
geometry and the number of axions change.
The bound also applies to any orthonormal subset, including mass
eigenstates with zero QCD coupling. 
When additional potentials mix only the QCD-neutral directions,
they redistribute the individual photon couplings but preserve
the sum of their squares at fixed moduli and fixed coupling
coefficients. 
If they also mix these directions with the QCD axion, the basis
used in Eq.~\eqref{eq:alp-total-photon-coupling} need not consist of
mass eigenstates.

\begin{figure}[tb]
 \centering
\includegraphics[width=0.92\textwidth]{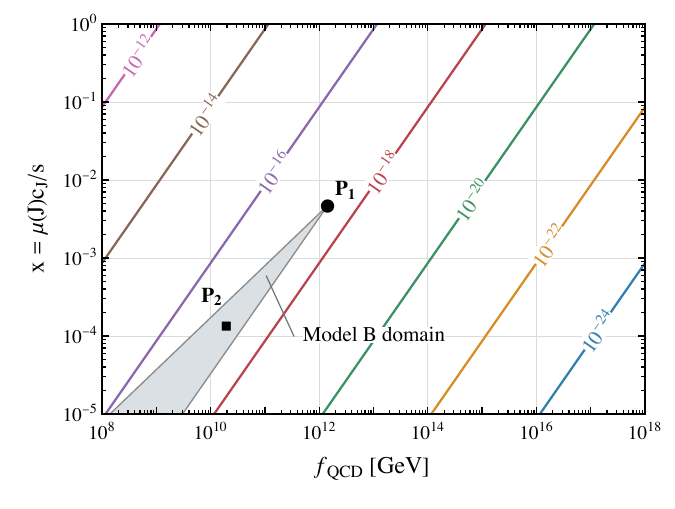}
 \caption{Upper bound on the photon coupling $g_\perp$ of QCD-neutral axions in the $(f_{\rm QCD},x)$ plane for $W=0$, using the leading bulk kinetic metric. The lines show constant
 values of $\alpha_{\rm em}x/(2\pi f_{\rm QCD})$, labeled in
 $\mathrm{GeV}^{-1}$, with $\alpha_{\rm em}^{-1}=137.036$.
The gray region corresponds to the symmetric HYM family $t^i=t$ of Model~B (section~\ref{sec:model-b}), with the illustrative cuts $t\geq5$ and $0<g_s\leq0.3$. The black circle $P_1$ and black square $P_2$ correspond to $(t,g_s)=(5,0.3)$ and $(10,0.1)$, respectively. These are the two parameter choices listed in Table~\ref{tab:modelb-general-bound}.}
 \label{fig:alp-bound-fqcd-x}
\end{figure}

Using $x=\mu(J)c_J/s$, the geometric bound becomes
\begin{align}
 |g_{j\gamma\gamma}|
 &\leq g_\perp
 \leq \frac{\alpha_{\rm em}}{2\pi F_*}\,s x.
 \label{eq:alp-geometric-bound-x}
\end{align}
For the leading bulk metric with $n_0=1$,
Eq.~\eqref{eq:qcd-decay-constant} gives
$f_{\rm QCD}^{-2}
=F_*^{-2}(s^2+n_K^T\mathcal H^{-1}n_K)$. 
It follows that
\begin{align}
 g_\perp
 &\leq \frac{\alpha_{\rm em}}{2\pi f_{\rm QCD}}\,x
\simeq 1.16\times10^{-21}
 \left(\frac{x}{10^{-2}}\right)
 \left(\frac{10^{16}\,\mathrm{GeV}}{f_{\rm QCD}}\right)
 \,\mathrm{GeV}^{-1}.
 \label{eq:alp-normalized-bound-xmax}
\end{align}
Here the numerical coefficient uses $\alpha_{\rm em}^{-1}=137.036$. 
Figure~\ref{fig:alp-bound-fqcd-x} shows the upper bound in
Eq.~\eqref{eq:alp-normalized-bound-xmax} as a function of
$f_{\rm QCD}$ and $x$. At fixed $f_{\rm QCD}$, a smaller $x$
gives a stronger bound on the photon coupling. Since
$|g_{j\gamma\gamma}|\leq g_\perp$, this bound also applies to
each QCD-neutral ALP. The values marked on the lines correspond to upper limits. 
Note that the values of
$f_{\rm QCD}$ and $x$ must be calculated for each compactification and need not vary independently.
For illustration, the gray region shows a range of Model~B backgrounds, while the marked points $P_1$ and $P_2$ indicate two reference points. Model~B is introduced in section~\ref{sec:model-b}.

The bound can also be evaluated directly from $\mu(J)c_J$. 
Using $F_*^2=M_P^2/(8\pi^2)$, the last term in
Eq.~\eqref{eq:alp-geometric-photon-bound} becomes
\begin{align}
 \frac{\alpha_{\rm em}}{2\pi F_*}\mu(J)c_J
 &\simeq
 4.2\times10^{-18}
 \left(
 \frac{\mu(J)c_J}{10^3}
 \right)
 {\rm GeV}^{-1}.
 \label{eq:alp-geometric-photon-bound-numerical}
\end{align}
For the CICY geometries listed in
Table~\ref{tab:cicy-geometric-inputs} at $J=5J_0$, where $J_0=\sum_i\omega_i$ in each ambient hyperplane basis,
this gives upper
bounds of order $10^{-18}\ {\rm GeV}^{-1}$. These values are
illustrative and scale linearly with the overall K\"ahler parameter
$\lambda$ for $J=\lambda J_0$.

These bounds constrain the photon couplings but not the ALP
masses, which depend on non-QCD potentials.
Figure~\ref{fig:general-cy-mass-photon} assumes that QCD determines
both the mass and the axion direction and therefore does not
give a general bound for QCD-neutral ALPs.

\subsection{Comparison with supersymmetric heterotic compactifications}
\label{sec:susy-comparison}

We compare supersymmetric and non-supersymmetric heterotic
compactifications with respect to two properties. The first concerns the relation
between the QCD and electromagnetic axion couplings. If these couplings
are proportional on the physical axion space, their ratio is independent
of the axion kinetic matrix. The second concerns the axion potentials
that determine the physical mass eigenstates and whether they allow a
bound on the photon coupling relative to the axion mass. These two
properties are logically independent.

We first consider the $E_8\times E_8$ embeddings studied in
Refs.~\cite{Agrawal:2024ejr,Reig:2025dqb}, where the hypercharge
embedding determines whether the QCD and electromagnetic couplings have
the same axion dependence. We then turn to supersymmetric $SO(32)$
compactifications, where quartic gauge invariants can already generate
nonuniversal axion couplings. These examples allow us to distinguish the
features specific to the present non-supersymmetric construction from
those that also occur in supersymmetric heterotic compactifications.

\subsubsection{\texorpdfstring{$E_8\times E_8$}{E8 x E8} embeddings and axion-coupling relations}
\label{subsec:susy-embedding-comparison}

For the embeddings considered in Ref.~\cite{Agrawal:2024ejr} in which
the Standard Model gauge factors arise from a single $E_8$ factor,
color, weak isospin, and hypercharge couple to the same combination of
the universal axion $a$ and the K\"ahler axions $b_i$. 
Omitting possible charged-matter phases, the interactions in the normalization of
section~\ref{sec:physical-framework} are
\begin{align}
 \mathcal L_{\rm top}\supset{}&
 \frac{\alpha_s}{8\pi}k_3
 \left(a+\sum_i\nu_i^{E_8} b_i\right)
 G^a_{\mu\nu}\widetilde G^{a\mu\nu}
+\frac{\alpha_{\rm em}}{8\pi}(k_2+k_Y)
 \left(a+\sum_i\nu_i^{E_8} b_i\right)
 F_{\mu\nu}\widetilde F^{\mu\nu},
 \label{eq:susy-common-axion-lagrangian}
\end{align}
where the coefficients
$\nu_i^{E_8}$ are determined by integrals of internal gauge and
tangent-bundle characteristic classes.
The numerical values of the $\nu_i^{E_8}$ then cancel from the ratio of the
electromagnetic and QCD couplings, giving
\begin{align}
 \frac EN=\frac{k_2+k_Y}{k_3}.
 \label{eq:susy-proportional-level-ratio}
\end{align}
For $k_3=k_2=1$ and $k_Y=5/3$, this gives $E/N=8/3$.
In the language of section~\ref{sec:physical-framework}, the required
property is the proportionality of the physical electromagnetic and
QCD coupling vectors. Once this proportionality holds, kinetic mixing
among the axions cannot change the ratio. 
The non-supersymmetric $SO(16)_{\rm v}\times SO(16)_{\rm h}$
theory is not included in the analysis of
Ref.~\cite{Agrawal:2024ejr}.

Ref.~\cite{Reig:2025dqb} considers embeddings for which the
proportionality above need not hold. Color and weak isospin arise
from the first $E_8$, while hypercharge receives contributions from
Abelian factors in both $E_8$ groups. The photon can therefore couple
to axion combinations associated with the second gauge factor that
do not couple to QCD. This provides a source of electromagnetic
axion couplings independent of the QCD coupling. 
The line bundles must be chosen so that this hypercharge combination
remains massless. In the case with one relevant Abelian factor from
each $E_8$, the associated line bundles have proportional first
Chern classes,
\begin{align}
 c_1(L_i)&=\ell_i \eta_Y,
 \qquad\mathrm{with}\qquad
 \ell_i\in\mathbb Z,
 \label{eq:susy-proportional-hypercharge-bundles}
\end{align}
where $i=1,2$ and $\eta_Y\in H^2(X,\mathbb Z)$. Their internal
St\"uckelberg charges are then proportional, allowing a linear
combination of the two Abelian generators to have vanishing internal
axion charges~\cite{Blumenhagen:2006ux}. The full masslessness
condition must also include the universal and any localized axions,
as in $Qy=0$ in section~\ref{subsec:embedding-generators}.

In addition to the axion combinations that couple universally to the
gauge factors descending from each $E_8$, the line bundles generate
an extra contribution to the Abelian axion couplings. Following
Ref.~\cite{Reig:2025dqb}, this contribution depends on
\begin{align}
 \varphi
 &=
 \sum_j b_j\int_X\omega_j^{\rm ref}\wedge \eta_Y^2,
 \label{eq:susy-additional-photonic-direction}
\end{align}
where $\omega_j^{\rm ref}$ are the internal two-forms in the expansion
$B_2=\sum_j b_j\omega_j^{\rm ref}+\cdots$.
This combination enters the couplings to the Abelian factors
participating in hypercharge, but not those to the non-Abelian
gauge sectors in that construction. 
Hence, this axion combination contributes to the photon coupling but not to the QCD coupling. 
These statements describe the topological interactions, not the
canonically normalized mass eigenstates.

In our direct-flux construction, the independent electromagnetic
axion dependence has a different origin. Hypercharge has no hidden
component, and all Standard Model gauge factors lie in the same
$SO(16)_{\rm v}$. Nevertheless, the quartic terms in
Eq.~\eqref{eq:general-cartan-coupling} distinguish the fluxes carried
by the different visible blocks. For the embedding of
section~\ref{subsec:embedding-proportionality}, the difference is
$P^{(4)}=A^2-B^2-L^2$. Subject to the conditions in
section~\ref{subsec:embedding-physical-rank}, a nonzero class gives
independent physical QCD and electromagnetic couplings. Thus
hypercharge need not extend over both ten-dimensional gauge factors
for this nonproportionality to occur. In either construction, the
mass eigenstates and their positions relative to the QCD line
also depend on the axion potentials.

\subsubsection{Nonuniversal couplings in supersymmetric
\texorpdfstring{$SO(32)$}{SO(32)} compactifications}
\label{subsec:susy-so32-nonuniversality}

Nonuniversal axion couplings also occur in supersymmetric $SO(32)$
compactifications~\cite{Blumenhagen:2005pm}. We first explain their
flux dependence and then use a Pati-Salam embedding to compare
the QCD and electromagnetic couplings with those in the
non-supersymmetric $SO(16)_{\rm v}\times SO(16)_{\rm h}$
heterotic compactifications.

In the supersymmetric $SO(32)$ heterotic string, the ten-dimensional
Green-Schwarz interaction also contains quartic gauge-field terms.
Schematically,
\begin{align}
 S_{\rm GS}
 &\supset
 \int B_2\wedge \mathscr X_8(F,R),
\end{align}
where the gauge-field part of $\mathscr X_8$ contains quartic
invariants such as $\operatorname{Tr}F^4$ and
$(\operatorname{Tr}F^2)^2$~\cite{Blumenhagen:2005pm}.
As in the non-supersymmetric $SO(16)_{\rm v}\times SO(16)_{\rm h}$
theory discussed in section~\ref{subsec:bulk-gs}, these quartic
terms generate flux-dependent axion couplings after dimensional
reduction. In particular, expanding
$F=\overline F+T\mathcal F$ produces terms proportional to
\begin{align}
 \operatorname{Tr}(\overline F^2T^2)\,
 \mathcal F\wedge\mathcal F,
\end{align}
which couple the K\"ahler axions to the four-dimensional gauge
topological density.

To see this explicitly, let us consider a $U(N)$ subgroup with an
internal Abelian background
$\overline F=2\pi A\,\mathbf1_N$, where $A$ is normalized as in
section~\ref{subsec:10d-action}. For an unbroken generator
$T\in\mathfrak{su}(N)$, we have
$[\overline F,T]=0$ and
$\operatorname{tr}_N(\overline F T)=0$. Here
$\operatorname{tr}_N$ denotes the trace over the fundamental
representation of $U(N)$. Nevertheless,
\begin{align}
 \operatorname{tr}_N(\overline F^2T^2)
 &=(2\pi)^2A^2\operatorname{tr}_N(T^2),
 \label{eq:commuting-quartic-example}
\end{align}
which is generally nonzero. The conjugate representation contained
in $SO(32)$ gives the same contribution. Thus an unbroken
non-Abelian gauge factor can have a K\"ahler-axion coupling that
depends on the internal flux associated with the corresponding
$U(N)$ subgroup.

The common axion dependence for $SO(32)$ in Ref.~\cite{Agrawal:2024ejr} is derived using the
vanishing of the mixed quartic trace. For the Abelian background
above, however, Eq.~\eqref{eq:commuting-quartic-example} shows that
$[\overline F,T]=0$ and
$\operatorname{tr}_N(\overline F T)=0$ do not imply
$\operatorname{tr}_N(\overline F^2T^2)=0$. Commutativity is therefore
not sufficient to establish a common axion combination for all
unbroken gauge factors in these line-bundle backgrounds. This
distinction applies to the Green-Schwarz coefficients obtained
from dimensional reduction. Whether the physical mass eigenstates satisfy the bound derived for
proportional QCD and electromagnetic couplings must still be determined
after low-energy matching and the axion potentials are taken into account.

The same flux dependence appears in the holomorphic gauge kinetic
functions derived in Ref.~\cite{Blumenhagen:2005pm}. For an
$SU(N_i)$ factor preserved by an Abelian background in a
$U(N_i)$ subgroup, the gauge kinetic function including the
one-loop correction is
\begin{align}
 f_{SU(N_i)}
 &=2S+\sum_k T_k
 \left[
  (\overline f_i^2)_k
  -\frac1{48}(\operatorname{tr}\overline R^2)_k
 \right].
 \label{eq:susy-so32-gauge-function}
\end{align}
Here $S$ and $T_k$ are the dilaton and K\"ahler-modulus
superfields, whose imaginary parts are axions. The symbol
$\overline f_i$ denotes the internal Abelian flux associated
with the $U(N_i)$ subgroup. The coefficients
$(\overline f_i^2)_k$ and
$(\operatorname{tr}\overline R^2)_k$ follow the convention of
Ref.~\cite{Blumenhagen:2005pm}, in which the corresponding internal
four-forms are expanded after factoring out $(2\pi)^2$.
The term $2S$ is the tree-level contribution, while the terms
proportional to $T_k$ arise at one loop. Their imaginary parts
encode the same flux-dependent K\"ahler-axion couplings obtained
from the Green-Schwarz reduction. 
The curvature term in
Eq.~\eqref{eq:susy-so32-gauge-function} originates from the
mixed gauge-gravity term in the ten-dimensional $SO(32)$
Green-Schwarz eight-form~\cite{Blumenhagen:2005pm}.
Both the supersymmetric $SO(32)$ and $E_8\times E_8$ theories
contain mixed gauge-gravity and purely gravitational terms
in their Green-Schwarz eight-forms~\cite{Blaszczyk:2015zta}.
These curvature terms are absent from
Eq.~\eqref{eq:x8-polynomial} for the non-supersymmetric
$SO(16)_{\rm v}\times SO(16)_{\rm h}$ theory.
In both supersymmetric theories, the K\"ahler-axion couplings
to gauge fields generated by the mixed term are proportional
to the embedding levels. This contribution therefore cancels from
$C_\gamma-[(k_2+k_Y)/k_3]C_3$.
For the $SO(32)$ line-bundle backgrounds discussed below,
the nonuniversal axion couplings instead arise from
flux-dependent quartic gauge contractions.

The imaginary part of each gauge kinetic function determines the
axion coupling to the corresponding gauge topological density.
For two such unitary subgroups, the common dilaton and curvature terms
cancel in the difference:
\begin{align}
 f_{SU(N_i)}-f_{SU(N_j)}
 &=\sum_k T_k
 \left[
  (\overline f_i^2)_k-(\overline f_j^2)_k
 \right].
 \label{eq:susy-unitary-function-difference}
\end{align}
Different flux-square coefficients therefore give different
K\"ahler-axion couplings. To determine how these differences
affect the QCD and electromagnetic couplings, we must specify
the Standard Model embedding.

A Pati-Salam embedding is useful for this purpose and connects
the discussion to intersecting D-brane models. Supersymmetric
$SO(32)$ heterotic compactifications are related by S-duality
to Type I compactifications, where line-bundle backgrounds
correspond to magnetic fluxes on D9-branes
\cite{Blumenhagen:2005pm}. In toroidal examples, T-duality maps
these configurations to intersecting D6-branes in Type IIA.
This connection motivates the study of Pati-Salam-like
$SO(32)$ line-bundle models, including those in
Ref.~\cite{Otsuka:2018oyf}.

Let $C_4$, $C_L$, and $C_R$ denote the Green-Schwarz
axion-coupling vectors for the level-one factors
$SU(4)_C$, $SU(2)_L$, and $SU(2)_R$, respectively, in the
normalization of section~\ref{sec:physical-framework}.
For the standard hypercharge embedding
$Y=T_R^3+(B-L)/2$, with no contribution from additional
Abelian generators, the interactions inherited by the
Standard Model subgroup satisfy
\begin{align}
 C_3&=C_4,
 \qquad
 C_2=C_L,
 \qquad
 C_Y=C_R+\frac23C_4.
 \label{eq:susy-ps-coefficient-matching}
\end{align}
The factor $2/3$ follows from the normalization of $(B-L)/2$
inside $SU(4)_C$. Using $C_\gamma=C_2+C_Y$, we obtain
\begin{align}
 C_\gamma-\frac83C_3
 &=C_L+C_R-2C_4.
 \label{eq:susy-ps-photon-difference}
\end{align}
The relation $C_\gamma=(8/3)C_3$ therefore holds in this sector
when $C_L+C_R=2C_4$. Equal coupling vectors for all three factors
are sufficient, but Pati-Salam gauge symmetry does not impose
this relation. The flux-dependent terms in
Eq.~\eqref{eq:susy-so32-gauge-function} can make
$C_L+C_R-2C_4$ nonzero.

This can be checked in a concrete model of
Ref.~\cite{Otsuka:2018oyf}. In that construction, the fluxes
associated with $SU(4)_C$, $SU(2)_L$, and $SU(2)_R$ are determined
by $L_2$, $L_1$, and $L_4$, respectively, with
$L_5\simeq L_4^{-1}$. The common dilaton and curvature terms
cancel in $C_L+C_R-2C_4$. Thus the bulk $8/3$ relation requires
\begin{align}
 c_1(L_1)^2+c_1(L_4)^2-2c_1(L_2)^2
 &=0
 \qquad\text{in }H^4(M,\mathbb R),
 \label{eq:susy-ps-bundle-relation}
\end{align}
where $M$ denotes the covering Calabi-Yau threefold used in Ref.~\cite{Otsuka:2018oyf}.

Following Ref.~\cite{Otsuka:2018oyf}, we take the coefficients
in $c_1(L_a)=m_a^i\omega_i$ to be 
$m_1=(-3,0,0,3)$, $m_2=(1,0,-1,2)$, and
$m_4=(0,-1,1,4)$.
Using the intersection numbers given there, we find
\begin{align}
 \left(
 \int_M\omega_i\wedge
 \left[c_1(L_1)^2+c_1(L_4)^2-2c_1(L_2)^2\right]
 \right)_{i=1}^{4}
 =(22,4,-74,-12).
 \label{eq:susy-ps-example-difference}
\end{align}
The class in Eq.~\eqref{eq:susy-ps-bundle-relation} is therefore
nonzero. This example does not obey the $8/3$ relation for the
bulk Green-Schwarz coefficients.

The same example also shows why these coefficients cannot
immediately be identified with the physical couplings.
The internal axion shift under $U(1)_1$ is proportional to $m_1$.
The contraction of $m_1$ with the vector in
Eq.~\eqref{eq:susy-ps-example-difference} is $-102$, so the bulk
coupling difference is not invariant under this shift.
Its universal-axion component vanishes and cannot cancel this
variation. This does not indicate an inconsistency: the
Green-Schwarz interactions participate in anomaly cancellation.
It means that the corresponding matter contributions must be
included before the difference can be treated as a physical
axion coupling. The additional $U(1)_{B-L}$ that remains in
this model must also be taken into account when identifying
the low-energy photon. We therefore do not infer a physical
value of $E/N$ from
Eq.~\eqref{eq:susy-ps-example-difference}.

In the direct-flux $SO(16)_{\rm v}\times SO(16)_{\rm h}$
construction, the Standard Model gauge factors arise from
$SO(16)_{\rm v}$ without an intermediate Pati-Salam symmetry.
The non-supersymmetric Green-Schwarz reduction gives
$P^{(4)}=A^2-B^2-L^2$ for the electromagnetic-color difference.
Model~B in section~\ref{sec:model-b} provides a concrete example
in which the QCD coefficient and this difference satisfy
$C_3^TQ=P^TQ=0$, with $P\neq0$. Since $(C_3)_0=1$ and $P_0=0$,
the two coupling vectors remain linearly independent after the
St\"uckelberg mechanism. In contrast to the Pati-Salam example
above, both Green-Schwarz coefficients are already invariant under
the gauged axion shifts. Finite low-energy contributions can still
modify the corresponding physical couplings, and their absence is
assumed in the numerical analysis of Model~B. 
The fluxes that determine these QCD and electromagnetic axion
couplings are simultaneously constrained by the HYM equations and
the Bianchi identity. These conditions lead to the bounds derived
in section~\ref{subsec:general-bounds}. 
Flux-dependent nonuniversal Green-Schwarz couplings can occur
in both theories, despite the different curvature dependence
of their eight-forms.

\section{Explicit compactifications}
\label{sec:models}

We first evaluate the geometric quantities of
section~\ref{subsec:general-bounds} for a specified set of CICYs.
We then study two explicit compactifications to determine their
axion couplings and discuss how the axion potentials affect the
relation between the axion mass and photon coupling. 
Section~\ref{sec:model-a}
presents an $SU(5)$-like construction with proportional Green-Schwarz
couplings. Section~\ref{sec:model-b} presents a Standard-Model-like
compactification obtained by direct flux breaking of $SO(16)_{\rm v}$, with independent QCD and electromagnetic Green-Schwarz
coupling vectors. For each model, we specify the
conditions under which these coefficients determine the physical
couplings.

\subsection{Geometric bounds for CICYs}
\label{subsec:cicy-geometric-inputs}

We now evaluate the geometric quantities that determine the bound on
$E/N$ derived in section~\ref{subsec:general-bounds} for several CICYs~\cite{Candelas:1987kf,Anderson:2017aux}. We use the reference
K\"ahler form $J_0=\sum_i\omega_i$ in the ambient hyperplane basis.
It sets $t^i=1$, so $c_{J_0}=\sum_i c_{2i}$ is the value of $c_J$
at this reference point. 
This choice is needed because $\mu c_J$ is not invariant under an overall rescaling:
$\mu(\lambda J_0)c_{\lambda J_0}=\lambda\mu(J_0)c_{J_0}$, although the combination $\mu c_J/\mathcal V^{1/3}$ does not depend on this
rescaling. In terms of this combination, the parameter in
Eq.~\eqref{eq:general-bound-x} is
\begin{align}
 x&=\frac{\mu c_J}{\mathcal V^{1/3}}
       \left(\frac{g_s}{s}\right)^{2/3}.
 \label{eq:cicy-scale-free-x}
\end{align}
This separates the dependence on the geometry from that on the overall size and the string coupling.

We evaluate $\mu(J_0)$ directly for several CICYs in
Table~\ref{tab:cicy-geometric-inputs}, including the geometry of
Model~B in section~\ref{sec:model-b}. The intersection numbers follow from the ambient-space
integration in Eq.~\eqref{eq:modelb-intersection-rule}. For a CICY
with multidegrees $d_{i\alpha}$ in $\prod_i\mathbb P^{n_i^{\rm amb}}$, the
second Chern class is~\cite{Candelas:1987kf,Anderson:2017aux}
\begin{align}
 c_2(TX)&=\frac12\sum_{ij}
 \left[\sum_\alpha d_{i\alpha}d_{j\alpha}
       -(n_i^{\rm amb}+1)\delta_{ij}\right]\omega_i\wedge\omega_j.
 \label{eq:cicy-second-chern}
\end{align}
For $\xi=\xi^i\omega_i$, let
$Z_i(\xi)=\kappa_{ijk}\xi^j\xi^k$ and
$\mathcal H_0=\mathcal H(J_0)$. The ratio in
Eq.~\eqref{eq:general-bound-mu} is unchanged by rescaling $\xi$,
so we may impose $q_{J_0}(\xi)=1$. The definition then becomes
\begin{align}
 \mu(J_0)^2
 &=\max_{\substack{\tau_i(J_0)\xi^i=0\\q_{J_0}(\xi)=1}}
 Z_i(\xi)(\mathcal H_0^{-1})^{ij}Z_j(\xi).
 \label{eq:cicy-direct-maximum}
\end{align}
This is the maximum of a quartic polynomial subject to the
primitivity condition
$\int_X J_0^2\wedge\xi=0$ and the quadratic normalization
$q_{J_0}(\xi)=1$. The normalized primitive classes form a compact
set because $q_{J_0}$ is positive definite on the primitive
subspace. We maximize over
$\xi\in H^{1,1}(X,\mathbb R)$, allowing arbitrary real
coefficients in the chosen cohomology basis. The class giving the
maximum therefore need not correspond to an integral, quantized
flux background.

For the first two geometries in Table~\ref{tab:cicy-geometric-inputs},
the primitive subspace is one-dimensional. Writing
$\xi=\xi^i\omega_i$ in the ambient hyperplane basis, the primitivity
condition fixes the direction of $\xi$ up to an overall normalization.
The corresponding primitive directions are proportional to
$(1,-1)$ and $(-5,2)$, respectively. Since the ratio defining
$\mu(J_0)$ is invariant under an overall rescaling of $\xi$, these
directions give $\mu=1/\sqrt3$ and $\mu=1$, respectively.
For the tetraquadric and the complete intersection in
$(\mathbb P^1)^5$, primitivity at $J_0$ is $\sum_i\xi^i=0$.
Writing $S_p=\sum_i(\xi^i)^p$, the quantity to be maximized is
\begin{align}
 \frac{\lVert\xi^2\rVert_{\mathcal H_0^{-1}}^2}
      {q_{J_0}(\xi)^2}
 &=\begin{cases}
 -\dfrac16+2\dfrac{S_4}{S_2^2},&h^{1,1}=4,\\[6pt]
 \dfrac1{27}+\dfrac{40}{27}\dfrac{S_4}{S_2^2},&h^{1,1}=5.
 \end{cases}
 \label{eq:cicy-symmetric-direct-ratios}
\end{align}
For $h$ real components with zero sum,
$S_4/S_2^2\le(h^2-3h+3)/[h(h-1)]$ for $h=4,5$, with equality
for a vector proportional to $(h-1,-1,\ldots,-1)$.
Both maxima in Eq.~\eqref{eq:cicy-symmetric-direct-ratios} are
therefore one. The full primitive-space maximum for Model~B is
derived in section~\ref{subsec:modelb-couplings}.

To relate these values to Figure~\ref{fig:general-cy-mass-photon},
let us vary only the overall K\"ahler scale and write
$J=\lambda J_0$. 
Since $\mathcal V(\lambda J_0)=\lambda^3\mathcal V(J_0)$ and
$c_{\lambda J_0}=\lambda c_{J_0}$, we obtain
\begin{align}
 x&=\frac{\mu(J_0)c_{J_0}}{\mathcal V(J_0)}
       \frac{g_s^2}{\lambda^2}.
 \label{eq:cicy-fixed-shape-x}
\end{align}
For $\lambda\ge\lambda_{\min}>0$ and $0<g_s\le g_{\max}\le1$,
this expression is largest at $\lambda=\lambda_{\min}$ and
$g_s=g_{\max}$. Thus
\begin{align}
 x&\le x_{\max},
 \qquad\mathrm{with}\qquad
 x_{\max}\equiv
 \frac{\mu(J_0)c_{J_0}}{\mathcal V(J_0)}
 \frac{g_{\max}^2}{\lambda_{\min}^2}.
 \label{eq:cicy-xmax}
\end{align}
Table~\ref{tab:cicy-geometric-inputs} gives the geometric coefficient
in this expression and an example with $\lambda_{\min}=5$ and
$g_{\max}=0.3$. 
All five entries use the exact maxima of
Eq.~\eqref{eq:cicy-direct-maximum} evaluated at the corresponding
$J_0$. For the tetraquadric, we take
$J_0=\sum_i\omega_i$, so that all K\"ahler parameters satisfy
$t^i=1$. This choice is used for the geometric comparison in
Table~\ref{tab:cicy-geometric-inputs} and need not coincide with
the HYM locus of an explicit compactification.

\begin{table}[t]
 \centering
 \small
 \renewcommand{\arraystretch}{1.2}
 \begin{tabularx}{\textwidth}{@{}>{\raggedright\arraybackslash}Xcccc@{}}
 \toprule
 Geometry & $h^{1,1}$ & $\mathcal V(J_0)$
 & $\displaystyle\frac{\mu(J_0)c_{J_0}}{\mathcal V(J_0)}$
 & $x_{\max}$\\
 \midrule
 $(3,3)$ in $\mathbb P^2\times\mathbb P^2$
 &2&$3$&$13.856$&$4.99\times10^{-2}$\\
 $(2,4)$ in $\mathbb P^1\times\mathbb P^3$
 &2&$7/3$&$29.143$&$1.05\times10^{-1}$\\
 $(2,2,2,2)$ in $(\mathbb P^1)^4$
 &4&$8$&$12$&$4.32\times10^{-2}$\\
 Two $(1,1,1,1,1)$ equations in $(\mathbb P^1)^5$
 &5&$20$&$6$&$2.16\times10^{-2}$\\
 Model~B, Eq.~\eqref{eq:modelb-configuration}
 &6&$138$&$1.305$&$4.70\times10^{-3}$\\
 \bottomrule
 \end{tabularx}
 \caption{Maximum values of $x$ along the one-parameter family
$J=\lambda J_0$, with $\lambda\ge\lambda_{\min}$ and
$0<g_s\le g_{\max}$, where $J_0=\sum_i\omega_i$ in each ambient
basis. Along this family, the ratios of the K\"ahler parameters are
kept fixed while the overall scale is varied. The fourth column
determines $x_{\max}$ through Eq.~\eqref{eq:cicy-xmax}. The last
column uses $\lambda_{\min}=5$ and $g_{\max}=0.3$. The values of
$\mu(J_0)$ are $1/\sqrt3$, $1$, $1$, $1$, and
$\sqrt{(3346+161\sqrt{21})/5875}$, respectively.}
 \label{tab:cicy-geometric-inputs}
\end{table}

For the sample regions in Table~\ref{tab:cicy-geometric-inputs}, all
five values satisfy $x_{\max}<1$. The $x_{\max}=1$ interval in
Figure~\ref{fig:general-cy-mass-photon} is therefore conservative for
these regions whenever the flux and low-energy assumptions of
section~\ref{subsec:general-bounds} are satisfied.

\subsection{Model A: an \texorpdfstring{$SU(5)$}{SU(5)}-like model with proportional Green-Schwarz couplings}
\label{sec:model-a}

We first study the non-supersymmetric line-bundle model discussed in
Ref.~\cite{GrootNibbelink:2015lme}. We focus on its $SU(5)$ embedding,
for which the QCD and electromagnetic Green-Schwarz axion-coupling
vectors satisfy
$C_\gamma=(8/3)C_3$.

The $SU(5)$ embedding of this model differs from the direct Standard
Model embedding studied in section~\ref{sec:embedding-criterion}. 
The general Cartan reduction of section~\ref{subsec:bulk-gs} and the
physical criterion of section~\ref{sec:physical-framework} apply to
both. Let $\widetilde X$ denote the tetraquadric covering space and
$X=\widetilde X/\mathbb Z_2$ its freely acting quotient. The tetraquadric
is a hypersurface of multidegree $(2,2,2,2)$ in $(\mathbb P^1)^4$.
A Wilson line on $X$ breaks the visible $SU(5)$ to the Standard Model
subgroup. The quotient spectrum contains three chiral generations.
In the following, the subscript $\mathrm A$ on $C_{3,\mathrm A}$, $Q_{\mathrm A}$, $n_{\mathrm A}$, and $e_{\mathrm A}$ labels
Model~A rather than an axion component.

To evaluate the Green-Schwarz coefficients, we first specify the
covering-space geometry and the flux embedding. The nonzero intersection
numbers and second-Chern-class components on $\widetilde X$ are
\begin{align}
 \kappa_{ijk}=2
 \quad\text{for mutually distinct }i,j,k,
 \qquad
 c_{2i}=24.
 \label{eq:a-topology}
\end{align}
Here $i=1,\ldots,4$, and intersection numbers with repeated indices vanish.
In the same divisor order, the flux vectors in
Ref.~\cite{GrootNibbelink:2015lme} have visible and hidden components:
\begin{align}
 V^{(\mathrm{vis})}
 &=
 \begin{pmatrix}
 1&-1&-1&-1&1&3&1&1\\
 0&0&0&0&0&-1&0&-1\\
 -1&1&1&1&-1&0&-2&-1\\
 0&0&0&0&0&-1&1&2
 \end{pmatrix},
 \label{eq:a-visible-flux}
\end{align}
and
\begin{align}
 V^{(\mathrm{hid})}
 &=
 \begin{pmatrix}
 0&0&0&0&0&0&0&0\\
 1&0&-1&1&1&0&0&0\\
 -1&0&0&0&0&0&-1&0\\
 -1&0&2&-2&-2&0&1&0
 \end{pmatrix}.
 \label{eq:a-hidden-flux}
\end{align}
The row index labels the divisor, and the column index labels a Cartan
generator. 
Let $\omega_i^{\rm cov}$ denote the restrictions of the ambient
hyperplane classes to $\widetilde X$, in the divisor order used above.
On the covering space, the fluxes are
$f_I=V_i^I\omega_i^{\rm cov}$. Inserting them into
Eq.~\eqref{eq:bianchi} gives
\begin{align}
 \left(\int_{\widetilde X}\omega_i^{\rm cov}\wedge S_{\rm v}\right)_{i=1}^4
 &=(-16,-48,-20,-44),
\nonumber\\
 \left(\int_{\widetilde X}\omega_i^{\rm cov}\wedge S_{\rm h}\right)_{i=1}^4
 &=(-32,0,-28,-4).
 \label{eq:a-hidden-flux-squares}
\end{align}
Their sum is $-48$ in each component, so
$W_i=24-48/2=0$. The Bianchi identity is therefore satisfied
without five-brane charge, and no NS5-branes are required in this
compactification. 
The leading HYM equations in Eq.~\eqref{eq:hym} are solved at
\begin{align}
 (t^1,t^2,t^3,t^4)
 =
 t\left(1,\frac14,1,1\right),
 \qquad
 t>0.
 \label{eq:a-hym}
\end{align}
At this point the divisor volumes are
\begin{align}
 (\tau_1,\tau_2,\tau_3,\tau_4)
 &=3t^2(1,2,1,1).
 \label{eq:a-divisor-volumes-check}
\end{align}
They satisfy $V_i^I\tau_i=0$ for every visible and hidden Cartan component.
All four K\"ahler parameters are positive.

The hypercharge generator $Y_{\mathrm A}=\sum_I(y_{\mathrm A})_I H_I$ is
\begin{align}
 y_{\mathrm A}=
 \left(
 -\frac13,\frac13,\frac13,-\frac12,\frac12,0,0,0;
 0^8
 \right)^T,
 \label{eq:a-hypercharge}
\end{align}
leading to its level:
\begin{align}
 2y_{\mathrm A}^Ty_{\mathrm A}=\frac53.
\end{align}
The relative signs reflect the choice of Cartan basis and do not change
the standard $SU(5)$ hypercharge embedding. The flux matrices above give
$V_i^I(y_{\mathrm A})_I=0$. The universal-axion charge obtained from
Eq.~\eqref{eq:universal-gauging} also obeys $Q_0{}^I(y_{\mathrm A})_I=0$.
Thus $Q_{\mathrm A}y_{\mathrm A}=0$, as required by Eq.~\eqref{eq:massless-hypercharge}.

We now evaluate the coefficients on the geometric quotient. Let
$\pi:\widetilde X\to X$ be the covering map. In real cohomology, 
we choose the basis forms $\omega_i$ on $X$ so that
$\pi^*\omega_i=\omega_i^{\rm cov}$. For a descended six-form $\eta^{(6)}$,
\begin{align}
 \int_X\eta^{(6)}&=\frac12\int_{\widetilde X}\pi^*\eta^{(6)}.
 \label{eq:a-covering-integral}
\end{align}
The factor $1/2$ therefore multiplies the internal integrals in the
Green-Schwarz coefficients. It does not change the universal coefficient,
which is normalized to one. Using the local axion coordinates
\begin{align}
 \Theta_{\mathrm A}=(a_0,b^1,b^2,b^3,b^4)^T,
\end{align}
we write the color interaction as
\begin{align}
 \mathcal L_{{\rm GS},\mathrm A}^{(4)}
 &\supset
 \frac{\alpha_s}{8\pi}(C_{3,\mathrm A}^T\Theta_{\mathrm A})
 G^a_{\mu\nu}\widetilde G^{a\mu\nu}.
 \label{eq:a-color-lagrangian}
\end{align}
The first five visible fluxes are $(A,-A,-A,-A,A)$, with
$A=\omega_1^{\rm cov}-\omega_3^{\rm cov}$. 
In a Cartan basis adapted to the $SU(5)$ subgroup,
Eq.~\eqref{eq:nonabelian-row} applies to the common internal flux
carried by the $SU(5)$ directions. The K\"ahler-axion coefficients
of the color Green-Schwarz coupling on the covering space are
\begin{align}
 \widetilde C_{i3}
 &=
 \int_{\widetilde X}
 \omega_i^{\rm cov}\wedge
 \left(B_{\rm v}-A^2\right),
 \qquad i=1,\ldots,4,
 \label{eq:modela-cover-color-coefficients}
\end{align}
which gives
\begin{align}
 \left(
 \widetilde C_{13},
 \widetilde C_{23},
 \widetilde C_{33},
 \widetilde C_{43}
 \right)^T
 &=
 (0,-4,-1,-3)^T.
\end{align}
Applying Eq.~\eqref{eq:a-covering-integral} gives
\begin{align}
 C_{3,\mathrm A}
 =
 \left(
 1,0,-2,-\frac12,-\frac32
 \right)^T.
 \label{eq:a-color-row}
\end{align}
Here the quotient refers to the Calabi-Yau geometry and should not be
confused with the physical axion quotient discussed in
Section~\ref{sec:physical-axion}.

The remaining visible Green-Schwarz coefficients satisfy
\begin{align}
 C_{2,\mathrm A}=C_{3,\mathrm A},
 \qquad
 C_{Y,\mathrm A}=\frac53C_{3,\mathrm A}.
 \label{eq:a-proportionality}
\end{align}
The same factor $1/2$ enters the integral defining $Q_0{}^I$.
The internal charges $Q_i{}^I=V_i^I$ are unchanged in these coordinates. This distinction is needed when the Green-Schwarz variation
is compared with the anomaly of the quotient theory.

The proportionality in Eq.~\eqref{eq:a-proportionality} follows from the
$SU(5)$ gauge embedding and does not require a particular Calabi-Yau
metric. In a basis in which the visible $SU(5)$ gauge group carries a common
flux $A$, the internal four-forms
for the color and weak couplings are both $B_{\rm v}-A^2$.

The traceless hypercharge generator satisfies
\begin{align}
 \ell_{\rm v}(Y)=0,
 \qquad
 U_{\rm v}(Y)=\frac56A^2.
\end{align}
Eq.~\eqref{eq:general-cartan-coupling} then gives
\begin{align}
 \mathcal C(Y)=\frac53(B_{\rm v}-A^2).
\end{align}
A flat Wilson line changes which gauge generators survive in four
dimensions but does not change this local Green-Schwarz contraction.
Consequently,
\begin{align}
 P_{\mathrm A}
 \equiv
 C_{\gamma,\mathrm A}-\frac83C_{3,\mathrm A}
 =0.
 \label{eq:a-zero-p}
\end{align}
This is an equality between coefficients in the Green-Schwarz action and
is independent of the K\"ahler metric.

There is nevertheless an additional issue before
$C_{3,\mathrm A}$ can be identified with a physical QCD axion coupling.
The charge matrix $Q_{\mathrm A}$ obtained from
Eqs.~\eqref{eq:internal-axion-charges} and \eqref{eq:universal-gauging}
has rank four. Thus four of the five bulk axion directions are gauged.
Only one local physical direction remains in the bulk axion sector.
This counting does not include the charged-matter phases that may be
needed to determine its QCD and electromagnetic couplings.\footnote{An explicit heterotic line-bundle model in which
the QCD axion arises from bundle-modulus phases was constructed
in Ref.~\cite{Buchbinder:2014qca}.}
For the Abelian generator
\begin{align}
 T_c=(1,-1,-1,0^5;0^8)^T,
\end{align}
the charge vector in the quotient coordinates is
\begin{align}
 Q_{\mathrm A} T_c&=\left(\frac32,3,0,-3,0\right)^T.
 \label{eq:a-gauged-direction}
\end{align}
Its contraction with Eq.~\eqref{eq:a-color-row} gives
\begin{align}
 C_{3,\mathrm A}^TQ_{\mathrm A} T_c=3.
 \label{eq:a-mixed-variation}
\end{align}
Under this transformation, $C_{3,\mathrm A}^T\Theta_{\mathrm A}$ shifts by $3\lambda_c$.
Thus $C_{3,\mathrm A}$ does not by itself define a gauge-invariant coupling on
the physical axion space, according to
Eq.~\eqref{eq:physical-gauge-invariance}. Its variation participates in the cancellation
of the mixed color anomaly. Removing the axion directions associated
with massive Abelian gauge fields therefore does not eliminate this
Green-Schwarz variation; the charged-matter contribution must be
included in the anomaly-matched low-energy coupling.

Eq.~\eqref{eq:a-zero-p} should consequently be interpreted first as
a relation in the Green-Schwarz sector. If the low-energy matter and
threshold matching preserves the same $SU(5)$ relation, the completed
physical couplings satisfy
\begin{align}
 e_{\mathrm A}=\frac83n_{\mathrm A}.
\end{align}
Provided that the completed QCD coupling is nonzero,
Eq.~\eqref{eq:qcd-observables-invariant} then gives
\begin{align}
 \left(\frac EN\right)_{\rm eff}=\frac83
\end{align}
for any positive-definite axion kinetic matrix in the QCD-dominated
regime of section~\ref{sec:physical-framework}. 
We do not determine the axion decay constant because the low-energy QCD coupling and its kinetic normalization have not been
calculated. With the proportional physical couplings above and
$M_{\rm other}^2\succeq0$, the QCD-line bound in
Eq.~\eqref{eq:proportional-mass-coupling-bound} also applies without
specifying this decay constant.

We next study a direct-flux background with independent QCD and
electromagnetic Green-Schwarz coupling vectors.

\subsection{Model B: Standard-Model-like model with two independent Green-Schwarz couplings}
\label{sec:model-b}

Model~B breaks the visible $SO(16)_{\rm v}$ directly to a Standard-Model-like gauge
sector by internal line-bundle flux. The gauge levels are $k_3=k_2=1$
and $k_Y=5/3$. Hypercharge is the only massless visible Abelian gauge
field. The net complex chiral spectrum consists of three Standard Model
families. The background satisfies the Bianchi identity with $W=0$, so
no localized NS5-brane axions enter the following calculation.

\subsubsection{Geometry, flux, and chiral spectrum}
\label{subsec:modelb-background}

The compactification uses a favorable complete-intersection Calabi-Yau
threefold~\cite{Candelas:1987kf,Anderson:2017aux} with
$(h^{1,1},h^{2,1})=(6,24)$,
\begin{align}
 X=\left[
 \begin{array}{c|ccccccccc}
 \mathbb P^2&1&1&1&0&0&0&0&0&0\\
 \mathbb P^2&1&0&0&1&1&0&0&0&0\\
 \mathbb P^2&0&1&0&0&0&1&1&0&0\\
 \mathbb P^2&0&0&1&0&0&0&0&1&1\\
 \mathbb P^2&0&0&0&1&0&1&0&1&0\\
 \mathbb P^2&0&0&0&0&1&0&1&0&1
 \end{array}\right],
 \qquad c_{2i}=36\quad(i=1,\ldots,6).
 \label{eq:modelb-configuration}
\end{align}
The six ambient hyperplane classes $\widehat\omega_i$ restrict to the basis $\omega_i$ of
$H^{1,1}(X)$. Each column of the configuration matrix specifies the
multidegree of one defining polynomial. If its entries are denoted by
$d_{r\alpha}$, the intersection numbers can be calculated from
\begin{align}
 \kappa_{ijk}
 &=\int_{(\mathbb P^2)^6}\widehat\omega_i\widehat\omega_j\widehat\omega_k
 \prod_{\alpha=1}^{9}\left(\sum_{r=1}^{6}d_{r\alpha}\widehat\omega_r\right).
 \label{eq:modelb-intersection-rule}
\end{align}
Here $\widehat\omega_i^3=0$ and $\int_{(\mathbb P^2)^6}\widehat\omega_1^2\cdots \widehat\omega_6^2=1$.
This determines the triple intersection numbers
$\kappa_{ijk}=\int_X\omega_i\wedge\omega_j\wedge\omega_k$, which
enter the index, Bianchi-identity, and axion-kinetic calculations
below. We restrict to the region $t^i>0$ in the K\"ahler-moduli
space defined by the ambient-space divisor basis.

The visible and hidden fluxes are chosen as
\begin{align}
 f_{\rm v}=(A,A,A,B,B,A+L,A-L,F),
 \qquad f_{\rm h}=0,
 \label{eq:modelb-flux}
\end{align}
where
\begin{align}
 \begin{split}
 A&=(-1,1,1,1,-1,-1),\qquad B=(-1,1,0,0,0,0),\\
 L&=(0,0,-1,1,-1,1),\qquad F=(-1,-1,1,1,-1,1).
 \end{split}
 \label{eq:modelb-classes}
\end{align}
Each six-component array gives the coefficients of the corresponding
two-form in the basis $\omega_i$; for example,
$A=-\omega_1+\omega_2+\omega_3+\omega_4-\omega_5-\omega_6$.
All four classes are integral. 
For each basis form $\omega_i$, the visible-sector Cartan flux
vector has integer components whose sum is even. 
With the vanishing hidden flux, the quantization condition of
section~\ref{subsec:10d-action} is then satisfied for both vector
and spinor string states.

The hypercharge generator is
$y_B=(-\tfrac13,-\tfrac13,-\tfrac13,0,0,\tfrac12,\tfrac12,0;0^8)^T$ and
has $k_Y=5/3$. Its internal axion charge vanishes because
\begin{align}
 \sum_I(y_B)_If_I
 &=-A+\frac12(A+L)+\frac12(A-L)=0.
 \label{eq:modelb-internal-hypercharge}
\end{align}
Using the triple intersection numbers, we find
$\int_X A L^2=0$. The universal axion therefore has no
St\"uckelberg coupling to hypercharge, in agreement with
Eq.~\eqref{eq:embedding-full-y}. Direct evaluation of
Eq.~\eqref{eq:universal-gauging} gives the stronger result
$Q_0{}^I=0$ for every Cartan component.

Using Eqs.~\eqref{eq:modelb-flux} and
\eqref{eq:modelb-classes}, the internal axion charges for the eight
visible Cartan generators are
\begin{align}
 Q_i{}^I=V_i{}^I
 &=
 \begin{pmatrix}
 -1&-1&-1&-1&-1&-1&-1&-1\\
  1& 1& 1& 1& 1& 1& 1&-1\\
  1& 1& 1& 0& 0& 0& 2& 1\\
  1& 1& 1& 0& 0& 2& 0& 1\\
 -1&-1&-1& 0& 0&-2& 0&-1\\
 -1&-1&-1& 0& 0& 0&-2& 1
 \end{pmatrix}.
 \label{eq:modelb-internal-charge-matrix}
\end{align}
This matrix has rank four, so its kernel has dimension four.
Three directions in the kernel are the Cartan generators of
$SU(3)_C\times SU(2)_L$, while the remaining direction is hypercharge.
The roots satisfying $\sum_I\alpha_If_I=0$ are precisely
$\mathbf e_i-\mathbf e_j$ with $i\ne j$ and $i,j\in\{1,2,3\}$, together with
$\pm(\mathbf e_4-\mathbf e_5)$. Thus there is no additional visible non-Abelian
enhancement. The hidden $SO(16)_{\rm h}$ remains unbroken because $f_{\rm h}=0$.
The massless continuous gauge algebra is therefore
\begin{align}
 \mathfrak g_{\rm massless}
 &=\mathfrak{su}(3)_C\oplus\mathfrak{su}(2)_L\oplus\mathfrak u(1)_Y
 \oplus\mathfrak{so}(16)_{\rm h}.
 \label{eq:modelb-massless-algebra}
\end{align}

The chiral spectrum is obtained by applying Eq.~\eqref{eq:line-index}
to the weights of the ten-dimensional fermion representations. We then
sum the indices for states with the same representation under
$\mathfrak g_{\rm massless}$, with the Majorana-Weyl identification
explained in section~\ref{sec:consistency}.
For example, the visible spinor weight
\begin{align}
 w_Q&=\frac12(1,-1,-1,1,-1,1,-1,1;0^8)
 \label{eq:modelb-quark-weight}
\end{align}
is a highest weight of $(\mathbf3,\mathbf2)_{1/6}$.
Its associated first Chern class has coefficients
$(0,-1,-1,1,-1,2)$, so
$\int_Xq_{w_Q}^3=18$ and $\int_Xc_2(TX)\wedge q_{w_Q}=0$.
Eq.~\eqref{eq:line-index} therefore gives $\chi(X,L_{w_Q})=3$.
The resulting net complex spectrum is
\begin{align}
 \mathcal I_{\rm complex}
 =3\bigl[
 (\mathbf3,\mathbf2)_{1/6}
 +(\overline{\mathbf3},\mathbf1)_{-2/3}
 +(\overline{\mathbf3},\mathbf1)_{1/3}
 +(\mathbf1,\mathbf2)_{-1/2}
 +(\mathbf1,\mathbf1)_1
 \bigr].
 \label{eq:modelb-net-spectrum}
\end{align}
Every other complex representation of the massless gauge algebra has zero
net index. The possible pseudoreal $(\mathbf1,\mathbf2;\mathbf{16})_0$
sector has $\chi(X,L_B)=0$, where $c_1(L_B)=B$. Each such multiplet
contains sixteen weak doublets, so it cannot produce an odd number of
$SU(2)_L$ Weyl doublets~\cite{Witten:1982fp}. The index calculation does not determine the
individual cohomologies or the masses of vector-like pairs. In
particular, Eq.~\eqref{eq:modelb-net-spectrum} is a statement about net
chirality, not a determination of the complete light spectrum.

The trivial hidden bundle also leaves a hidden $\mathbf{128}$ fermion
at leading order~\cite{Blaszczyk:2015zta}. The calculation below assumes
that the hidden-sector potential is subleading to QCD. The conditions
for this assumption are discussed in
section~\ref{subsec:modelb-potentials}.

\subsubsection{HYM conditions and the Bianchi identity}
\label{subsec:modelb-geometry}

We next check the geometric conditions used in section~\ref{sec:consistency}.
The four independent flux classes in Eq.~\eqref{eq:modelb-classes}
require $A^i\tau_i=B^i\tau_i=L^i\tau_i=F^i\tau_i=0$.
Using their explicit components gives the linear system:
\begin{align}
 \begin{pmatrix}
 -1&1&1&1&-1&-1\\
 -1&1&0&0&0&0\\
 0&0&-1&1&-1&1\\
 -1&-1&1&1&-1&1
 \end{pmatrix}
 \begin{pmatrix}\tau_1\\\tau_2\\\tau_3\\\tau_4\\\tau_5\\\tau_6\end{pmatrix}
 &=0.
 \label{eq:modelb-hym-linear-system}
\end{align}
Its solution is
\begin{align}
 \tau_1=\tau_2=\tau_3=\tau_6,
 \qquad
 \tau_4=\tau_5.
 \label{eq:modelb-hym-family}
\end{align}
In the region $t^i>0$, these conditions are equivalent to
\begin{align}
 t^1=t^2=t^3=t^6&=u,\qquad t^4=t^5=v,\qquad u,v>0.
 \label{eq:modelb-hym-parameters}
\end{align}
The converse follows from symmetries of the volume
$\mathcal V(t)$. The volume is unchanged under the exchanges
$t^1\leftrightarrow t^6$ and $t^2\leftrightarrow t^3$, as well as
under the simultaneous exchanges
$t^1\leftrightarrow t^2$, $t^3\leftrightarrow t^6$, and
$t^4\leftrightarrow t^5$. The gradient is
\begin{align}
 \partial_i[-\log\mathcal V]
 &=-\frac{\tau_i}{\mathcal V}.
 \label{eq:modelb-volume-gradient}
\end{align}
The relations
$\tau_1=\tau_2=\tau_3=\tau_6$ and $\tau_4=\tau_5$ therefore imply
that this gradient is invariant under the same exchanges.
Since $-\log\mathcal V$ is strictly convex in the positive
K\"ahler region, its gradient is injective. The K\"ahler parameters
must consequently satisfy
$t^1=t^2=t^3=t^6$ and $t^4=t^5$.

The intersection rule in Eq.~\eqref{eq:modelb-intersection-rule} gives
the volume on this two-parameter locus:
\begin{align}
 \mathcal V(u,v)&=34u^3+72u^2v+30uv^2+2v^3.
 \label{eq:modelb-hym-volume}
\end{align}
For the Bianchi identity, we need the four-form integrals rather than
the two-form coefficient arrays in Eq.~\eqref{eq:modelb-classes}.
They satisfy
\begin{align}
 \int_X\omega_i\wedge S_{\rm v}&=-72,\qquad i=1,\ldots,6,
 \label{eq:modelb-flux-square-integrals}
\end{align}
with $S_{\rm h}=0$. Hence $W_i=36-72/2=0$ for every $i$.
The nondegeneracy of the cohomological integrals discussed in
section~\ref{subsec:embedding-physical-rank} then gives
\begin{align}
 W=0.
 \label{eq:modelb-bianchi}
\end{align}
No NS5 source or localized tensor axion is therefore required in Model~B.

\subsubsection{Axion couplings and the predicted range of \texorpdfstring{$E/N$}{E/N}}
\label{subsec:modelb-couplings}

We now evaluate the axion coefficients using the same topological data.
The internal integrals needed in
Eqs.~\eqref{eq:nonabelian-row} and \eqref{eq:embedding-p-row} are
\begin{align}
 \left(\int_X\omega_i\wedge A^2\right)_{i=1}^6
 &=(-6,-6,-6,-6,-6,-6),
\nonumber\\
 \left(\int_X\omega_i\wedge B^2\right)_{i=1}^6
 &=(-2,-2,-5,-5,-5,-5),
\nonumber\\
 \left(\int_X\omega_i\wedge L^2\right)_{i=1}^6
 &=(-12,-12,-9,-9,-9,-9).
 \label{eq:modelb-l-square-integrals}
\end{align}
Since $B_{\rm v}=S_{\rm v}/6$ in this model,
$\int_X\omega_i\wedge(B_{\rm v}-A^2)=-12+6=-6$.
Similarly, $\int_X\omega_i\wedge(A^2-B^2-L^2)=8$ for every $i$.
The QCD Green-Schwarz vector and its departure from the $8/3$ relation are
\begin{align}
 C_3=(1;-6,-6,-6,-6,-6,-6)^T,
 \qquad
 P\equiv C_\gamma-\frac83C_3=(0;8,8,8,8,8,8)^T.
 \label{eq:modelb-gs-rows}
\end{align}
The semicolon separates the universal-axion component from the six
K\"ahler-axion components. Each flux class in
Eq.~\eqref{eq:modelb-classes} has vanishing sum of its divisor
coefficients, so $\sum_iV_i^I=0$ for every $I$. Together with
$Q_0{}^I=0$, this gives
\begin{align}
 (C_3^TQ)^I&=-6\sum_iV_i^I=0,
\nonumber\\
 (P^TQ)^I&=8\sum_iV_i^I=0.
 \label{eq:modelb-difference-invariance}
\end{align}
Thus both interactions are invariant under the gauged axion shifts.
Moreover, $P\ne0$ has zero universal component, whereas $(C_3)_0=1$.
The two vectors cannot be proportional. 
The analysis in section~\ref{subsec:embedding-physical-rank} then shows
that the QCD and electromagnetic coupling vectors are linearly
independent before specifying the axion kinetic metric.

For the numerical prediction we assume that finite contributions from
the low-energy matter sector do not change these two gauge-invariant
coefficient vectors. Thus $n=C_3$ and $p=e-(8/3)n=P$.
We also assume the QCD-dominated regime discussed above. The tree-level
bulk metric is the one used in Eq.~\eqref{eq:general-bound-leading-metric},
with $s=\mathcal V/g_s^2=\alpha_{3,\rm tree}^{-1}
=\alpha_{2,\rm tree}^{-1}$. 
Its common overall normalization cancels between the numerator and
denominator of Eq.~\eqref{eq:qcd-observables-invariant}.

There are seven bulk axions and four independent gauged shifts.
The physical bulk axion space therefore has dimension three. Since
$Q_0{}^I=0$, one coordinate is $a_0$. A local choice for the two
remaining coordinates is
\begin{align}
 b&=R_B\begin{pmatrix}\theta_1\\\theta_2\end{pmatrix},
 \qquad
 R_B=\begin{pmatrix}
 1&0\\1&0\\1&0\\0&1\\0&1\\1&0
 \end{pmatrix}.
 \label{eq:modelb-axion-representatives}
\end{align}
These equalities can be imposed by gauge transformations and do
not constrain the remaining physical axion degrees of freedom. 
Let $Q_K^{\rm ind}$ contain four independent columns spanning the
internal gauged shifts, for example the coefficient vectors $A,B,L,F$.
The HYM condition $(Q_K^{\rm ind})^T\tau(u,v)=0$ holds throughout this locus.
The two columns of $R_B$ are the derivatives of $t^i(u,v)$ with respect
to $u$ and $v$. Since $\mathcal H_{ij}=-\partial_j(\tau_i/\mathcal V)$,
differentiating $(Q_K^{\rm ind})^T(\tau/\mathcal V)=0$ with respect to these two
parameters gives
\begin{align}
 (Q_K^{\rm ind})^T\mathcal H R_B&=0
 \label{eq:modelb-horizontal-representatives}
\end{align}
on the HYM locus. The subtraction term in the quotient metric of
Eq.~\eqref{eq:physical-quotient-metric} therefore vanishes for axion
configurations satisfying these equalities. 
The physical internal metric is
\begin{align}
 \mathcal H_{\rm phys}
 &=R_B^T\mathcal H R_B
 =-\begin{pmatrix}
 \partial_u^2\log\mathcal V&\partial_u\partial_v\log\mathcal V\\
 \partial_u\partial_v\log\mathcal V&\partial_v^2\log\mathcal V
 \end{pmatrix},
 \label{eq:modelb-physical-internal-metric}
\end{align}
where $\mathcal V=\mathcal V(u,v)$ is given in
Eq.~\eqref{eq:modelb-hym-volume}.

In this gauge the Green-Schwarz interactions become
\begin{align}
 \mathcal L_{{\rm GS},B}^{(4)}\supset{}&
 \frac{\alpha_s}{8\pi}
 (a_0-24\theta_1-12\theta_2)
 G^a_{\mu\nu}\widetilde G^{a\mu\nu}
 \nonumber\\
 &+\frac{\alpha_{\rm em}}{8\pi}
 \left(\frac83a_0-32\theta_1-16\theta_2\right)
 F_{\mu\nu}\widetilde F^{\mu\nu}.
 \label{eq:modelb-physical-lagrangian}
\end{align}
Hence the internal coefficients are $n_{K,\rm phys}=R_B^Tn_K=(-24,-12)^T$ and
$p_{K,\rm phys}=R_B^Tp_K=(32,16)^T$. In the rest of this subsection these two-component
vectors refer to $(\theta_1,\theta_2)$; the six-component vectors in
Eq.~\eqref{eq:modelb-gs-rows} refer to the original axion coordinates.
Although $p_{K,\rm phys}=-(4/3)n_{K,\rm phys}$, the full vectors remain independent because
their universal components are different.

To evaluate the coupling ratio throughout the region
$u,v>0$ satisfying the HYM equations, we use the volume in
Eq.~\eqref{eq:modelb-hym-volume}. 
The combination entering the coupling ratio is
\begin{align}
 n_{K,\rm phys}^T\mathcal H_{\rm phys}^{-1}n_{K,\rm phys}
 &=24\,
 \frac{521u^4+760u^3v+540u^2v^2+214uv^3+35v^4}
 {59u^2+43uv+13v^2}.
 \label{eq:modelb-anisotropic-contraction}
\end{align}
Since $p_{K,\rm phys}=-(4/3)n_{K,\rm phys}$, the ratio throughout this locus is
\begin{align}
 \left(\frac EN\right)_{\rm eff}
 &=\frac83-\frac43
 \frac{n_{K,\rm phys}^T\mathcal H_{\rm phys}^{-1}n_{K,\rm phys}}
 {\mathcal V(u,v)^2/g_s^4+n_{K,\rm phys}^T\mathcal H_{\rm phys}^{-1}n_{K,\rm phys}}.
 \label{eq:modelb-anisotropic-ratio}
\end{align}
This explicitly connects the dependence on the kinetic metric to the
allowed K\"ahler moduli, as discussed in
section~\ref{sec:proportionality-criterion}.

At the symmetric point $u=v=t$, the volume is $\mathcal V=138t^3$.
The physical internal metric is
\begin{align}
 \mathcal H_{\rm phys}
 =\frac1{23t^2}
 \begin{pmatrix}34&12\\12&11\end{pmatrix}.
 \label{eq:modelb-isotropic-metric}
\end{align}
Its inverse gives $n_{K,\rm phys}^T\mathcal H_{\rm phys}^{-1}n_{K,\rm phys}=432t^2$ and
$p_{K,\rm phys}^T\mathcal H_{\rm phys}^{-1}n_{K,\rm phys}=-576t^2$.
Eq.~\eqref{eq:qcd-observables-invariant} therefore gives
\begin{align}
 \left(\frac EN\right)_{\rm eff}
 =\frac83-
 \frac{576t^2}{s^2+432t^2}
 =\frac83-
 \frac{16}{529\,t^4/g_s^4+12}.
 \label{eq:modelb-bulk-ratio}
\end{align}

The HYM locus extends to weak string coupling and large K\"ahler
parameters. An overall rescaling of the K\"ahler form preserves the
vanishing-slope conditions. It also leaves the flux quantization,
Bianchi identity, chiral indices, and Green-Schwarz coefficients
unchanged. Thus the same rank-two coupling structure is present for
$g_s\ll1$ and $u,v\gg1$.

Eqs.~\eqref{eq:modelb-bulk-ratio} and
\eqref{eq:modelb-anisotropic-ratio} determine the leading value of
$E/N$ at any specified point in this regime, under the low-energy and
QCD-dominance assumptions. On the symmetric locus $u=v=t$,
expanding Eq.~\eqref{eq:modelb-bulk-ratio} at weak coupling and large
radius gives
\begin{align}
 \frac83-\left(\frac EN\right)_{\rm eff}
 &=\frac{16g_s^4}{529t^4+12g_s^4}
 =\frac{16}{529}\frac{g_s^4}{t^4}
 +O\!\left(\frac{g_s^8}{t^8}\right).
 \label{eq:modelb-controlled-expansion}
\end{align}
This expansion describes the leading kinetic result. The departure
becomes smaller at weaker coupling or larger radius, but the
Green-Schwarz coupling vectors remain linearly independent. The
limiting value $8/3$ therefore reflects the relative kinetic weights of
the universal and K\"ahler axions rather than proportionality of their
coupling vectors.

To specify a weak-coupling, large-radius region, let us choose
$t_{\rm min}\gg1$ and $0<g_{\rm max}\ll1$ and require
\begin{align}
 u,v\ge t_{\rm min},\qquad 0<g_s\le g_{\rm max}.
 \label{eq:modelb-controlled-region}
\end{align}
To find the largest departure from $8/3$ in the region defined by Eq.~\eqref{eq:modelb-controlled-region}, we examine its dependence
on $u$, $v$, and $g_s$. Eqs.~\eqref{eq:modelb-anisotropic-contraction}
and \eqref{eq:modelb-anisotropic-ratio} give
\begin{align}
 \partial_u\left[\frac83-\left(\frac EN\right)_{\rm eff}\right]&<0,
 \qquad
 \partial_v\left[\frac83-\left(\frac EN\right)_{\rm eff}\right]<0,
 \qquad
 \partial_{g_s}\left[\frac83-\left(\frac EN\right)_{\rm eff}\right]>0
 \label{eq:modelb-monotonicity-gs}
\end{align}
for $u,v,g_s>0$. Thus the departure from $8/3$ is largest at
$u=v=t_{\rm min}$ and $g_s=g_{\rm max}$. Using
Eq.~\eqref{eq:modelb-bulk-ratio} at this point, we obtain
\begin{align}
 \frac83-\frac{16}{529\,t_{\rm min}^4/g_{\rm max}^4+12}
 \le
 \left(\frac EN\right)_{\rm eff}
 <\frac83.
 \label{eq:modelb-stretched-bulk-bound}
\end{align}
This expression gives the range for any chosen lower limit on the
K\"ahler parameters and upper limit on the string coupling within the
weak-coupling, large-radius regime.

The same moduli also determine the tree-level gauge coupling. Since all
coefficients of $\mathcal V(u,v)$ are positive, the region in
Eq.~\eqref{eq:modelb-controlled-region} gives
\begin{align}
 s=\frac{\mathcal V(u,v)}{g_s^2}
 &\ge\frac{138\,t_{\rm min}^3}{g_{\rm max}^2}.
 \label{eq:modelb-controlled-gauge-coupling}
\end{align}
A comparison with measured gauge couplings requires their evolution to
the compactification scale together with the CP-even threshold
corrections. These corrections are not determined by the Green-Schwarz
coefficients and must therefore be computed explicitly before such a
comparison can be made.\footnote{Ref.~\cite{Benabou:2026jtv} imposes the unified gauge
coupling when constraining the volume and computing the QCD
axion mass in heterotic compactifications.}

We can now compare Model~B with the general bound using the same
geometric data. At $u=v=t$, $c_J=216t$, and $\mu(tJ_1)=\mu(J_1)$,
where $J_1=\sum_{i=1}^6\omega_i$. We compute this value directly
from Eq.~\eqref{eq:cicy-direct-maximum} using the full metric of all six K\"ahler axions. 
At $J_1$, $\tau_i=69$, so primitivity requires $\sum_i\xi^i=0$. 
The triple intersection numbers are unchanged under permutations
of the divisor labels $1,5,6$, under permutations of $2,3,4$,
and under interchange of these two sets.
We decompose the coefficients of $\xi$ as 
$(\xi^1,\xi^5,\xi^6)=a_{\xi}(1,1,1)+\eta$ and
$(\xi^2,\xi^3,\xi^4)=-a_{\xi}(1,1,1)+\zeta$, with
$\sum_i\eta_i=\sum_i\zeta_i=0$. Let
$S_{\eta}=\sum_i\eta_i^2$ and $S_{\zeta}=\sum_i\zeta_i^2$. Direct substitution
gives $q_{J_1}(\xi)=36a_{\xi}^2+15(S_{\eta}+S_{\zeta})$ and
\begin{align}
 \lVert\xi^2\rVert_{\mathcal H_1^{-1}}^2
 &=432a_{\xi}^4+\frac{3456}{5}a_{\xi}^2(S_{\eta}+S_{\zeta})
   +\frac{1233}{10}(S_{\eta}+S_{\zeta})^2-\frac{828}{5}S_{\eta}S_{\zeta}
 \nonumber\\
 &\quad+\frac{1656}{5}a_{\xi}
 \left(\sum_i\eta_i^3-\sum_i\zeta_i^3\right),
 \label{eq:modelb-direct-norm-polynomial}
\end{align}
where $\mathcal H_1=\mathcal H(J_1)$. For three real components
with zero sum, $|\sum_i\eta_i^3|\le S_{\eta}^{3/2}/\sqrt6$, and the
same inequality holds for $\zeta$ and $S_{\zeta}$.
At fixed $a_{\xi}^2$ and $S_{\eta}+S_{\zeta}$, the term proportional to $S_{\eta}S_{\zeta}$ is
nonpositive, while
$S_{\eta}^{3/2}+S_{\zeta}^{3/2}\le(S_{\eta}+S_{\zeta})^{3/2}$. 
Equality holds in both inequalities when one zero-sum vector vanishes and
the other is proportional to $(2,-1,-1)$, with its sign chosen
to make the cubic term nonnegative.

Up to permutations and an overall rescaling, a maximizing class
can therefore be taken as
$\xi=(r_{\xi}+2,-r_{\xi},-r_{\xi},-r_{\xi},r_{\xi}-1,r_{\xi}-1)$ with $r_{\xi}\ge0$.
The case $S_{\eta}+S_{\zeta}=0$ is included as $r_{\xi}\to\infty$.
Thus the maximum over the full primitive subspace reduces to
\begin{align}
 \mu(J_1)^2
 &=\max_{r_{\xi}\ge0}
 \frac{40r_{\xi}^4+384r_{\xi}^2+184r_{\xi}+411}{30(2r_{\xi}^2+5)^2}.
 \label{eq:modelb-direct-one-variable}
\end{align}
The derivative of the ratio is
$-92(r_{\xi}+1)(4r_{\xi}^2+2r_{\xi}-5)/[15(2r_{\xi}^2+5)^3]$.
It changes from positive to negative at
$r_{\xi}=(\sqrt{21}-1)/4$, giving the exact value
\begin{align}
 \mu(tJ_1)&=\mu(J_1)
 =\sqrt{\frac{3346+161\sqrt{21}}{5875}}
 \simeq0.83.
 \label{eq:modelb-mu-upper}
\end{align}
Using $s=138t^3/g_s^2$, the parameter entering the general bound is
therefore
\begin{align}
 x&=\frac{216t\,\mu(J_1)}{s}
 =\frac{36}{23}\sqrt{\frac{3346+161\sqrt{21}}{5875}}
   \frac{g_s^2}{t^2}.
 \label{eq:modelb-general-bound-x}
\end{align}

\begin{table}[t]
 \centering
 \begin{tabular}{c|ccccc}
 \hline
  & $t$ & $g_s$ & $-L(x)$ & $\Delta_{\rm Model\,B}$ & $U(x)$\\
 \hline
 $P_1$ & $5$ & $0.3$ & $-7.36\times10^{-6}$
 & $-3.92\times10^{-7}$ & $9.20\times10^{-7}$\\
 $P_2$ & $10$ & $0.1$ & $-5.68\times10^{-9}$
 & $-3.02\times10^{-10}$ & $7.10\times10^{-10}$\\
 \hline
 \end{tabular}
 \caption{Comparison of the general geometric bound with Model~B on the
 symmetric HYM locus. Here $\Delta=(E/N)_{\rm eff}-8/3$. The Model~B
 column is calculated from Eq.~\eqref{eq:modelb-bulk-ratio}; the two
 adjacent columns use Eqs.~\eqref{eq:general-bound-U},
 \eqref{eq:general-bound-L}, and \eqref{eq:modelb-general-bound-x}.
 Both calculations use the same leading kinetic approximation and
 the low-energy assumptions specified above. The parameter points
 $P_1$ and $P_2$ are also shown in
 Fig.~\ref{fig:alp-bound-fqcd-x}.}
 \label{tab:modelb-general-bound}
\end{table}

Table~\ref{tab:modelb-general-bound} compares the general interval
with the leading Model~B result at two choices of moduli
satisfying the HYM equations. Both calculations use the same
moduli and low-energy assumptions. The corresponding values are
$s\simeq1.92\times10^5$ and $1.38\times10^7$.
The Model~B results lie within the general interval, which is
wider because the general derivation allows the coupling
coefficients to vary continuously, whereas the Model~B flux
fixes them. We have not checked whether these parameter values
reproduce the measured gauge couplings. 
On the symmetric HYM locus, the relation
$p_{K,\rm phys}=-(4/3)n_{K,\rm phys}$ gives
\begin{align}
 \lVert p_{K,\rm phys}\rVert_{\mathcal H_{\rm phys}^{-1}}
 +\lVert p_{K,\rm phys}+3n_{K,\rm phys}\rVert_{\mathcal H_{\rm phys}^{-1}}
 &=36\sqrt3\,t.
 \label{eq:modelb-correlated-norm-value}
\end{align}
The quotient identity of section~\ref{sec:physical-axion} allows this
quantity, evaluated with the physical axion metric, to be compared
directly with the geometric upper bound in
Eq.~\eqref{eq:general-bound-correlated-norm}. Using
$\mu(J_1)c_J=216\mu(J_1)t$, we find
\begin{align}
 \frac{
 \lVert p_{K,\rm phys}\rVert_{\mathcal H_{\rm phys}^{-1}}
 +\lVert p_{K,\rm phys}+3n_{K,\rm phys}\rVert_{\mathcal H_{\rm phys}^{-1}}
 }{
 \mu(J_1)c_J
 }
 &\simeq 0.346.
\end{align}
The sum of norms for Model~B is therefore well below the
geometric upper bound. This helps explain why the general
interval for $E/N$ is wider than the range obtained for the
Model~B flux.

We next examine how corrections can affect the small departure in
Eq.~\eqref{eq:modelb-controlled-expansion}. The higher-order terms shown
there come from expanding the leading formula; they do not include
string-loop or $\alpha'$ corrections. Let $G_0$ be the leading physical
metric and $\mathcal R_0$ the corresponding value of $E/N$. At fixed
$n_{\rm phys}$ and $e_{\rm phys}$, a small change $\delta G$ gives
\begin{align}
 \delta\mathcal R
 &=-\frac{(e_{\rm phys}-\mathcal R_0n_{\rm phys})^T
 G_0^{-1}\delta G\,G_0^{-1}n_{\rm phys}}
 {n_{\rm phys}^TG_0^{-1}n_{\rm phys}}.
 \label{eq:modelb-kinetic-linear-change}
\end{align}
This is the first-order variation of Eq.~\eqref{eq:qcd-observables-invariant}.
The relative K\"ahler contribution to the QCD normalization is
\begin{align}
 \zeta_B
 &\equiv\frac{\sqrt{n_{K,\rm phys}^T\mathcal H_{\rm phys}^{-1}n_{K,\rm phys}}}{s}
 =\frac{2\sqrt3}{23}\frac{g_s^2}{t^2}
 \qquad (u=v=t).
 \label{eq:modelb-small-kinetic-weight}
\end{align}
The leading departure is
$\Delta_0=\mathcal R_0-8/3=-4\zeta_B^2/[3(1+\zeta_B^2)]$.
In coordinates canonically normalized with $G_0$, let
$\mathsf E=G_0^{-1/2}\delta G\,G_0^{-1/2}$ and
$\varepsilon_{\rm kin}=\lVert\mathsf E\rVert_{\rm op}$, where the norm
is the largest absolute value of an eigenvalue of $\mathsf E$. A generic correction that mixes
the universal axion with the QCD-coupled K\"ahler combination changes
$E/N$ at order $\varepsilon_{\rm kin}\zeta_B$. Thus
$\varepsilon_{\rm kin}\ll\zeta_B$ is sufficient to resolve the leading
departure. Without this mixing, small diagonal corrections only change
the departure by a relative amount of order $\varepsilon_{\rm kin}$.

For Model~B, the relevant mixing can be reduced to a single combination.
As an example allowed by the axion shift symmetries, let
\begin{align}
 \delta\widehat G_{0i}&=\frac{\gamma_i}{s^2},
 \qquad i=1,2,
 \label{eq:modelb-mixing-example}
\end{align}
where the indices refer to $(\theta_1,\theta_2)$. From
Eq.~\eqref{eq:modelb-isotropic-metric},
\begin{align}
 \mathcal H_{\rm phys}^{-1}n_{K,\rm phys}&=-12t^2(1,1)^T.
 \label{eq:modelb-mixing-projection}
\end{align}
Substituting in Eq.~\eqref{eq:modelb-kinetic-linear-change} gives
\begin{align}
 \frac{\delta\mathcal R_{\rm mix}}{\Delta_0}
 &=\frac{\gamma_1+\gamma_2}{36}
   \frac{1-\zeta_B^2}{1+\zeta_B^2}.
 \label{eq:modelb-mixing-relative-change}
\end{align}
Only $\gamma_1+\gamma_2$ enters this first-order correction. At fixed
$\gamma_i$, it has the same $g_s^4/t^4$ scaling as the leading departure,
so increasing the radius or reducing $g_s$ does not suppress this
relative correction. We have not calculated $\gamma_i$ from string
amplitudes. 
Eq.~\eqref{eq:modelb-mixing-relative-change} identifies the
combination that controls corrections to the leading
$g_s^4/t^4$ term in
$8/3-(E/N)_{\rm eff}$. The effect of a non-QCD potential on the
light axion direction is discussed in
section~\ref{subsec:modelb-potentials}.

\subsubsection{Photon couplings of the two QCD-neutral axions}
\label{subsec:modelb-alp-couplings}

We now evaluate the photon couplings of the two axions other than
the QCD axion and compare them with
Eq.~\eqref{eq:alp-total-geometric-bound}. We use the leading kinetic
metric and assume that the low-energy contributions do not change
$n_{\rm phys}$ and $e_{\rm phys}$, as in
section~\ref{subsec:modelb-couplings}. When non-QCD potentials are
neglected, both axions are massless. Their variations obey
$\delta a_0-24\delta\theta_1-12\delta\theta_2=0$, as follows from
Eq.~\eqref{eq:modelb-physical-lagrangian}.
In particular, the direction
$(\delta a_0,\delta\theta_1,\delta\theta_2)\propto(0,1,-2)$
has neither a QCD nor a photon coupling. After canonical normalization,
we can therefore choose two orthogonal fields $a_\gamma$ and $a_{\rm d}$
such that only $a_\gamma$ couples to photons. Their couplings are
$|g_{a_\gamma\gamma\gamma}|=g_\perp$ and
$g_{a_{\rm d}\gamma\gamma}=0$, where $g_\perp$ is defined in
Eq.~\eqref{eq:alp-total-photon-coupling}. The vanishing photon
coupling does not exclude other interactions of $a_{\rm d}$.

Using $p_{K,\rm phys}=-(4/3)n_{K,\rm phys}$ and
$G_{\rm phys}=F_*^2\operatorname{diag}(s^{-2},\mathcal H_{\rm phys})$
in Eq.~\eqref{eq:alp-total-photon-coupling}, we obtain
\begin{align}
 g_\perp
 &=\frac{\alpha_{\rm em}}{2\pi F_*}
 \frac{\frac43\sqrt{n_{K,\rm phys}^T\mathcal H_{\rm phys}^{-1}n_{K,\rm phys}}}
 {\sqrt{1+n_{K,\rm phys}^T\mathcal H_{\rm phys}^{-1}n_{K,\rm phys}/s^2}}.
 \label{eq:modelb-alp-photon-general}
\end{align}
At $u=v=t$, Eq.~\eqref{eq:modelb-isotropic-metric} gives
$n_{K,\rm phys}^T\mathcal H_{\rm phys}^{-1}n_{K,\rm phys}=432t^2$. Hence
\begin{align}
 g_\perp
 &=\frac{\alpha_{\rm em}}{2\pi F_*}
 \frac{16\sqrt3\,t}
 {\sqrt{1+\dfrac{12g_s^4}{529t^4}}}.
 \label{eq:modelb-alp-photon-isotropic}
\end{align}
With $c_J=216t$ and the value of $\mu(J_1)$ in
Eq.~\eqref{eq:modelb-mu-upper}, its ratio to the geometric bound is
\begin{align}
 \frac{g_\perp}
 {\dfrac{\alpha_{\rm em}}{2\pi F_*}\mu(J_1)c_J}
\simeq\frac{0.153886}
 {\sqrt{1+\dfrac{12g_s^4}{529t^4}}}<0.154.
 \label{eq:modelb-alp-bound-comparison}
\end{align}
Thus the combined photon coupling of the two axions is less than
$15.4\%$ of the geometric upper bound on this symmetric family of HYM
solutions. 
The gray region in Figure~\ref{fig:alp-bound-fqcd-x} is obtained by mapping the symmetric HYM family $u=v=t$ to the $(f_{\rm QCD},x)$ plane, using Eqs.~\eqref{eq:qcd-decay-constant} and~\eqref{eq:modelb-general-bound-x}. For this illustration, we restrict to $t\geq5$ and $0<g_s\leq0.3$. These numerical cuts specify the range shown in the figure and are not fixed by the HYM equations. The marked points $P_1$ and $P_2$ correspond to the two choices $(t,g_s)=(5,0.3)$ and $(10,0.1)$ in Table~\ref{tab:modelb-general-bound}, respectively. The point $P_1$ maximizes $x$ within this range, while $P_2$ lies in the interior of the shaded region. At the point $P_1$, we find
$g_\perp\simeq5.9\times10^{-19}\,\mathrm{GeV}^{-1}$, whereas the
geometric upper bound is $3.8\times10^{-18}\,\mathrm{GeV}^{-1}$.
These values use $F_*^2=M_P^2/(8\pi^2)$,
$M_P=2.435\times10^{18}\,\mathrm{GeV}$, and
$\alpha_{\rm em}^{-1}=137.036$.
They compare the two coupling calculations at the same values of the
K\"ahler moduli and string coupling. Whether these parameter values
reproduce the measured gauge couplings is not examined here.

The same coefficient relation also connects the combined ALP coupling
to the QCD axion. Combining Eqs.~\eqref{eq:qcd-decay-constant},
\eqref{eq:modelb-anisotropic-ratio}, and
\eqref{eq:modelb-alp-photon-general} gives
\begin{align}
 g_\perp
 &=\frac{\alpha_{\rm em}}{2\pi f_{\rm QCD}}
 \sqrt{
 \left[\frac83-\left(\frac EN\right)_{\rm eff}\right]
 \left[\left(\frac EN\right)_{\rm eff}-\frac43\right]
 }.
 \label{eq:modelb-alp-qcd-relation}
\end{align}
This relation holds for the leading metric 
for all $u,v>0$ satisfying the HYM equations. The quantities $f_{\rm QCD}$ and
$(E/N)_{\rm eff}$ refer to the direction determined by the QCD
potential. No ALP mass enters this relation.

Additional potentials can give masses to the two QCD-neutral axions
and mix them. If they do not mix with the QCD axion, the photon
couplings of the two mass eigenstates can be written as
$g_{1\gamma\gamma}=g_\perp\cos\beta$ and
$g_{2\gamma\gamma}=g_\perp\sin\beta$, up to field-sign choices,
where $\beta$ is the mixing angle. The sum of the squared photon
couplings remains $g_\perp^2$, and the bound in
Eq.~\eqref{eq:modelb-alp-bound-comparison} continues to hold at
fixed moduli and fixed coupling coefficients. Determining the
individual masses and photon couplings requires the additional
potentials. 
If those potentials also mix them with the QCD axion, the mass eigenstates
can have nonzero QCD couplings and must be obtained from the full
mass matrix. The light axion direction can then differ from the
QCD-only direction, as discussed in
section~\ref{subsec:modelb-potentials}.

\subsubsection{Effects of additional axion potentials}
\label{subsec:modelb-potentials}

We next examine the axion mass and photon coupling in Model~B using
the general results of section~\ref{sec:mass-constraints}. We first identify
conditions under which non-QCD contributions do not change the leading
QCD-only result.

For Model~B, QCD dominance is an additional assumption because the
hidden $SO(16)_{\rm h}$ gauge factor is unbroken. At leading order, the trivial
hidden bundle leaves a massless $\mathbf{128}$ fermion. Its anomalous
phase rotation can remove the hidden vacuum angle if the fermion is
exactly massless and the rotation is not explicitly broken by other
interactions, as in the massless-fermion argument for a confining gauge
theory~\cite{Leutwyler:PhysRevD46}. The vanishing of a mass in the
leading spectrum does not establish these properties in the full
effective theory. We therefore assume that the hidden-sector potential
for the light direction of interest is subleading to QCD, rather than
deriving that assumption from the index calculation.

The positivity assumption in section~\ref{sec:mass-constraints}
was used for the mass-coupling bounds. For the following perturbative estimate, $M_{\rm other}^2$ need not be
positive semidefinite, but the full mass matrix must be stable.
Let us keep the leading physical metric $G_0$ fixed and use
$\nu=G_0^{-1/2}n_{\rm phys}$. We write
$\widehat u_0=\nu/\lVert\nu\rVert$ and
$m_0^2=\chi_{\rm QCD}\lVert\nu\rVert^2$. At a common CP-conserving vacuum,
a non-QCD Hessian with
$\varepsilon_V=\lVert M_{\rm other}^2\rVert_{\rm op}/m_0^2\ll1$
changes the QCD eigenvector by
\begin{align}
 \delta \widehat u_{\rm QCD}
 &=\frac{(\mathbf1-\widehat u_0\widehat u_0^T)M_{\rm other}^2\widehat u_0}{m_0^2}
   +O(\varepsilon_V^2).
 \label{eq:comparison-potential-direction-change}
\end{align}
For a normalized mass eigenvector $\widehat u$ with a nonzero QCD coupling, the
coefficient ratio is
\begin{align}
 \mathcal R_{\rm eig}(\widehat u)&=
 \frac{e_{\rm phys}^TG_0^{-1/2}\widehat u}
      {n_{\rm phys}^TG_0^{-1/2}\widehat u}.
 \label{eq:comparison-eigenstate-ratio}
\end{align}
Expanding this expression about $\widehat u=\widehat u_0$ gives, at first order for the
leading Model~B metric,
\begin{align}
 |\delta\mathcal R_V|
 &\le\frac{4\zeta_B}{3(1+\zeta_B^2)}\varepsilon_V.
 \label{eq:comparison-potential-precision}
\end{align}
Hence $\varepsilon_V\ll\zeta_B$ is a sufficient condition for this
change to be smaller than $|\Delta_0|$, not merely smaller than unity.
The condition is stronger than QCD dominance. It is unnecessary when
$(\mathbf1-\widehat u_0\widehat u_0^T)M_{\rm other}^2\widehat u_0=0$, since the direction then does
not rotate. A displacement of the vacuum away from the CP-conserving
point must be checked separately. 
If both internal combinations are instead heavy and decouple,
the remaining universal-axion ratio tends to $8/3$, as follows from
its QCD and electromagnetic coefficients.

Worldsheet instantons~\cite{Wen:1985jz,Svrcek:2006yi} must also be compared with the QCD scale of the
same background. Their exponential factors have the form $e^{-a_r^{\rm inst}t}$ when the ratios of the K\"ahler parameters are fixed. 
The index $r$ labels the instanton sector, and
$a_r^{\rm inst}t$ is its real, dimensionless classical action.
The coefficient $a_r^{\rm inst}>0$ is fixed by the wrapped
curve at the chosen K\"ahler space.
An asymptotically free color spectrum
with one-loop coefficient $b_3^{\rm RG}>0$ gives
$\Lambda_{\rm QCD}\sim M_{\rm c}e^{-2\pi s/b_3^{\rm RG}}$ up to threshold and
mass factors. Here $M_{\rm c}$ is the compactification scale. Since
$s=138t^3/g_s^2$ on the symmetric locus, large radius or weak string
coupling does not by itself establish QCD dominance. Fermion zero modes
or cancellations can remove an instanton term, but this requires a
calculation for the actual non-supersymmetric bundle.

Neglecting the non-QCD corrections discussed above,
$m_a=\sqrt{\chi_{\rm QCD}}/f_{\rm QCD}$ and
Eq.~\eqref{eq:qcd-photon-coupling} give
\begin{align}
 \frac{g_{a\gamma\gamma}}{m_a}
 =
 \frac{\alpha_{\rm em}}
      {2\pi\sqrt{\chi_{\rm QCD}}}
 \left[
 \frac{
 e_{\rm phys}^TG_{\rm phys}^{-1}n_{\rm phys}
 }{
 n_{\rm phys}^TG_{\rm phys}^{-1}n_{\rm phys}
 }
 -
 \mathcal C_{\rm QCD}
 \right].
 \label{eq:comparison-qcd-ratio}
\end{align}
Here $\chi_{\rm QCD}$ is the QCD topological susceptibility and
$\mathcal C_{\rm QCD}$ is the usual low-energy contribution from light
QCD degrees of freedom~\cite{GrillidiCortona:2015jxo}.
Experimental plots conventionally use $|g_{a\gamma\gamma}|$.

If a hidden-sector potential becomes important, the physical mass
eigenstates and their coupling ratios can differ from the QCD-only
result. Determining the size of this contribution for Model~B is left
for future work.

\section{Conclusions}
\label{sec:conclusions}

We have studied the photon couplings of QCD axions and
axion-like particles in non-supersymmetric
$SO(16)_{\rm v}\times SO(16)_{\rm h}$ heterotic line-bundle
compactifications. Direct flux breaking can produce independent
QCD and electromagnetic Green-Schwarz coupling vectors even when
all Standard Model gauge factors lie in the visible $SO(16)_{\rm v}$
with $k_3=k_2=1$ and $k_Y=5/3$. If this independence survives the
St\"uckelberg mechanism and low-energy matching, $E/N$ can depend
on the axion kinetic matrix. Proportional physical coupling vectors
instead fix the ratio in the QCD-dominated regime. We demonstrate
the independent coupling structure in a line-bundle background
with three generations of Standard Model fermions and no five-brane
sources.

For the direct Standard Model embedding, the HYM equations and
the Bianchi identity give the correlated geometric bound in
Eq.~\eqref{eq:general-bound-correlated-norm}. For backgrounds
without five-brane sources, this leads to the upper and lower
bounds on $E/N$ in Eq.~\eqref{eq:general-bound-W0-envelope}.
These bounds use the leading bulk kinetic metric and assume
that low-energy matching leaves the flux-derived coupling
vectors unchanged. Their size is controlled by the dimensionless geometric parameter $x=\mu(J)c_J/s$, which includes the geometric quantity $\mu(J)$, the pairing of the Kähler form with the second Chern class $c_J$, and the inverse gauge coupling $s$ at tree level. The allowed departure from $8/3$ is of order $x^2$ for $x\ll1$. We evaluate $\mu(J_0)$ and the resulting
$x_{\max}$ for several CICY geometries along
$J=\lambda J_0$. 
For the local NS5-brane contribution considered in
section~\ref{subsec:gs-ns5}, the source terms contribute to the
electromagnetic and color coefficients in the same $8/3$ ratio.
They therefore do not change the departure from the relation
$C_\gamma=(8/3)C_3$, although they modify the QCD coefficients and
introduce localized axions. 
The corresponding bounds depend on the assumptions about the kinetic
terms of the localized axions and their mixing with the bulk axions
described in section~\ref{subsec:general-bound-Wnonzero}.

We obtain two complementary bounds on axion-photon couplings. 
First,
when the axion combination in the low-energy photon interaction
is a linear combination of those appearing in the positive
quadratic terms of the potential, the assumptions of
section~\ref{subsec:general-mass-constraints} give an upper bound
on the photon coupling relative to the axion mass that is
independent of the axion kinetic metric. 
Second, the same flux constraints give a geometric upper bound
on the combined photon coupling of all QCD-neutral axions.
In the absence of five-brane sources, and under the same kinetic and low-energy matching assumptions, this bound is $g_\perp\leq\alpha_{\rm em}\mu(J)c_J/(2\pi F_*)$. 
It also implies
$g_\perp\leq\alpha_{\rm em}x/(2\pi f_{\rm QCD})$.
Thus the same parameter $x$ constrains both the departure of
$E/N$ from $8/3$ and the photon couplings of QCD-neutral ALPs.
These bounds do not require the potentials that generate the
ALP masses.

These general results are illustrated by the two explicit models.
Model~A gives $E/N=8/3$ when the charged-matter contributions preserve
the $SU(5)$ relation. Model~B instead has independent QCD and
electromagnetic Green-Schwarz couplings. Provided that low-energy
matching does not change these coefficients and that QCD gives the
dominant potential for the QCD axion, the leading kinetic metric gives
$(E/N)_{\rm eff}<8/3$, approaching $8/3$ at weak coupling and large
radius. The two coupling vectors remain linearly independent in this
limit. Corrections to this result can arise from the axion kinetic
terms and additional axion potentials, while comparison with measured
gauge couplings also requires renormalization-group evolution and
CP-even threshold corrections.

Model~B also provides an explicit application of the geometric bound
to QCD-neutral ALPs. The geometric upper bound applies to the square
root of the sum of their squared photon couplings, and the two
QCD-neutral axion directions in Model~B satisfy this bound on the
symmetric HYM locus in the leading kinetic approximation. The combined
coupling can also be expressed in terms of the QCD-axion decay constant
and $(E/N)_{\rm eff}$. 
Mixing between the two QCD-neutral directions changes the
individual photon couplings but leaves the sum of their
squares unchanged at fixed moduli; their masses depend on
additional non-QCD potentials. Model~B therefore illustrates both the departure of the QCD-axion coupling ratio from $8/3$ and the geometric constraint on photon
couplings of QCD-neutral ALPs within the same compactification. 
The comparison with supersymmetric $SO(32)$ compactifications
shows that flux-dependent nonuniversal axion couplings are not
specific to the absence of supersymmetry. The bounds derived
in this work follow from the gauge embedding, the flux
background, and the compactification conditions.

A systematic study of these axion-photon couplings over a broader set
of Calabi-Yau compactifications is left for future work.

\section*{Acknowledgements}

H.O. was supported by JSPS KAKENHI Grant Numbers JP25H01539 and JP26K07087. 

\appendix

\section{Reduction of the local NS5-brane anomaly-inflow terms}
\label{app:ns5-reduction}

We derive the gauge couplings in section~\ref{subsec:gs-ns5} using the
local tensor sector described in 
Refs.~\cite{Basile:2023knk,Basile:2025mnj}. 
We use the local
anomaly couplings of this proposal, not a microscopic determination of
the full five-brane spectrum. The gravitational and normal-bundle terms
do not enter the visible gauge coefficients calculated here.

In the characteristic-class convention of Refs.~\cite{Basile:2023knk,Basile:2025mnj},
\begin{align}
 \mathfrak c_a&=-\frac{1}{2(2\pi)^2}\operatorname{tr}_{16}F_a^2,
 \qquad a={\rm v},{\rm h},
 \label{eq:app-ns5-chern-convention}
\end{align}
so $\mathfrak c_a|_X=-S_a$. The Bianchi four-form is
\begin{align}
 X_4&=\frac12\left[p_1+\mathfrak c_{\rm v}+\mathfrak c_{\rm h}\right],
 \label{eq:app-ns5-bianchi-class}
\end{align}
where $p_1$ is the first Pontryagin class of the
ten-dimensional tangent bundle.
Since $p_1(TX)=-2c_2(TX)$, $[X_4|_X]=-[W]$. The eight-form in the
anomaly convention is related to the polynomial of the main text by
\begin{align}
 Y_8&=-\frac{\mathscr X_8}{24(2\pi)^4}.
 \label{eq:app-ns5-polynomial-dictionary}
\end{align}
The proposed NS5-brane worldvolume spectrum, including its
chiral tensor, has anomaly polynomial $I_8^{\rm chiral}$.
The anomaly factorization, with the normal-bundle terms omitted, is
\begin{align}
 I_8^{\rm chiral}-Y_8&=X_4Y_4+\frac12 Z_4^2,
 \label{eq:app-ns5-factorization}
\end{align}
where
\begin{align}
 Y_4=\frac{\mathfrak c_{\rm v}+\mathfrak c_{\rm h}+3p_1}{48},
\qquad
 Z_4=\frac{\mathfrak c_{\rm v}-\mathfrak c_{\rm h}}4.
 \label{eq:app-ns5-z4}
\end{align}
We choose the orientation and tensor-axion sign to agree with
$\delta\Theta=Q\boldsymbol\lambda$ in the main text.

The first term in Eq.~\eqref{eq:app-ns5-factorization} is cancelled by
\begin{align}
 S_{{\rm loc},\ell}&=-\frac{1}{2\pi\alpha'}
 \int_{M_4\times C_{\ell}}B_2\wedge Y_4.
 \label{eq:app-ns5-bulk-counterterm}
\end{align}
This normalization agrees with the bulk term
$-\int B_2\wedge Y_8/(2\pi\alpha')$.
Let $\int_{C_{\ell}}\eta_{\ell}=1$ and $m_{\ell I}=\int_{C_{\ell}}f_I$. The required
components are
\begin{align}
 B_2|_{M_4\times C_{\ell}}
 &=B_2^{(4)}+2\pi\alpha'W_{\ell,i}b^i\eta_{\ell},
 \label{eq:app-ns5-bulk-pullback}
\end{align}
and
\begin{align}
 \left.\mathfrak c_a\right|_{\mathcal F^I}
 &=-\frac{2}{2\pi}f_I\wedge\mathcal F^I,
 \qquad I\in a.
 \label{eq:app-ns5-linear-chern}
\end{align}
Here $\mathcal F^I$ and $\mathcal F_a$ are the four-dimensional field
strengths before canonical normalization. Integration over $C_{\ell}$ gives
\begin{align}
 S_{{\rm loc},\ell}^{(4)}\supset{}&
 \int_{M_4}\frac{W_{\ell,i}b^i}{96(2\pi)^2}
 \left(\operatorname{tr}_{16}\mathcal F_{\rm v}^2
       +\operatorname{tr}_{16}\mathcal F_{\rm h}^2\right)
+\frac{1}{(2\pi)^2\alpha'}\int_{M_4}B_2^{(4)}\wedge
       \sum_I\frac{m_{\ell I}}{24}\mathcal F^I.
 \label{eq:app-ns5-bulk-reduction}
\end{align}
For the level-one color factor, the conversion to canonical fields is
\begin{align}
 \frac{1}{4(2\pi)^2}
 \left.\operatorname{tr}_{16}\mathcal F_{\rm v}^2\right|_{SU(3)_C}
 &=\frac{\alpha_s}{8\pi}
 G^b_{\mu\nu}\widetilde G^{b\mu\nu}\,d^4x.
 \label{eq:app-ns5-density-normalization}
\end{align}
Thus the first term in Eq.~\eqref{eq:app-ns5-bulk-reduction} gives
\begin{align}
 \delta C_{i3}&=\sum_{\ell}\frac{W_{\ell,i}}{24}=\frac{W_i}{24}.
 \label{eq:app-ns5-color-shift}
\end{align}
The second term has the same two-form normalization as the bulk term
linear in $\mathcal F^I$. Dualizing $B_2^{(4)}$ to $a_0$ gives
\begin{align}
 \delta Q_0{}^I&=\frac1{24}\sum_{\ell} m_{\ell I}.
 \label{eq:app-ns5-universal-shift}
\end{align}

For the chiral tensor, let the worldvolume two-form have internal
component $\sigma_{\ell}\eta_{\ell}$ and field strength satisfying
$d\mathfrak h_{3,\ell}=2\pi Z_4$. The part with two spacetime indices and
two indices on $C_{\ell}$ is
\begin{align}
 \int_{C_{\ell}}Z_{4,(2,2)}
 &=-\frac{1}{2(2\pi)}\left(
 \sum_{I\in{\rm v}}m_{\ell I}\mathcal F^I
 -\sum_{I\in{\rm h}}m_{\ell I}\mathcal F^I\right).
 \label{eq:app-ns5-tensor-mixed}
\end{align}
It gives
\begin{align}
 D\sigma_{\ell}&=d\sigma_{\ell}
 -\frac12\sum_{I\in{\rm v}}m_{\ell I}A^I
 +\frac12\sum_{I\in{\rm h}}m_{\ell I}A^I,
 \label{eq:app-ns5-tensor-derivative}
\end{align}
which reproduces the tensor charges in
Eq.~\eqref{eq:ns5-charge-shifts}. 
The first and second entries of each subscript give the form
degrees on $M_4$ and $C_{\ell}$, respectively. They do not denote
Hodge types. 
The six-dimensional self-duality condition identifies the external
two-form with the dual of $\sigma_{\ell}$; 
they describe the same four-dimensional degree of freedom. In the reduced anomaly,
\begin{align}
 \int_{C_{\ell}}\frac12 Z_4^2
 &=Z_{4,(4,0)}\wedge\int_{C_{\ell}}Z_{4,(2,2)},
 \label{eq:app-ns5-tensor-anomaly-reduction}
\end{align}
the two equal cross terms in $Z_4^2$ cancel the factor $1/2$. The scalar
interaction in this self-dual tensor convention is
\begin{align}
 S_{\sigma_{\ell}}^{(4)}
 &=-\int_{M_4}\sigma_{\ell} Z_{4,(4,0)}
 =\int_{M_4}\frac{\sigma_{\ell}}{8(2\pi)^2}
 \left(\operatorname{tr}_{16}\mathcal F_{\rm v}^2
       -\operatorname{tr}_{16}\mathcal F_{\rm h}^2\right).
 \label{eq:app-ns5-tensor-interaction}
\end{align}
Comparison with Eq.~\eqref{eq:app-ns5-density-normalization} yields
$C_{\sigma_{\ell}3}=1/2$. Counting an additional independent two-form, or
omitting the self-duality condition, would not give this normalization.
For a general Cartan generator the two contributions are
\begin{align}
 \delta C_i(T)&=\frac{W_i}{24}[k_{\rm v}(T)+k_{\rm h}(T)],
 \label{eq:app-ns5-general-bulk}
\end{align}
and
\begin{align}
 C_{\sigma_{\ell}}(T)&=\frac12[k_{\rm v}(T)-k_{\rm h}(T)].
 \label{eq:app-ns5-general-tensor}
\end{align}
For the visible Standard Model embedding this gives
$\delta P_i=P_{\sigma_{\ell}}=0$. The derivation fixes these local anomaly
couplings, not the global tensor period or the source kinetic matrix.

\bibliographystyle{JHEP}
\bibliography{references}
\end{document}